\documentclass[aps,prc,twocolumn,superscriptaddress,showpacs,preprintnumbers,amsmath,amssymb,floatfix]{revtex4}

\usepackage{graphicx}
\usepackage{dcolumn}
\usepackage{bm}
\usepackage{longtable}
\usepackage{color}

\begin{document}

\preprint{Jefferson Lab Experiment E89-003 Summary Article}

\title{The dynamics of the quasielastic $\bm{^{16}}$O$\bm{(e,e^{\prime}p)}$ reaction at $\bm{Q^{2}}$ $\approx$ 0.8 (GeV/$\bm{c}$)$\bm{^{2}}$}

\author{K.G. Fissum}
\altaffiliation{Corresponding author; \texttt{kevin.fissum@nuclear.lu.se}}
\affiliation{Massachusetts Institute of Technology, Cambridge, Massachusetts, 02139, USA}
\affiliation{University of Lund, Box 118, SE-221 00 Lund, Sweden}
\author{M. Liang}
\affiliation{Thomas Jefferson National Accelerator Facility, Newport News, Virginia, 23606, USA}
\author{B.D. Anderson}
\affiliation{Kent State University, Kent, Ohio, 44242, USA}
\author{K.A. Aniol}
\affiliation{California State University Los Angeles, Los Angeles, California, 90032, USA}
\author{L. Auerbach}
\affiliation{Temple University, Philadelphia, Pennsylvania, 19122, USA}
\author{F.T. Baker}
\affiliation{University of Georgia, Athens, Georgia, 30602, USA}
\author{J. Berthot}
\affiliation{IN2P3, F-63177 Aubi\`ere, France}
\author{W. Bertozzi}
\affiliation{Massachusetts Institute of Technology, Cambridge, Massachusetts, 02139, USA}
\author{P.-Y. Bertin}
\affiliation{IN2P3, F-63177 Aubi\`ere, France}
\author{L. Bimbot}
\affiliation{Institut de Physique Nucl\'eaire, F-91406 Orsay, France}
% ...

\collaboration{The Jefferson Lab Hall A Collaboration}
\noaffiliation

\author{J.M. Ud\'{\i}as and J.R. Vignote\\{\it Universidad Complutense de Madrid, E-28040 Madrid, Spain}\\}
\noaffiliation

\author{J. Ryckebusch and D. Debruyne\\{\it Ghent University, B-9000 Ghent, Belgium}\vspace*{4mm}}
\noaffiliation

\date{\today}

\begin{abstract}
The physics program in Hall A at Jefferson Lab commenced in the summer of 1997 
with a detailed investigation of the $^{16}$O$(e,e^{\prime}p)$ reaction in 
quasielastic, constant $(q,\omega)$ kinematics at $Q^{2}$ $\approx$ 0.8 
(GeV/$c$)$^{2}$, $q \approx$ 1 GeV/$c$, and $\omega \approx$ 445 MeV.  Use of a
self-calibrating, self-normalizing, thin-film waterfall target enabled a 
systematically rigorous measurement.  Five-fold differential cross-section data
for the removal of protons from the $1p$-shell have been obtained for 0 $<$ 
$p_{\rm miss}$ $<$ 350 MeV/$c$.  Six-fold differential cross-section data for 
0 $<$ $E_{\rm miss}$ $<$ 120 MeV were obtained for 0 $<$ $p_{\rm miss}$ $<$ 340
MeV/$c$.  These results have been used to extract the $A_{LT}$ asymmetry and 
the $R_{L}$, $R_{T}$, $R_{LT}$, and $R_{L+TT}$ effective response functions 
over a large range of $E_{\rm miss}$ and $p_{\rm miss}$.  Detailed comparisons 
of the $1p$-shell data with Relativistic Distorted-Wave Impulse Approximation 
(\textsc{rdwia}), Relativistic Optical-Model Eikonal Approximation 
(\textsc{romea}), and Relativistic Multiple-Scattering Glauber Approximation 
(\textsc{rmsga}) calculations indicate that two-body currents stemming from 
Meson-Exchange Currents (MEC) and Isobar Currents (IC) are not needed to 
explain the data at this $Q^{2}$.  Further, dynamical relativistic effects are 
strongly indicated by the observed structure in $A_{LT}$ at $p_{\rm miss}$ 
$\approx$ 300 MeV/$c$.  For 25 $<$ $E_{\rm miss}$ $<$ 50 MeV and $p_{\rm miss}$
$\approx$ 50 MeV/$c$, proton knockout from the $1s_{1/2}$-state dominates, and 
\textsc{romea} calculations do an excellent job of explaining the data.  
However, as $p_{\rm miss}$ increases, the single-particle behavior of the 
reaction is increasingly hidden by more complicated processes, and for 280 $<$ 
$p_{\rm miss}$ $<$ 340 MeV/$c$, \textsc{romea} calculations together with 
two-body currents stemming from MEC and IC account for the shape and transverse
nature of the data, but only about half the magnitude of the measured cross 
section.  For 50 $<$ $E_{\rm miss}$ $<$ 120 MeV and 145 $<$ $p_{\rm miss}$ $<$ 
340 MeV/$c$, $(e,e^{\prime}pN)$ calculations which include the contributions of
central and tensor correlations (two-nucleon correlations) together with MEC 
and IC (two-nucleon currents) account for only about half of the measured cross
section.  The kinematic consistency of the $1p$-shell normalization factors 
extracted from these data with respect to all available 
$^{16}$O$(e,e^{\prime}p)$ data is also examined in detail.  Finally, the 
$Q^{2}$-dependence of the normalization factors is discussed.

\end{abstract}


\maketitle

\section{\label{sec:intro}Introduction}

Exclusive and semi-exclusive $(e,e^{\prime}p)$ in quasielastic (QE) kinematics 
\footnote{
Kinematically, an electron scattered through angle $\theta_e$ transfers 
momentum $\bm q$ and energy $\omega$ with $Q^2 = \bm q\hspace*{0.5mm}^2 - 
\omega^2$.  The ejected proton has mass $m_p$, momentum $\bm{{p}_p}$, energy 
$E_p$, and kinetic energy $T_p$.  In QE kinematics, $\omega \approx 
Q^{2}/2m_{p}$.  The cross section is typically measured as a function of 
missing energy $E_{\rm miss} = \omega - T_p - T_{B}$ and missing momentum 
$p_{\rm miss} = \vert \bm q - \bm{p_p} \vert$.  $T_{B}$ is the kinetic energy 
of the residual nucleus.  The lab polar angle between the ejected proton and 
virtual photon is $\theta_{pq}$ and the azimuthal angle is $\phi$.  
$\theta_{pq} > 0^\circ$ corresponds to $\phi = 180^\circ$, $\theta_p > 
\theta_q$, and $+p_{\rm miss}$.  $\theta_{pq} < 0^\circ$ corresponds to 
$\phi = 0^\circ$, $\theta_p$ $<$ $\theta_q$, and $-p_{\rm miss}$.
}
has long been used as a precision tool for the study of nuclear electromagnetic
responses (see Refs. \cite{frullani84,kelly96,boffi96,kelly97}).  Cross-section
data have provided information used to study the single-nucleon aspects of 
nuclear structure and the momentum distributions of protons bound inside the 
nucleus, as well as to search for non-nucleonic degrees of freedom and to 
stringently test nuclear theories.  Effective response-function separations 
\footnote{
In the One-Photon Exchange Approximation, the unpolarized $(e,e^{\prime}p)$ 
cross section can be expressed as the sum of four independent response 
functions:  $R_{L}$ (longitudinal), $R_{T}$ (transverse), $R_{LT}$ 
(longitudinal-transverse interference, and $R_{TT}$ (transverse-transverse 
interference).  See also Eq. (\ref{equation:sixfold}).
} 
have been used to extract detailed information about the different reaction 
mechanisms contributing to the cross section since they are selectively 
sensitive to different aspects of the nuclear current.  

Some of the first $(e,e^{\prime}p)$ energy- and momentum-distribution 
measurements were made by Amaldi {\it et al.} \cite{amaldi67}.  These results, 
and those which followed (see Refs. \cite{mougey76,frullani84,kelly96}), were 
interpreted within the framework of single-particle knockout from nuclear 
valence states, even though the measured cross-section data was as much as 40\% 
lower than predicted by the models of the time.  The first relativistic 
calculations for $(e,e^{\prime}p)$ bound-state proton knockout were performed 
by Picklesimer, Van Orden, and Wallace 
\cite{picklesimer85,picklesimer87,picklesimer89}.  Such Relativistic 
Distorted-Wave Impulse Approximation (\textsc{rdwia}) calculations are 
generally expected to be more accurate at higher $Q^{2}$, since QE 
$(e,e^{\prime}p)$ is expected to be dominated by single-particle interactions 
in this regime of four-momentum transfer.  

Other aspects of the structure as well as of the reaction mechanism have 
generally been studied at higher missing energy ($E_{\rm miss}$).  While it is 
experimentally convenient to perform measurements spanning the valence-state 
knockout and higher $E_{\rm miss}$ excitation regions simultaneously, there is 
as of yet no rigorous, coherent theoretical picture that uniformly explains the
data for all $E_{\rm miss}$ and all missing momentum ($p_{\rm miss}$).  In the 
past, the theoretical tools used to describe the two energy regimes have been 
somewhat different.  M\"{u}ther and Dickhoff \cite{muther94} suggest that the 
regions are related mainly by the transfer of strength from the valence states 
to higher $E_{\rm miss}$.

The nucleus $\rm ^{16}O$ has long been a favorite of theorists, since it has a 
doubly closed shell whose structure is thus easier to model than other nuclei.
It is also a convenient target for experimentalists.  While the knockout of 
$1p$-shell protons from $\rm ^{16}O$ has been studied extensively in the past 
at lower $Q^{2}$, few data were available at higher $E_{\rm miss}$ for any 
$Q^{2}$ in 1989, when this experiment was first conceived.

\subsection{$\bm{1p}$-shell knockout}

The knockout of $1p$-shell protons in $^{16}$O$(e,e^{\prime}p)$ was studied by 
Bernheim {\it et al.} \cite{bernheim82} and Chinitz {\it et al.} 
\cite{chinitz91} at Saclay, Spaltro {\it et al.} \cite{spaltro93} and Leuschner
{\it et al.} \cite{leuschner94} at NIKHEF, and Blomqvist {\it et al.} 
\cite{blomqvist95} at Mainz at $Q^{2}$ $<$ 0.4 (GeV/$c$)$^{2}$.  In these 
experiments, cross-section data for the lowest-lying fragments of each shell 
were measured as a function of $p_{\rm miss}$, and normalization factors 
(relating how much lower the measured cross-section data were than predicted) 
were extracted.  These published normalization factors ranged between 0.5 and 
0.7, but Kelly \cite{kelly96,kelly97} has since demonstrated that the Mainz 
data suggest a significantly smaller normalization factor (see also Table 
\ref{table:kellyspecs}).  

Several calculations exist (see Refs. 
\cite{udias93,hummel94,udias95,caballero98a,udias99,udias00}) which demonstrate
the sensitivity 
\footnote{
In the nonrelativistic Plane-Wave Impulse Approximation, the transverse 
amplitude in the $R_{LT}$ response is uniquely determined by the convection 
current.  At higher $Q^{2}$, it is well-known that the convection current 
yields small matrix elements.  As a result, the nonrelativistic Impulse 
Approximation (\textsc{ia}) contributions which dominate $R_{L}$ and $R_{T}$ 
are suppressed in $R_{LT}$ (and thus $A_{LT}$).  Hence, these observables are 
particularly sensitive to any mechanisms beyond the \textsc{ia}, such as 
channel coupling and relativistic and two-body current mechanisms 
\cite{ryckebuschtech01}.
}
of the longitudinal-transverse interference response function $R_{LT}$ and the 
corresponding left-right asymmetry $A_{LT}$
\footnote{
$A_{LT} \equiv \frac
{\sigma(\phi=0^{\circ})-\sigma(\phi=180^{\circ})}
{\sigma(\phi=0^{\circ})+\sigma(\phi=180^{\circ})}$.  $A_{LT}$ is a particularly
useful quantity for experimentalists because it is systematically much less 
challenging to extract than either an absolute cross section or an effective
response function.
}
to `spinor distortion' (see Section \ref{sec:RDWIA-formalism}), especially for 
the removal of bound-state protons.  Such calculations predict that proper 
inclusion of these dynamical relativistic effects is needed to simultaneously 
reproduce the cross-section data, $A_{LT}$, and $R_{LT}$.

\begin{figure}
\resizebox{0.47\textwidth}{!}{\includegraphics{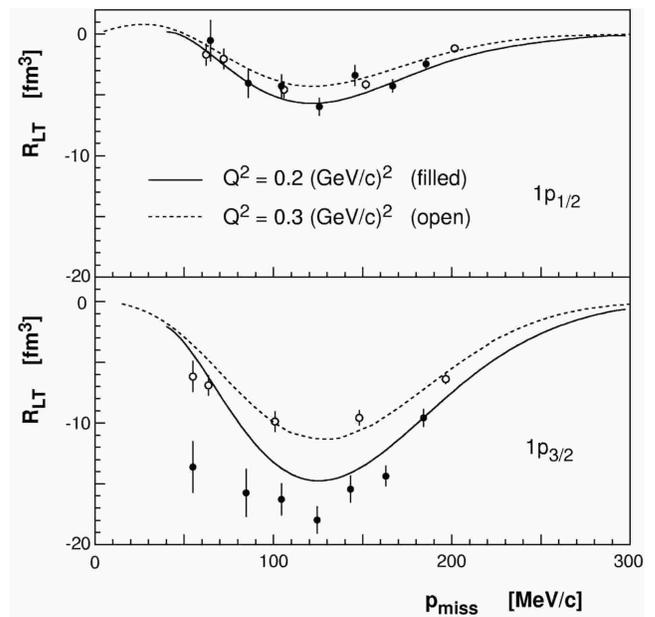}}
\caption{\label{fig:chinspal}
Longitudinal-transverse interference effective responses $R_{LT}$ as a function
of $p_{\rm miss}$ for the removal of protons from the $1p$-shell of 
$\rm ^{16}O$.  The open and filled circles were extracted from QE data obtained
by Chinitz {\it et al.} at $Q^{2}$ = 0.3 (GeV/$c$)$^{2}$ and Spaltro {\it et 
al.} at $Q^{2}$ = 0.2 (GeV/$c$)$^{2}$, respectively.  The dashed ($Q^{2}$ $=$ 
0.3 (GeV/$c$)$^{2}$) and solid ($Q^{2}$ $=$ 0.2 (GeV/$c$)$^{2}$) curves are 
modern \textsc{rdwia} calculations.  Overall, agreement is good, and improves 
with increasing $Q^{2}$.
}
\end{figure}

Fig. \ref{fig:chinspal} shows the effective response $R_{LT}$ as a function of 
$p_{\rm miss}$ for the removal of protons from the $1p$-shell of $\rm ^{16}O$ 
for the QE data obtained by Chinitz {\it et al.} at $Q^{2}$ = 0.3 
(GeV/$c$)$^{2}$ (open circles) and Spaltro {\it et al.} at $Q^{2}$ = 0.2 
(GeV/$c$)$^{2}$ (solid circles) together with modern \textsc{rdwia} calculations
(see Sections \ref{sec:theory} and \ref{sec:results_this_work} for a complete 
discussion of the calculations).  The solid lines correspond to the 0.2 
(GeV/$c$)$^{2}$ data, while the dashed lines correspond to the $Q^{2}$ = 0.3 
(GeV/$c$)$^{2}$ data.  Overall, agreement is good, and as anticipated, improves
with increasing $Q^{2}$.

\subsection{Higher missing energies}

Few data are available for $^{16}$O$(e,e^{\prime}p)$ at higher $E_{\rm miss}$, 
and much of what is known about this excitation region is from studies of other
nuclei such as $^{12}$C.  At MIT-Bates, a series of $^{12}$C$(e,e^{\prime}p)$ 
experiments have been performed at missing energies above the two-nucleon 
emission threshold (see Refs. 
\cite{lourie86,ulmer87,weinstein90,holtrop98,morrison99}).  The resulting 
cross-section data were much larger than the predictions of single-particle 
knockout models
\footnote{
$1s$-shell nucleons are generally knocked out from high-density regions of the 
target nucleus.  In these high-density regions, the \textsc{ia} is expected to 
be less valid than for knockout from the valence $p$-shell states lying near 
the surface.  In this region of `less-valid' \textsc{ia}, sizeable 
contributions to the $s$-shell cross-section data arise from two-nucleon 
current contributions stemming from MEC and IC.  In addition to affecting the 
single-nucleon knockout cross section, the two-nucleon currents can result in 
substantial multi-nucleon knockout contributions to the higher $E_{\rm miss}$ 
continuum cross section \cite{ryckebuschtech01}.
}.  
In particular, Ulmer {\it et al.} \cite{ulmer87} identified a marked increase 
in the transverse-longitudinal difference $S_{T} - S_{L}$ 
\footnote{
The transverse-longitudinal difference is $S_{T} - S_{L}$, where $S_{X} = 
\sigma_{\rm Mott}V_{X}R_{X}/\sigma^{X}_{ep}$, and $X~\epsilon~\{T,L\}$.  
$\sigma^{X}_{ep}$ represents components of the off-shell $ep$ cross section and 
may be calculated using the \textsc{cc1}, \textsc{cc2}, or \textsc{cc3} 
prescriptions of de Forest \cite{deforest83}.
}.
A similar increase has subsequently been observed by Lanen {\it et al.} for 
$^{6}$Li \cite{lanen90}, by van der Steenhoven {\it et al.} for $^{12}$C 
\cite{steenhoven88}, and most recently by Dutta {\it et al.} for $^{12}$C 
\cite{dutta00}, $^{56}$Fe, and $^{197}$Au \cite{dutta99}.  The transverse 
increase exists over a large range of four-momentum transfers, though the 
excess at lower $p_{\rm miss}$ seems to decrease with increasing $Q^{2}$.  
Theoretical attempts by Takaki \cite{takaki89}, the Ghent Group 
\cite{ryckebusch97}, and Gil {\it et al.} \cite{gil97} to explain the data at 
high $E_{\rm miss}$ using two-body knockout models coupled to Final-State 
Interactions (FSI) have not succeeded.  Even for QE kinematics, this transverse 
increase which starts at the two-nucleon knockout threshold seems to be a 
strong signature of multinucleon currents.

\section{\label{sec:expt}Experiment}

\begin{figure}
\resizebox{0.45\textwidth}{!}{\includegraphics{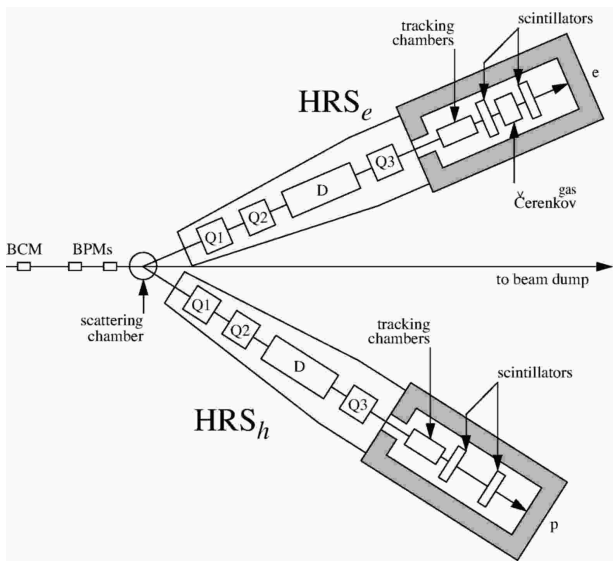}}
\caption{\label{fig:halla}
The experimental infrastructure in Hall A at Jefferson Lab at the time of this 
experiment.  The electron beam passed through a beam-current monitor (BCM) and 
beam-position monitors (BPMs) before striking a waterfall target located in the
scattering chamber. Scattered electrons were detected in the HRS$_{e}$, while 
knocked-out protons were detected in the HRS$_{h}$.  Non-interacting electrons
were dumped. The spectrometers could be rotated about the central pivot.
}
\end{figure}

This experiment \cite{proposal89,fissum97}, first proposed by Bertozzi 
{\it et al.} in 1989, was the inaugural physics investigation performed in Hall
A \cite{halla00} (the High Resolution Spectrometer Hall) at the Thomas 
Jefferson National Accelerator Facility (JLab) \cite{jlab00}.  An overview of 
the apparatus in the Hall at the time of this measurement is shown in Fig. 
\ref{fig:halla}.  For a thorough discussion of the experimental infrastructure 
and its capabilities, the interested reader is directed to the paper by Alcorn 
{\it et al.} \cite{hallanim}.  For the sake of completeness, a subset of the 
aforementioned information is presented here.

\subsection{\label{subsec:beam}Electron beam}

Unpolarized 70 $\mu$A continuous electron beams with energies of 0.843, 1.643,
and 2.442 GeV (corresponding to the virtual photon polarizations shown in Table
\ref{table:kinematics}) were used for this experiment.  Subsequent analysis of 
the data demonstrated that the actual beam energies were within 0.3\% of the 
nominal values \cite{gao98}.
\begin{table}
\caption{\label{table:kinematics}
The QE, constant $(q,\omega)$ kinematics employed in this measurement.  At each
beam energy, $q$ $\approx$ 1 GeV/$c$.}
\begin{ruledtabular}
\begin{tabular}{rrrr}
$E_{\rm beam}$ & $\theta_{e}$ & virtual photon &                        $\theta_{pq}$ \\
         (GeV) & ($^{\circ}$) &   polarization &                         ($^{\circ}$) \\
\hline
         0.843 &       100.76 &          0.21  &                             0, 8, 16 \\
         1.643 &        37.17 &          0.78  &                            0, $\pm$8 \\
         2.442 &        23.36 &          0.90  & 0,$\pm$2.5, $\pm$8, $\pm$16, $\pm$20 \\
\end{tabular}
\end{ruledtabular}
\end{table}
The typical laboratory $\pm$4$\sigma$ beam envelope at the target was 0.5 mm 
(horizontal) by 0.1 mm (vertical).  Beam-current monitors \cite{bcms00} 
(calibrated using an Unser monitor \cite{unser81}) were used to determine the 
total charge delivered to the target to an accuracy of 2\% \cite{ulmer98}.  
Beam-position monitors (BPMs) \cite {bpmsa,bpmsb} were used to ensure the 
location of the beam at the target was no more than 0.2 mm from the beamline 
axis, and that the instantaneous angle between the beam and the beamline axis 
was no larger than 0.15 mrad.  The readout from the BCM and BPMs was 
continuously passed into the data stream \cite{epics00}.  Non-interacting 
electrons were dumped in a well-shielded, high-power beam dump \cite{dump00} 
located roughly 30 m from the target.

\subsection{\label{subsec:target}Target}

A waterfall target \cite{garibaldi92} positioned inside a scattering chamber 
located at the center of the Hall provided the H$_{2}$O used for this study 
of $^{16}$O.  The target canister was a rectangular box 20 cm long $\times$ 15 
cm wide $\times$ 10 cm high containing air at atmospheric pressure.  The beam 
entrance and exit windows to this canister were respectively 50 $\mu$m and 75 
$\mu$m gold-plated beryllium foils.  Inside the canister, three thin, parallel, 
flowing water films served as targets.  This three-film configuration was 
superior to a single film 3$\times$ thicker because it reduced the 
target-associated multiple scattering and energy loss for particles originating
in the first two films and it allowed for the determination of the film in 
which the scattering vertex was located, thereby facilitating a better overall 
correction for energy loss.  The films were defined by 2 mm $\times$ 2 mm 
stainless-steel posts.  Each film was separated by 25 mm along the direction of
the beam, and was rotated beam right such that the normal to the film surface 
made an angle of 30$^{\circ}$ with respect to the beam direction.  This 
geometry ensured that particles originating from any given film would not 
intersect any other film on their way into the spectrometers.  

The thickness of the films could be changed by varying the speed of the water 
flow through the target loop via a pump.  The average film thicknesses were 
fixed at (130 $\pm$ 2.5\%) mg/cm$^{2}$ along the direction of the beam 
throughout the experiment, which provided a good trade-off between resolution 
and target thickness.  The thickness of the central water film was determined 
by comparing $^{16}$O$(e,e^{\prime})$ cross-section data measured at $q$ 
$\approx$ 330 MeV/$c$ obtained from both the film and a (155 $\pm$ 1.5\%) 
mg/cm$^{2}$ BeO target foil placed in a solid-target ladder mounted beneath the
target canister.  The thicknesses of the side films were determined by 
comparing the concurrently measured $^{1}$H$(e,e)$ cross section obtained from 
these side films to that obtained from the central film.  Instantaneous 
variations in the target-film thicknesses were monitored throughout the entire 
experiment by continuously measuring the $^{1}$H$(e,e)$ cross section.  

\subsection{\label{subsec:hrsdet}Spectrometers and detectors}

The base apparatus used in the experiment was a pair of optically identical 
4 GeV/$c$ superconducting High Resolution Spectrometers (HRS) \cite{hrs00}.
These spectrometers have a nominal 9\% momentum bite and a FWHM momentum 
resolution $\Delta p / p$ of roughly 10$^{-4}$.  The nominal laboratory 
angular acceptance is $\pm$25 mrad (horizontal) by $\pm$50 mrad (vertical).  
Scattered electrons were detected in the Electron Spectrometer (HRS$_{e}$), 
and knocked-out protons were detected in the Hadron Spectrometer (HRS$_{h}$) 
(see Fig. \ref{fig:halla}).  Before the experiment, the absolute momentum 
calibration of the spectrometers was determined to $\Delta p / p$ = 1.5 
$\times$ 10$^{-3}$ \cite{gao98}.  Before and during the experiment, both the 
optical properties and acceptances of the spectrometers were studied 
\cite{liyanage98}.  Some optical parameters are presented in Table 
\ref{table:optics}.
\begin{table}
\caption{\label{table:optics}
Selected results from the optics commissioning.}
\begin{ruledtabular}
\begin{tabular}{rrr}
                   &             resolution &               reconstruction \\
         parameter &                 (FWHM) &                     accuracy \\
\hline
out-of-plane angle &              6.00 mrad &               $\pm$0.60 mrad \\
    in-plane angle &              2.30 mrad &               $\pm$0.23 mrad \\
  $y_{\rm target}$ &                2.00 mm &                 $\pm$0.20 mm \\
    $\Delta p / p$ & 2.5 $\times$ 10$^{-4}$ &                            - \\
\end{tabular}
\end{ruledtabular}
\end{table}
During the experiment, the locations of the spectrometers were surveyed to an 
accuracy of 0.3 mrad at every angular location \cite{liang99029}.  The status 
of the magnets was continuously monitored and logged \cite{epics00}. 

The detector packages were located in well-shielded detector huts built on 
decks located above each spectrometer (approximately 25 m from the target and 
15 m above the floor of the Hall).  The bulk of the instrumentation electronics
was also located in these huts, and operated remotely from the Counting House.
The HRS$_{e}$ detector package consisted of a pair of thin scintillator planes 
\cite{trig00} used to create triggers, a Vertical Drift Chamber (VDC) package 
\cite{fissum01a,fissum01b} used for particle tracking, and a Gas \v{C}erenkov 
counter \cite{iodice98} used to distinguish between $\pi^{-}$ and electron 
events.  Identical elements, except for the Gas \v{C}erenkov counter, were also
present in the HRS$_{h}$ detector package.  The status of the various detector 
subsystems was continuously monitored and logged \cite{epics00}.  The 
individual operating efficiencies of each of these three devices was $>$99\%.  

\subsection{\label{subsec:daq}Electronics and data acquisition}

For a given spectrometer, a coincidence between signals from the two 
trigger-scintillator planes indicated a `single-arm' event.  Simultaneous 
HRS$_{e}$ and HRS$_{h}$ singles events were recorded as `coincidence' events.  
The basic trigger logic \cite{daq00} allowed a prescaled fraction of single-arm
events to be written to the data stream.  Enough HRS$_{e}$ singles were taken 
for a 1\% statistics $^{1}$H$(e,e)$ cross-section measurement at each 
kinematics.  Each spectrometer had its own VME crate (for scalers) and FASTBUS 
crate (for ADCs and TDCs).  The crates were managed by readout controllers 
(ROCs).  In addition to overseeing the state of the run, a trigger supervisor 
(TS) generated the triggers which caused the ROCs to read out the crates on an 
event-by-event basis.  The VME (scaler) crate was also read out every ten 
seconds.  An event builder (EB) collected the resulting data shards into 
events.  An analyzer/data distributer (ANA/DD) analyzed and/or sent these 
events to the disk of the data-acquisition computer.  The entire 
data-acquisition system was managed using the software toolkit \texttt{CODA} 
\cite{coda14}.  

Typical scaler events were about 0.5 kb in length.  Typical single-arm events 
were also about 0.5 kb, while typical coincidence events were about 1.0 kb.  
The acquisition deadtime was monitored by measuring the TS output-to-input 
ratio for each event type.  The event rates were set by varying the prescale 
factors and the beam current such that the DAQ computer was busy at most only 
20\% of the time.  This resulted in a relatively low event rate (a few kHz), at
which the electronics deadtime was $<$1\%.  Online analyzers \cite{dhistdspy} 
were used to monitor the quality of the data as it was taken.  Eventually, the 
data were transferred to magnetic tape.  The ultimate data analysis was 
performed on the DEC-8400 CPU farm ABACUS \cite{abacus} at the Massachusetts 
Institute of Technology using the analysis package \textsc{espace} 
\cite{espace99}.

\section{\label{sec:analysis}Analysis}

The interested reader is directed to the Ph.D. theses of Gao \cite{gao99} and 
Liyanage \cite{liyanage99} for a complete discussion of the data analysis.  For
the sake of completeness, a subset of the aforementioned information is 
presented here.

\begin{figure}
\resizebox{0.47\textwidth}{!}{\includegraphics{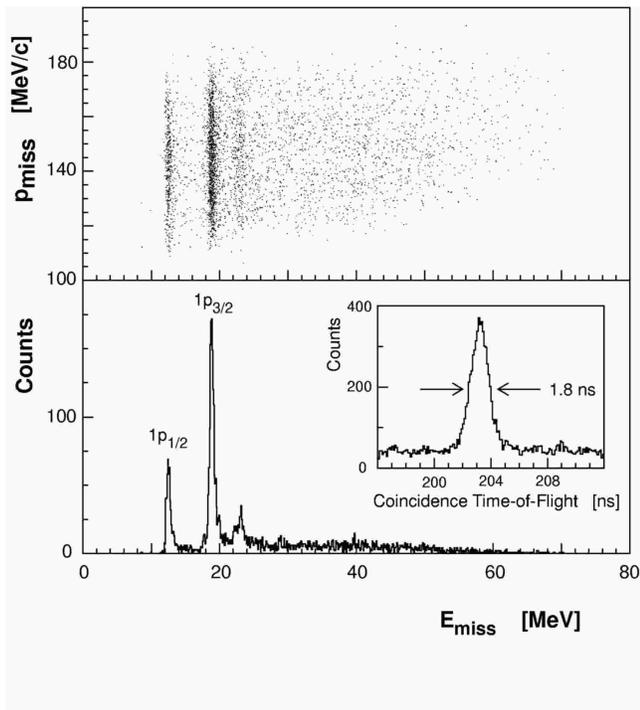}}
\caption{\label{fig:yield}
Yield spectrum obtained at $E_{\rm beam}$ = 0.843 GeV and $\theta_{pq}$ = 
+8$^{\circ}$, corresponding to $p_{\rm miss}$ $=$ 148 MeV/$c$.  Pion rejection 
has been performed, and all timing corrections have been applied.  The top 
panel shows a scatterplot of $p_{\rm miss}$ versus $E_{\rm miss}$.  The dark 
vertical bands project into the peaks located at 12.1 and 18.3 MeV in the 
bottom panel.  These peaks correspond to protons knocked-out of the 
1$p_{1/2}$- and 1$p_{3/2}$-states of $^{16}$O, respectively.  The 
$E_{\rm miss}$ resolution was roughly 0.9 MeV FWHM, which did not allow for 
separation of the $2s_{1/2}1d_{5/2}$-doublet located at $E_{\rm miss}$ $=$ 
17.4 MeV from the $1p_{3/2}$-state at 18.3 MeV.  The bump located at roughly 
23 MeV is a negative-parity doublet which was not investigated.  The insert 
shows the corresponding optimized coincidence TOF peak which has a FWHM of 
1.8 ns.  The signal-to-noise ratio was about 8:1 in these kinematics.
}
\end{figure}

\subsection{\label{subsec:pid}Timing corrections and particle identification}

Identification of coincidence $(e,e^{\prime}p)$ events was in general a 
straightforward process.  Software corrections were applied to remove timing 
variations induced by the trigger-scintillator circuit and thus sharpen all 
flight-time peaks.  These included corrections to proton flight times due to 
variations in the proton kinetic energies, and corrections for variations in 
the electron and proton path lengths through the spectrometers.  Pion rejection
was performed using a flight-time cut for $\pi^{+}$s in the HRS$_{h}$ and the 
Gas \v{C}erenkov for $\pi^{-}$s in the HRS$_{e}$. A sharp, clear, coincidence 
Time-of-Flight (TOF) peak with a FWHM of 1.8 ns resulted (see Fig. 
\ref{fig:yield}).  High-energy correlated protons which punched through the 
HRS$_{h}$ collimator ($<$10\% of the prompt yield) were rejected by requiring 
both spectrometers to independently reconstruct the coincidence-event vertex in
the vicinity of the same water film.  The resulting prompt-peak yields for each
water film were corrected for uncorrelated (random) events present in the 
peak-time region on a bin-by-bin basis as per the method suggested by 
Owens \cite{owens90}.  These per-film yields were then normalized individually.

\subsection{\label{subsec:norms}Normalization}

The relative focal-plane efficiencies for each of the two spectrometers were 
measured independently for each of the three water films at every spectrometer 
excitation used in the experiment.  By measuring the same single-arm cross 
section at different locations on the spectrometer focal planes, variations
in the relative efficiencies were identified.  The position variation across 
the focal plane was investigated by systematically shifting the central 
excitation of the spectrometer about the mean momentum setting in a series of 
discrete steps such that the full momentum acceptance was `mapped'.  A smooth, 
slowly varying dip-region cross section was used instead of a single discrete 
peak for continuous coverage of the focal plane.  The relative-efficiency 
profiles were unfolded from these data using the program {\sc releff} 
\cite{releff} by Baghaei.  For each water film, solid-angle cuts were then 
applied to select the flat regions of the angular acceptance.  These cuts 
reduced the spectrometer apertures by roughly 20\% to about 4.8 msr.  Finally, 
relative-momentum cuts were applied to select the flat regions of momentum 
acceptance.  These cuts reduced the spectrometer momentum acceptance by roughly
22\% to $-$3.7\% $<$ $\delta$ $<$ 3.3\%.  The resulting acceptance profile of 
each spectrometer was uniform to within 1\%.

The absolute efficiency at which the two spectrometers operated in coincidence 
mode was given by

\begin{equation}
\epsilon = \epsilon_e\cdot\epsilon_p\cdot\epsilon_{\rm coin},
\end{equation}

\noindent
where $\epsilon_e$ was the single-arm HRS$_{e}$ efficiency, $\epsilon_p$ was 
the single-arm HRS$_{h}$ efficiency, and $\epsilon_{\rm coin}$ was the 
coincidence-trigger efficiency.  The quantity 
($\epsilon_p\cdot\epsilon_{\rm coin}$) was measured at $\theta_{pq}$ $=$ 
0$^{\circ}$ at $E_{\rm beam}$ = 0.843 GeV using the $^{1}$H$(e,e)$ reaction.  A 
0.7 msr collimator was placed in front of the HRS$_{e}$.  In these kinematics, 
the cone of recoil protons fit entirely into the central flat-acceptance region
of the HRS$_{h}$.  The number of $^{1}$H$(e,e)$ events where the proton was 
also detected was compared to the number of $^{1}$H$(e,e)$ events where the 
proton was not detected to yield a product of efficiencies 
($\epsilon_p\cdot\epsilon_{\rm coin}$) of 98.9\%.  The 1.1\% effect was due to 
proton absorption in the waterfall target exit windows, spectrometer windows, 
and the first layer of trigger scintillators.  Since the central field of the 
HRS$_{h}$ was held constant throughout the entire experiment, this measurement 
was applicable to each of the hadron kinematics employed.  A similar method was
used to determine the quantity ($\epsilon_e\cdot\epsilon_{\rm coin}$) at each 
of the three HRS$_{e}$ field settings.  Instead of a collimator, software cuts 
applied to the recoil protons were used to ensure that the cone of scattered 
electrons fit entirely into the central flat-acceptance region of the 
HRS$_{e}$.  This product of efficiencies was $>$99\%.  Thus, the coincidence 
efficiency $\epsilon_{\rm coin}$ was firmly established at nearly 100\%.  A 
nominal systematic uncertainty of $\pm 1.5\%$ was attributed to $\epsilon$.

\begin{figure}
\resizebox{0.47\textwidth}{!}{\includegraphics{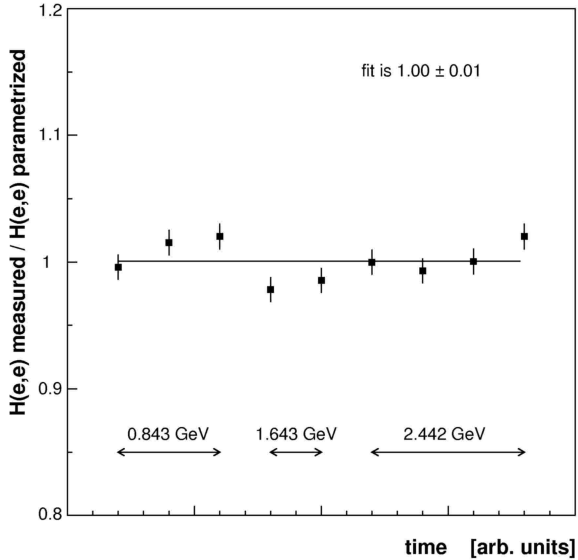}}
\caption{\label{fig:lumeff}
Measured $^{1}$H$(e,e)$ cross-section data normalized to the absolute 
predictions of a parametrization at similiar $Q^{2}$. Statistical error bars 
are shown.  The data shown were taken over the course of a three-month run 
period. The different data points for each $E_{\rm beam}$ represent different 
HRS$_h$ angular settings.
}
\end{figure}

The quantity ($L\cdot\epsilon_e$), where $L$ is the luminosity (the 
product of the effective target thickness and the number of incident electrons)
was determined to $\pm$4\% by comparing the measured $^{1}$H$(e,e)$ cross 
section for each film at each of the electron kinematics to a parametrization 
established at a similar $Q^{2}$ by Simon {\it et al.} \cite{mainzffa} and 
Price {\it et al.} \cite{mainzffb} (see Fig. \ref{fig:lumeff}).  The results 
reported in this paper have all been normalized in this fashion.  As a 
consistency check, a direct absolute calculation of ($L\cdot\epsilon_e$) 
using information from the BCMs, the calibrated thicknesses of the water films,
and the single-arm HRS$_{e}$ efficiency agrees within uncertainty.

At every kinematics, a Monte Carlo of the phase-space volume subtended by each 
experimental bin was performed.  For each water foil, $N_{0}$ software 
$(e,e^{\prime}p)$ events were generated, uniformly distributed over the 
scattered-electron and knocked-out proton momenta ($p_{e}$, $p_{p}$) and 
in-plane and out-of-plane angles ($\phi_{e}$, $\theta_{e}$, $\phi_{p}$, 
$\theta_{p}$).  For each of these events, all of the kinematic quantities were 
calculated.  The flat-acceptance cuts determined in the analysis of the 
relative focal-plane efficiency data were then applied, as were all other cuts 
that had been performed on the actual data.  The pristine detection volume 
$\Delta V_{b}$($E_{\rm miss}$, $p_{\rm miss}$, $\omega$, $Q^2$) subtended by a 
bin $b$($\Delta E_{\rm miss}$, $\Delta p_{\rm miss}$, $\Delta \omega$, 
$\Delta Q^2$) containing $N_{b}$ pseudoevents was thus

\begin{eqnarray}
\Delta V_{b}(E_{\rm miss}, p_{\rm miss}, \omega, Q^2) = \nonumber \\ & &
\hspace*{-30mm}
\frac{N_{b}}{N_{0}} 
~[(\Delta p_{e} \cdot \Delta \Omega_{e}) \cdot
(\Delta p_{p} \cdot \Delta \Omega_{p})] \; ,
\end{eqnarray}

\noindent
where the quantity $(\Delta p_{e} \cdot \Delta \Omega_{e}) \cdot (\Delta p_{p} 
\cdot \Delta \Omega_{p})$ was the total volume sampled over in the Monte Carlo 
(purposely set larger than the experimental acceptance in all dimensions
\footnote{
When necessary, the differential dependencies of the measured cross-section 
data were changed to match those employed in the theoretical calculations.  The
pristine detection volume $\Delta V_{b}$($E_{\rm miss}$, $p_{\rm miss}$, 
$\omega$, $Q^{2}$) was changed to a weighted detection volume by weighting each
of the trials with the appropriate Jacobian(s).
}). 
The pseudodata were binned exactly as the real data, and uniformly on both 
sides of $\bm{q}$.  At each kinematics, the bin with the largest volume 
$\Delta V_{\rm max}$ was located.  Only bins subtending volumes larger than 
50\% of $\Delta V_{\rm max}$ were analyzed further.  

Corrections based on the TS output-to-input ratio were applied to the data to 
account for the acquisition deadtime to coincidence events.  On average, these 
corrections were roughly 20\%.  An acquisition Monte Carlo by Liang
\cite{liang98} was used to cross-check these corrections and establish the 
absolute uncertainty in them at 2\%.

Corrections to the per-film cross-section data for electron radiation before 
and after scattering were calculated on a bin-by-bin basis in two ways:  first 
using a version of the code \textsc{radcor} by Quint \cite{radcor00a} modified 
by Florizone \cite{radcor00b}, and independently, the prescriptions of Borie 
and Dreschel \cite{mceep00a} modified by Templon {\it et al.} \cite{mceep00b} 
for use within the simulation package \textsc{mceep} written by Ulmer 
\cite{mceep00c}.  The two approaches agreed to within the statistical 
uncertainty of the data and amounted to $<$55\% of the measured cross section 
for the bound states, and $<$15\% of the measured cross section for the 
continuum.  Corrections for proton radiation at these energies are much less 
than 1\% and were not performed.  

\subsection{\label{subsec:sigma}Cross section}

The radiatively corrected average cross section in the bin
$b$($\Delta E_{\rm miss}$, $\Delta p_{\rm miss}$, $\Delta \omega$, 
$\Delta Q^2$) was calculated according to
 
\begin{eqnarray}
\label{d6sigma}
\left<\frac{d^{6}\sigma}{d \omega~d\Omega_{e}~
d E_{\rm miss}~d\Omega_{p}}\right>_{b} = \nonumber \\ & &
\hspace*{-20mm}\frac{R_{^{16}{\rm O}(e,e^{\prime}p)}}{(L\cdot\epsilon_e)(\epsilon_p\cdot\epsilon_{\rm coin})}
\left(\frac{Y_{b}}{\Delta V_{b}}\right),
\end{eqnarray}

\noindent
where $Y_{b}$ was the total number of real events which were detected in 
$b$($\Delta E_{\rm miss}$, $\Delta p_{\rm miss}$, $\Delta \omega$, 
$\Delta Q^2$), ${\Delta V_{b}}$ was the phase-space volume, and 
$R_{^{16}{\rm O}(e,e^{\prime}p)}$ was a correction applied to account for 
events which radiated in or out of $\Delta V_{b}$.  The average cross section
was calculated as a function of $E_{\rm miss}$ for a given kinematic setting 
\footnote{
The difference between cross-section data averaged over the reduced 
spectrometer acceptances and calculated for a small region of the central 
kinematics was no more than 1\%.  Thus, the finite acceptance of the 
spectrometers was not an issue.
}.
Bound-state cross-section data for the $1p$-shell were extracted by integrating 
over the appropriate range in $E_{\rm miss}$, weighting with the appropriate 
Jacobian
\footnote{
This Jacobian is given by 
$\frac{\partial E_{\rm miss}}{\partial p_p}$ = $\frac{p_p}{E_p}$ 
+ $\frac{\bm{{p}_p} \cdot \bm{{p}_{B}}}{p_p E_{B}}$,
where $E_{B}$ = $\sqrt{p^2_{B}+m^2_{B}}$.
}.  
Five-fold differential cross-section data for QE proton knockout from the 
$1p$-shell of $^{16}$O are presented in Tables \ref{table:pshellpararesults} 
and \ref{table:pshellperpresults}.  Six-fold differential cross-section data
for QE proton knockout from $^{16}$O at higher $E_{\rm miss}$ are presented in 
Tables \ref{table:0843cont000deg} $-$ \ref{table:2442cont020deg}.

\subsection{\label{subsec:separations}Asymmetries and response functions}

In the One-Photon Exchange Approximation, the unpolarized six-fold differential
cross section may be expressed in terms of four independent response functions 
as (see Refs. \cite{picklesimer89,raskin89,kelly96})
\begin{widetext}
\begin{equation}
\label{equation:sixfold}
\frac{d^{6}\sigma}{d \omega~d\Omega_{e}~d E_{\rm miss}~d\Omega_{p}} = K~\sigma_{\rm Mott}
\left[
v_{L} R_{L} 
+ v_{T} R_{T}
+ v_{LT} R_{LT} \cos(\phi)
+ v_{TT} R_{TT} \cos(2 \phi) \right],
\end{equation}
\end{widetext}
where $K$ is a phase-space factor, $\sigma_{\rm Mott}$ is the Mott cross 
section, and the $v_{\rm i}$ are dimensionless kinematic factors 
\footnote{
The phase-space factor $K$ is given in Eq. (\ref{equation:K}), while 
$\sigma_{\rm Mott}$ = 
$\frac{\alpha^{2}\cos^{2}(\theta_{e}/2)}
{4\varepsilon^{2}_{i}\sin^{4}(\theta_{e}/2)}$.  
The dimensionless kinematic factors are as follows: 
$v_{L} = \frac{Q^{4}}{\bm{q}\hspace*{0.5mm}^{4}}$, 
$v_{T} = \frac{Q^{2}}{2\bm{q}\hspace*{0.5mm}^{2}}+\tan^{2}(\theta_{e}/2)$, 
$v_{LT} = \frac{Q^{2}}{\bm{q}\hspace*{0.5mm}^{2}}
\sqrt{\frac{Q^{2}}{\bm{q}\hspace*{0.5mm}^{2}}+
\tan^{2}(\theta_{e}/2)}$,
and $v_{TT} = \frac{Q^{2}}{2\bm{q}\hspace*{0.5mm}^{2}}$.}.  
Ideal response functions are not directly measureable because electron
distortion does not permit the azimuthal dependences to be separated exactly.  
The effective response functions which are extracted by applying Eq. 
(\ref{equation:sixfold}) to the data are denoted $R_{L}$ (longitudinal), 
$R_{T}$ (transverse), $R_{LT}$ (longitudinal-transverse), and $R_{TT}$ 
(transverse-transverse).  They contain all the information which may be 
extracted from the hadronic system using $(e,e^{\prime}p)$.  Note that the 
$v_{\rm i}$ depend only on ($\omega$, $Q^{2}$, $\theta_e$), while the response 
functions depend on ($\omega$, $Q^{2}$, $E_{\rm miss}$, $p_{\rm miss}$).

The individual contributions of the effective response functions may be 
separated by performing a series of cross-section measurements varying 
$v_{\rm i}$ and/or $\phi$, but keeping $q$ and $\omega$ constant 
\footnote{
The accuracy of the effective response-function separation depends on precisely
matching the values of $q$ and $\omega$ at each of the different kinematic 
settings.  This precise matching was achieved by measuring $^{1}$H$(e,ep)$ with
a pinhole collimator (in practice, the central hole of the sieve-slit 
collimator) placed in front of the HRS$_{e}$.  The proton momentum was thus 
$q$.  The $^{1}$H$(e,ep)$ proton momentum peak was determined to 
$\Delta p/p$ $=$ 1.5 $\times$ 10$^{-4}$, which allowed for an identical 
matching of $\Delta{q}/{q}$ between the different kinematic settings.
}.  
In the case where the proton is knocked-out of the nucleus in a direction 
parallel to $\bm{q}$ (`parallel' kinematics), the interference terms $R_{LT}$
and $R_{TT}$ vanish, and a Rosenbluth separation \cite{rosenbluth50} may be 
performed to separate $R_{L}$ and $R_{T}$.  In the case where the proton is 
knocked-out of the nucleus in the scattering plane with a finite angle 
$\theta_{pq}$ with respect to $\bm{q}$ (`quasiperpendicular' kinematics), the 
asymmetry $A_{LT}$ and the interference $R_{LT}$ may be separated by performing
symmetric cross-section measurements on either side of $\bm{q}$ ($\phi$ $=$ 
0$^{\circ}$ and $\phi$ $=$ 180$^{\circ}$).  The contribution of $R_{TT}$ cannot
be separated from that of $R_{L}$ with only in-plane measurements; however, by 
combining the two techniques, an interesting combination of response functions 
$R_{T}$, $R_{LT}$, and $R_{L+TT}$ 
\footnote{
$R_{L+TT} \equiv R_{L} + \frac{V_{TT}}{V_L}R_{TT}$.
} 
may be extracted.

For these data, effective response-function separations were performed where 
the phase-space overlap between kinematics permitted.  For these separations, 
bins were selected only if their phase-space volumes $\Delta V_{b}$ were all 
simultaneously 50\% of $\Delta V_{\rm max}$.  Separated effective response 
functions for QE proton knockout from the $1p$-shell of $^{16}$O are presented 
in Tables \ref{table:pshellrlrt}, \ref{table:pshellrlttrt}, and 
\ref{table:pshellaltrlt}.  Separated effective response functions for QE proton
knockout from the $^{16}$O continuum are presented in Tables 
\ref{table:contrlrt}, \ref{table:contrlttrt}, and \ref{table:contrlt}.
 
\subsection{\label{subsec:sysun}Systematic uncertainties}

The systematic uncertainties in the cross-section measurements were classified 
into two categories -- kinematic-dependent uncertainties and scale 
uncertainties.  For a complete discussion of how these uncertainties were 
evaluated, the interested reader is directed to a report by Fissum and Ulmer 
\cite{fissum02}.  For the sake of completeness, a subset of the aforementioned 
information is presented here.

In a series of simulations performed after the experiment, \textsc{mceep} was 
used to investigate the intrinsic behavior of the cross-section data when 
constituent kinematic parameters were varied over the appropriate 
experimentally determined ranges presented in Table \ref{table:mceep_input}.  
Based on the experimental data, the high-$E_{\rm miss}$ region was modelled as 
the superposition of a peak-like $1s_{1/2}$-state on a flat continuum.  
Contributions to the systematic uncertainty from this flat continuum were taken
to be small, leaving only those from the $1s_{1/2}$-state.  The 
$^{16}$O$(e,e^{\prime}p)$ simulations incorporated as physics input the 
bound-nucleon \textsc{rdwia} calculations detailed in Section \ref{sec:rdwia}, 
which were based on the experimental $1p$-shell data.

\begin{table}
\caption{\label{table:mceep_input}
Kinematic-dependent systematic uncertainties folded into the \textsc{mceep} 
simulation series.}
\begin{ruledtabular}
\begin{tabular}{rrr}
Quantity            &                           description &               $\delta$ \\
\hline
$E_{\rm beam}$      &                           beam energy & 1.6 $\times$ 10$^{-3}$ \\
$\phi_{\rm beam}$   &                   in-plane beam angle &                ignored\footnote
{As previously mentioned, the angle of incidence of the electron beam was 
determined using a pair of BPMs located upstream of the target (see Fig. 
\ref{fig:halla}).  The BPM readback was calibrated by comparing the location of
survey fiducials along the beamline to the Hall A survey fiducials.  Thus, in 
principle, uncertainty in the knowledge of the incident electron-beam angle 
should be included in this analysis.  However, the simultaneous measurement of 
the kinematically overdetermined $^{1}$H$(e,ep)$ reaction allowed for a 
calibration of the absolute kinematics, and thus an elimination of this 
uncertainty.  That is, the direction of the beam defined the axis relative to 
which all angles were measured via $^{1}$H$(e,ep)$.}
                                                                                     \\
$\theta_{\rm beam}$ &               out-of-plane beam angle &             2.0   mrad \\
$p_{e}$             &           scattered electron momentum & 1.5 $\times$ 10$^{-3}$ \\
$\phi_{e}$          &     in-plane scattered electron angle &             0.3   mrad \\
$\theta_{e}$        & out-of-plane scattered electron angle &             2.0   mrad \\
$p_{p}$             &                       proton momentum & 1.5 $\times$ 10$^{-3}$ \\
$\phi_{p}$          &                 in-plane proton angle &             0.3   mrad \\
$\theta_{p}$        &             out-of-plane proton angle &             2.0   mrad \\
\end{tabular}
\end{ruledtabular}
\end{table}

For each kinematics, the central water foil was considered, and 1M events were 
generated.  In evaluating the simulation results, the exact cuts applied in the
actual data analyses were applied to the pseudo-data, and the cross section 
was evaluated for the identical $p_{\rm miss}$ bins used to present the 
results.  The experimental constraints to the kinematic-dependent observables 
afforded by the overdetermined $^{1}$H$(e,ep)$ reaction were exploited to 
calibrate and constrain the experimental setup.  The in-plane electron and 
proton angles $\phi_{e}$ and $\phi_{p}$ were chosen as independent parameters.
When a known shift in $\phi_{e}$ was made, $\phi_{p}$ was held constant and the
complementary variables $E_{\rm beam}$, $p_{e}$, and $p_{p}$ were varied as 
required by the constraints enforced by the $^{1}$H$(e,ep)$ reaction.  
Similarly, when a known shift in $\phi_{p}$ was made, $\phi_{e}$ was held 
constant and the complementary variables $E_{\rm beam}$, $p_{e}$, and $p_{p}$ 
were varied as appropriate.  The overall constrained uncertainty was taken to 
be the quadratic sum of the two contributions.

The global convergence of the uncertainty estimate was examined for certain 
extreme kinematics, where 10M-event simulations (which demonstrated the same 
behavior) were performed.  The behavior of the uncertainty as a function of 
$p_{\rm miss}$ was also investigated by examining the uncertainty in the 
momentum bins adjacent to the reported momentum bin in exactly the same 
fashion.  The kinematically induced systematic uncertainty in the 
$^{16}$O$(e,e^{\prime}p)$ cross-section data was determined to be dependent 
upon $p_{\rm miss}$, with an average value of 1.4\%.  The corresponding 
uncertainties in the $^{1}$H$(e,e)$ cross-section data were determined to be 
negligible.

The scale systematic uncertainties which affect each of the cross-section 
measurements are presented in Table \ref{table:scale_un}.  As previously 
mentioned, the $^{16}$O$(e,e^{\prime}p)$ cross-section results reported in this
paper have been normalized by comparing simultaneously measured $^{1}$H$(e,e)$ 
cross-section data to a parametrization established at a similar $Q^{2}$.  
Thus, the first seven listed uncertainties simply divide out of the quotient, 
such that only the subsequent uncertainties affect the results.  The average 
systematic uncertainty associated with a $1p$-shell cross section was 5.6\%, 
while that for the continuum was 5.9\%.  The small difference was due to 
contamination of the high-$E_{\rm miss}$ data by collimator punch-through 
events.

\begin{table*}
\caption{\label{table:scale_un}
Summary of the scale systematic uncertainties contributing to the cross-section 
data.  The first seven entries do not contribute to the systematic 
uncertainties in the reported cross-section data as they contribute equally to 
the $^{1}$H$(e,e)$ cross-section data to which the $^{16}$O$(e,e^{\prime}p)$ 
are normalized.}
\begin{ruledtabular}
\begin{tabular}{rrr}
Quantity                                                &                                                       description & $\delta$ (\%)\\
\hline
$\eta_{\rm DAQ}$                                        &                              data acquisition deadtime correction &          2.0 \\
$\eta_{\rm elec}$                                       &                                   electronics deadtime correction &       $<$1.0 \\
$\rho t^{\prime}$                                       &                                        effective target thickness &          2.5 \\
$N_{e}$                                                 &                                      number of incident electrons &          2.0 \\
$\epsilon_{e}$                                          &                                     electron detection efficiency &          1.0 \\
$\Delta\Omega_{e}$\footnotemark[1]                      &                                             HRS$_{e}$ solid angle &          2.0 \\
$\epsilon_{e}\cdot\epsilon_{p}\cdot\epsilon_{\rm coin}$ &         product of electron, proton, and coincidence efficiencies &          1.5 \\ \hline
$L\cdot\epsilon_{e}$                                 & obtained from a form-factor parametrization of $^{1}{\rm H}(e,e)$ &          4.0 \\
$R_{^{16}{\rm O}(e,e^{\prime}p)}$\footnotemark[2]       &    radiative correction to the $^{16}{\rm O}(e,e^{\prime}p)$ data &          2.0 \\ 
$R_{^{1}{\rm H}(e,e)}$\footnotemark[2]                  &               radiative correction to the $^{1}{\rm H}(e,e)$ data &          2.0 \\
$\epsilon_{p}\cdot\epsilon_{\rm coin}$                  &                    product of proton and coincidence efficiencies &       $<$1.0 \\
$\Delta\Omega_{p}$\footnotemark[1]                      &                                             HRS$_{h}$ solid angle &          2.0 \\
punchthrough\footnotemark[3]                            &            protons which punched through the HRS$_{h}$ collimator &          2.0 \\
\end{tabular}
\end{ruledtabular}
\begin{minipage}[t]{\textwidth}
\footnotetext[1]
{The systematic uncertainties in the solid angles $\Delta\Omega_{e}$ and 
$\Delta\Omega_{p}$ were quantified by studying sieve-slit collimator optics 
data at each of the spectrometer central momenta employed.  The angular 
locations of each of the reconstructed peaks corresponding to the 7 $\times$ 7 
lattice of holes in the sieve-slit plate were compared to the locations 
predicted by spectrometer surveys, and the overall uncertainy was taken to be 
the quadratic sum of the individual uncertainties.
}
\footnotetext[2]
{At first glance, it may be surprising to note that the uncertainty due to the 
radiative correction to the data is included as a scale uncertainty.  In 
general, the radiative correction is strongly dependent on kinematics.  
However, the $1p$-shell data analysis, and for that matter any bound-state data
analysis, involves $E_{\rm miss}$ cuts.  These cuts to a large extent remove 
the strong kinematic dependence of the radiative correction, since only 
relatively small photon energies are involved.  In order to compensate for any 
remaining weak kinematic dependence, the uncertainty due to the radiative 
correction was slightly overestimated.
}
\footnotetext[3]
{High $E_{\rm miss}$ data only.}
\end{minipage}
\end{table*}

\begin{figure}
\resizebox{0.47\textwidth}{!}{\includegraphics{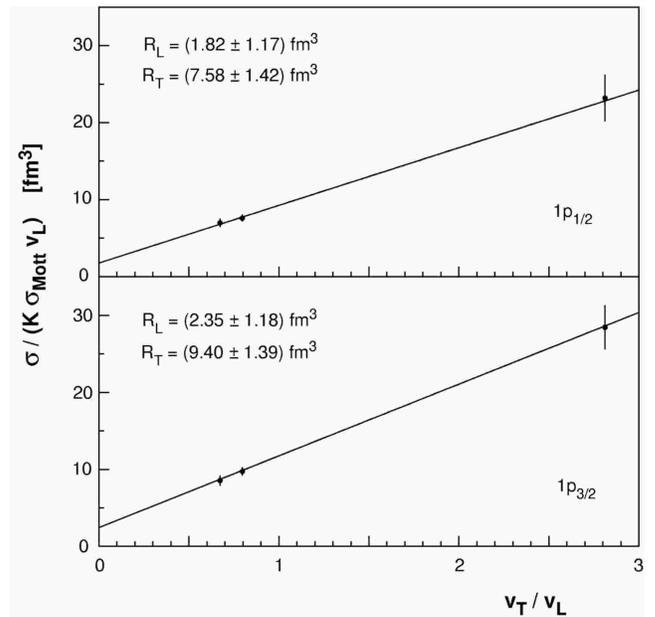}}
\caption{\label{fig:parqual}
Cross-section data for the removal of protons from the 1$p$-shell of $^{16}$O 
measured in parallel kinematics at three different beam energies as a function 
of the separation lever arm $v_{T}/v_{L}$.  The data points correspond to beam 
energies of 2.442, 1.643, and 0.843 GeV from left to right.  The effective 
response functions $R_{L}$ (offset) and $R_{T}$ (slope) have been extracted 
from the fitted line.  The uncertainties shown are statistical only.  The 
extremely linear behavior of the data (which persists even after corrections 
for Coulomb distortion are applied) indicates that the statistical 
uncertainties were dominant (see Section \ref{subsec:sysun} for a complete 
discussion).
}

\end{figure}

The quality of these data in terms of their associated systematic uncertainties 
was clearly demonstrated by the results obtained for the effective 
response-function separations.  In Fig. \ref{fig:parqual}, cross-section data
for the $1p$-shell measured in parallel kinematics at three different beam 
energies are shown as a function of the separation lever arm $v_{T}/v_{L}$.  
The values of the effective response functions $R_{L}$ (offset) and $R_{T}$ 
(slope) were extracted from the fitted line.  The extremely linear trend in the
data indicated that the magnitude of the systematic uncertainties was small, 
and that statistical uncertainties dominated.  This is not simply a test of the 
One-Photon Exchange Approximation employed in the data analysis as it has been 
demonstrated by Traini {\it et al.} \cite{traini88} and Ud{\'\i}as 
\cite{udiasthesis} that the linear behavior of the Rosenbluth plot persists 
even after Coulomb distortion is included.

\begin{figure}
\resizebox{0.47\textwidth}{!}{\includegraphics{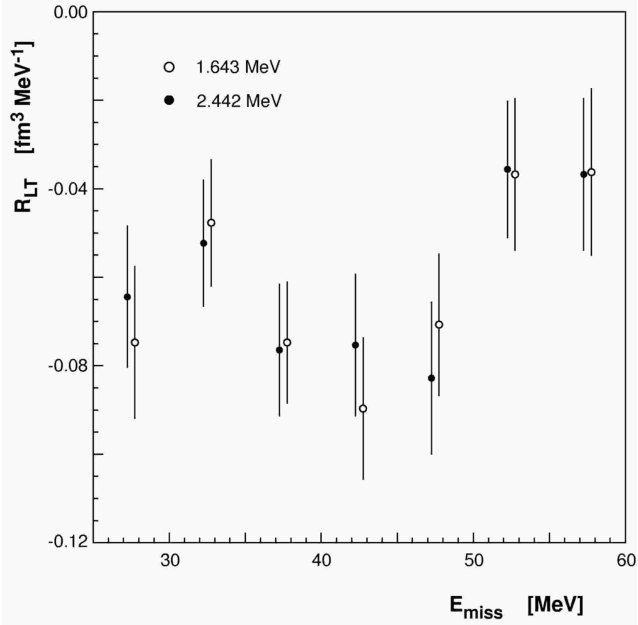}}
\caption{\label{fig:rltqual}
$R_{LT}$ for $\theta_{pq}$ $=$ $\pm8^{\circ}$ ($p_{\rm miss}$ $=$ 145 
MeV/$c$) as a function of $E_{\rm miss}$ for $E_{\rm beam}$ $=$ 1.643 GeV and 
2.442 GeV. Statistical uncertainties only are shown. The statistical agreement
over a broad range of $E_{\rm miss}$ emphasizes the systematic precision of the
measurement (see Section \ref{subsec:sysun} for a complete discussion). Note 
that the averages of these $R_{LT}$ values are presented as the $p_{\rm miss}$ 
$=$ 145 MeV/$c$ data in Fig. \ref{fig:nilanga34} and Table \ref{table:contrlt}.
}
\end{figure}

Given the applicability of the One-Photon Exchange Approximation at these 
energies, the quality of the data was also demonstrated by the results 
extracted from identical measurements which were performed in different 
electron kinematics.  The asymmetries $A_{LT}$ and effective response functions
$R_{LT}$ for QE proton knockout were extracted for both $E_{\rm beam}$ $=$ 
1.643 GeV and 2.442 GeV for $\theta_{pq}$ $=$ $\pm8^{\circ}$ ($p_{\rm miss}$ 
$=$ 148 MeV/$c$).  They agree within the statistical uncertainty.  Table
\ref{table:pshellaltrlt} presents the results at both beam energies for 
$1p$-shell knockout for $<Q^{2}>$ $=$ 0.800 (GeV/$c$)$^{2}$, $<\omega>$ $=$ 436 
MeV, and $<T_{p}>$ $=$ 427 MeV, while Fig. \ref{fig:rltqual} shows the 
results for 25 $<$ $E_{\rm miss}$ $<$ 60 MeV.  

\section{\label{sec:theory}Theoretical Overview}

In the following subsections, overviews of Relativistic Distorted-Wave Impulse 
Approximation (\textsc{rdwia}), Relativistic Optical-Model Eikonal 
Approximation (\textsc{romea}), and Relativistic Multiple-Scattering Glauber 
Approximation (\textsc{rmsga}) calculations are presented.

\subsection{\label{sec:rdwia}RDWIA}
                                                                                
Reviews of work on proton electromagnetic knockout using essentially
nonrelativistic approaches may be found in Refs. 
\cite{frullani84,kelly96,boffi96}.  As previously mentioned, the Relativistic 
Distorted-Wave Impulse Approximation (\textsc{rdwia}) was pioneered by 
Picklesimer, Van Orden, and Wallace 
\cite{picklesimer85,picklesimer87,picklesimer89} and subsequently developed in 
more detail by several groups (see Refs. 
\cite{mcdermott90,jin92,udias93,hedayati95,kelly99a,udias00,meucci01}).  In 
Section \ref{sec:RDWIA-formalism}, the \textsc{rdwia} formalism for direct 
knockout based upon a single-nucleon operator is outlined in sufficient detail 
that the most important differences with respect to nonrelativistic 
\textsc{dwia} may be identified.  In Section \ref{sec:RDWIA-tests}, a direct 
numerical comparison between two different implementations of \textsc{rdwia}
is presented.

\subsubsection{Formalism}
\label{sec:RDWIA-formalism}
                                                                                
The five-fold differential cross section for the exclusive $A(e,e^\prime N)B$ 
reaction leading to a discrete final state takes the form (see Ref. 
\cite{kelly96})
\begin{equation}
\label{equation:cross-section}
\frac{d^5\sigma}{d\varepsilon_f d\Omega_e d\Omega_N}
= 
K \frac{\varepsilon_f}{\varepsilon_i}
\frac{\alpha^2}{Q^4} \eta_{\mu\nu} {\cal W}^{\mu\nu}~,
\end{equation}
where
\begin{equation}
\label{equation:K}
K = {\cal R} \frac{p_N E_N}{(2\pi)^3}
\end{equation} 
is a phase-space factor, $k_i=(\varepsilon_i,{\bm{k}}_i)$ and 
$k_f=(\varepsilon_f,{\bm{k}}_f)$ are the initial and final electron momenta,
$p_A=(E_A,{\bm{p}}_A)$ and $p_B=(E_B,{\bm{p}}_B)$ are the initial and final 
target momenta, $p_N=(E_N,{\bm{p}}_N)$ is the ejected-nucleon momentum, 
${q} = {k}_i - {k}_f = (\omega,{\bm{q}}\hspace*{0.5mm})$ is the momentum 
transfer carried by the virtual photon, 
$Q^2 = -q_\mu q^\mu = {\bm{q}}\hspace*{0.5mm}^2 - \omega^2$ is the photon 
virtuality, and
\begin{equation}
\label{equation:recoil_factor}
{\cal R} 
=   
\left| 1 -
\frac{{\bm{v}}_N \cdot {\bm{v}}_B}{{\bm{v}}_N \cdot {\bm{v}}_N} 
\right|^{-1}
\end{equation}
(with $v_N=p_N/E_N$) is a recoil factor which adjusts the nuclear phase space 
for the missing-energy constraint.  In the One-Photon Exchange Approximation, 
the invariant electroexcitation matrix element is represented by the 
contraction of electron and nuclear response tensors of the form
\begin{eqnarray}
\eta_{\mu\nu}     &=& \langle j_\mu j_\nu^\dagger \rangle \\
{\cal W}_{\mu\nu} &=& \langle {\cal J}_\mu {\cal J}_\nu^\dagger \rangle~,
\end{eqnarray}
where $j^\mu$ is the electron current, ${\cal J}^\mu$ is a matrix element of 
the nuclear electromagnetic current, and the angled brackets denote averages 
over initial states and sums over final states.  

The reduced cross section is given by
\begin{equation}
\sigma_{\rm red} 
=
\frac{d^5 \sigma}{d\varepsilon _{f}d\Omega _{e}d\Omega _{N}} /
K \sigma_{eN}~,
\end{equation}
where
\begin{equation}
\sigma_{eN} 
= 
\frac{\varepsilon_f}{\varepsilon_i} \frac{\alpha^2}{Q^4}
\left( \eta_{\mu\nu} {\cal W}^{\mu\nu} \right)_{\rm PWIA}
\end{equation}
is the elementary cross section for electron scattering from a moving free
nucleon in the Plane-Wave Impulse Approximation (\textsc{pwia}).  The 
\textsc{pwia} response tensor is computed for a free nucleon in the final
state, and is given by
\begin{equation}
W^{\mu\nu}_{\rm PWIA} 
= 
\frac{1}{2} {\rm Trace}{J^\mu {J^\nu}^\dagger}~,
\end{equation}
where
\begin{equation}
J^\mu_{s_f,s_i} =  \sqrt{\frac{m^2}{\varepsilon_i \varepsilon_f}}
\bar{u}({{\bm{p}}}_f,s_f) \Gamma^\mu u({{\bm{p}}}_i,s_i)
\end{equation}
is the single-nucleon current between free spinors normalized to unit flux.  
The initial momentum (${{\bm p}}_i = {{\bm p}}_f - {{\bm q}}_{\rm eff}$) is 
obtained from the final ejectile momentum (${\bm{p}}_f$) and the effective 
momentum transfer (${\bm{q}}_{\rm eff}$) in the laboratory frame, and the 
initial energy is placed on shell.  The effective momentum transfer accounts
for electron acceleration in the nuclear Coulomb field and is discussed further
later in this Section.  

In the nonrelativistic \textsc{pwia} limit, $\sigma_{\rm red}$ reduces to the 
bound-nucleon momentum distribution, and the cross section given in Eq. 
(\ref{equation:cross-section}) may be expressed as the product of the 
phase-space factor $K$, the elementary cross section $\sigma_{eN}$, and the 
momentum distribution.  This is usually referred to `factorization'.  
Factorization is not strictly valid relativistically because the binding 
potential alters the relationship between lower and upper components of a Dirac
wave function -- see Ref. \cite{caballero98b}.  

In this Section, it is assumed that the nuclear current is represented by a 
one-body operator, such that
\begin{equation}
\label{equation:psi_minus}
{\cal J}^\mu
=
\int d^3 r \; \exp{(i \bm{t}}\cdot {\bm{r}})
\langle \bar{\Psi}^{(-)}({\bm{p}},{\bm{r}}\hspace*{0.5mm}) | \Gamma^\mu |
\phi({\bm{r}}\hspace*{0.5mm}) \rangle~,
\end{equation}
where $\phi$ is the nuclear overlap for single-nucleon knockout (often 
described as the bound-nucleon wave function), $\bar{\Psi}^{(-)}$ is the Dirac
adjoint of the time-reversed distorted wave, ${\bm{p}}$ is the relative 
momentum, and
\begin{equation}
\label{equation:recoil}
\bm{t} = \frac{E_B}{W} {\bm{q}}
\end{equation}
is the recoil-corrected momentum transfer in the barycentric frame.  Here 
$(\omega,{\bm{q}}\hspace*{0.5mm})$ and $E_B$ are the momentum transfer and the 
total energy of the residual nucleus in the laboratory frame respectively, and 
$W = \sqrt{ (m_A+\omega)^2 - {\bm{q}}\hspace*{0.5mm}^2 }$ is the invariant 
mass. 

De Forest \cite{deforest83} and Chinn and Picklesimer \cite{chinn92} have
demonstrated that the electromagnetic vertex function for a free nucleon can be
represented by any of three Gordon-equivalent operators 
\begin{subequations}
\label{equation:gamma}
\begin{eqnarray}
   \Gamma_1^\mu(\bm{p}_f,\bm{p}_i) \hspace*{-1.5mm}&=& \hspace*{-1.5mm}\gamma^\mu G_M(Q^2) -
     \frac{P^\mu}{2m}  F_2(Q^2)
\label{eq:gamma1} \\
   \Gamma_2^\mu(\bm{p}_f,\bm{p}_i) \hspace*{-1.5mm}&=& \hspace*{-1.5mm}\gamma^\mu F_1(Q^2) + i \sigma^{\mu\nu}
     \frac{q_\nu}{2m}  F_2(Q^2)  \\
\label{eq:gamma2}
   \Gamma_3^\mu(\bm{p}_f,\bm{p}_i) \hspace*{-1.5mm}&=& \hspace*{-1.5mm}\frac{P^\mu}{2m}F_1(Q^2) + i \sigma^{\mu\nu}
     \frac{q_\nu}{2m} G_M(Q^2)
\label{eq:gamma3}
\end{eqnarray}
\end{subequations}
where $P=(E_f+E_i,{\bm{p}_f}+{\bm{p}_i}\hspace*{0.5mm})$.  Note the 
correspondence with Eq. (\ref{equation:gamma_bar}) below.  Although 
$\Gamma_2$ is arguably the most fundamental because it is defined in terms of 
the Dirac and Pauli form factors $F_1$ and $F_2$, $\Gamma_1$ is often used 
because the matrix elements are easier to evaluate.  $\Gamma_3$ is rarely used 
but no less fundamental.  In all calculations presented here, the momenta in 
the vertex functions are evaluated using asymptotic laboratory kinematics 
instead of differential operators.  

Unfortunately, as bound nucleons are not on shell, an off-shell extrapolation 
(for which no rigorous justification exists) is required.  The de Forest 
prescription is employed, in which the energies of both the initial and the 
final nucleons are placed on shell based upon effective momenta, and the energy
transfer is replaced by the difference between on-shell nucleon energies in the
operator.  Note that the form factors are still evaluated at the $Q^2$
determined from the electron-scattering kinematics.  In this manner, three 
prescriptions
\begin{subequations}
\label{equation:gamma_bar}
\begin{eqnarray}
  \bar{\Gamma}_1^\mu &=& \gamma^\mu G_M(Q^2) -
     \frac{\bar{P}^\mu}{2m}  F_2(Q^2)
\label{eq:cc1} \\
  \bar{\Gamma}_2^\mu &=&  \gamma^\mu F_1(Q^2) + i \sigma^{\mu\nu}
     \frac{\bar{q}_\nu}{2m}  F_2(Q^2)  \\
\label{eq:cc2}
  \bar{\Gamma}_3^\mu &=& \frac{\bar{P}^\mu}{2m}F_1(Q^2) + i \sigma^{\mu\nu}
     \frac{\bar{q}_\nu}{2m} G_M(Q^2)~,
\label{eq:cc3}
\end{eqnarray}
\end{subequations}
are obtained, where
\begin{eqnarray*}
   \bar{q} &=& (E_f - \bar{E}_i, {\bm{q}}\hspace*{0.5mm})~,   \\
   \bar{P} &=& (E_f + \bar{E}_i, 2{\bm{p}_f} - {\bm{q}}\hspace*{0.5mm})~,
\end{eqnarray*}
and where $\bar{E}_i=\sqrt{m_N^2 + ({\bm{p}_f}-{\bm{q}}\hspace*{0.5mm})^2}$ is 
placed on shell based upon the externally observable momenta ${\bm{p}_f}$ and 
${\bm{q}}$ evaluated in the laboratory frame.  
When electron distortion is included, the local momentum transfer 
${\bm{q}} \rightarrow {\bm{q}}_{\rm eff}$
is interpreted as the effective momentum transfer with Coulomb distortion.
These operators are commonly named \textsc{cc1}, \textsc{cc2}, and 
\textsc{cc3}, and are no longer equivalent when the nucleons are off-shell.  
Furthermore, the effects of possible density dependence in the nucleon form 
factors can be evaluated by applying the Local Density Approximation (LDA) to 
Eq. (\ref{equation:gamma_bar}) -- see Refs. \cite{chinn92,kelly99a}.

The overlap function is represented as a Dirac spinor of the form
\begin{equation}
\phi_{\kappa m}({\bm{r}}) 
= 
\left(
\begin{array}{r}
f_{\kappa}(r) {\cal Y}_{\kappa m}(\hat{r}) \\
i g_{-\kappa}(r) {\cal Y}_{-\kappa m}(\hat{r})
\end{array} \right)~,
\end{equation}
where
\begin{equation}
{\cal Y}_{\kappa m}(\hat{r}) 
= 
\sum_{\nu,m_s}
\left< \ell \; \; \nu \; \; \frac{1}{2} \; \; m_s \mid j \; \; m \; \right>
Y_{\ell\nu}(\hat{r})\chi_{m_s}
\end{equation}
is the spin spherical harmonic and where the orbital and total angular momenta 
are respectively given by
\begin{subequations}
\begin{eqnarray}
\ell &=& S_\kappa(\kappa + \frac{1}{2}) - \frac{1}{2} \\
j &=& S_\kappa \kappa - \frac{1}{2}~,
\end{eqnarray}
\end{subequations}
with $S_\kappa={\rm sign}{(\kappa)}$.  
The functions $f_{\kappa}$ and $g_{\kappa}$ satisfy the usual coupled linear 
differential equations -- see for example Ref. \cite{rose61}.  The 
corresponding momentum wave function
\begin{equation}
\tilde{\phi}_{\kappa m}({\bm{p}_{m}}\hspace*{0.5mm}) 
=
\int d^3 r \; \exp{(-i {\bm{p}_{m}} \cdot {\bm{r}}\hspace*{0.5mm})} \phi_{\kappa m}({\bm{r}}\hspace*{0.5mm})
\end{equation}
then takes the form
\begin{equation}
\tilde{\phi}_{\kappa m}({\bm{p}_{m}}\hspace*{0.5mm}) 
= 
4\pi i^{-\ell} \left(
\begin{array}{r}
\tilde{f}_{\kappa}(p_{m}) {\cal Y}_{\kappa m}(\hat{p}_{m}) \\
-S_\kappa \tilde{g}_{-\kappa}(p_{m}) {\cal Y}_{-\kappa m}(\hat{p}_{m})
\end{array} \right)~,
\end{equation}
where
\begin{subequations}
\begin{eqnarray}
\tilde{f}_{\kappa}(p_{m}) &=& \int dr \; r^2  j_{\ell}(p_{m} r) f_{\kappa}(r) \\
\tilde{g}_{-\kappa}(p_{m}) &=& \int dr \; r^2  j_{\ell^\prime}(p_{m} r) g_{-\kappa}(r)~,
\end{eqnarray}
\end{subequations}
and where in the \textsc{pwia}, the initial momentum $\bm{p}_m$ would equal the 
experimental missing momentum $\bm{p}_{\rm miss}$.  Thus, the 
momentum distribution
\begin{equation}
\rho(p_{m}) 
= 
\frac{1}{2\pi^2} \left(
|\tilde{f}_{\kappa}(p_{m})|^2 + |\tilde{g}_{\kappa}(p_{m})|^2 \right)
\end{equation}
is obtained, normalized to
\begin{equation}
4\pi \int dp \; p_{m}^2 \rho(p_{m}) = 1
\end{equation}
for unit occupancy.  

Similarly, let
\begin{equation}
\Psi^{(+)}({\bm{p}},{\bm{r}}) 
= 
\sqrt{\frac{E + m}{2E}}
\left( \begin{array}{r}
\psi({\bm{r}}) \\  \zeta({\bm{r}}) \end{array} \right)
\end{equation}
represent a wave function of the $N+B$ system with an incoming Coulomb wave and
outgoing spherical waves open in all channels.  Specific details regarding
the boundary conditions may be found in 
Refs. \cite{satchler83,rawitscher97,kelly99b}.  

The Madrid \textsc{rdwia} calculations \cite{udias93} employ a partial-wave 
expansion of the first-order Dirac equation, leading to a pair of coupled 
first-order differential equations.  Alternatively, the \texttt{LEA} code 
\cite{LEA} by Kelly uses the Numerov algorithm to solve a single second-order 
differential equation that emerges from an equivalent Schr\"{o}dinger equation 
of the form
\begin{equation}
 \left[ \nabla^2 + k^2 - 2\mu \left( U^C + U^{LS} \bm{L} \cdot
\bm{\sigma} \right) \right] \xi = 0~,
\end{equation}
where $k$ is the relativistic wave number, $\mu$ is the reduced energy, and
\begin{subequations}
\label{equation:scheq}
\begin{eqnarray}
     U^C &=& \frac{E}{\mu} \left[ V + \frac{m}{E}S + \frac{S^2-V^2}{2E}
             \right] + U^D \\
     U^D &=& \frac{1}{2\mu} \left[ -\frac{1}{2r^2 D} \frac{d}{dr} (r^2 D^\prime)
      + \frac{3}{4} \left( \frac{D^\prime}{D} \right)^2 \right] \\
  U^{LS} &=& - \frac{1}{2\mu} \frac{D^\prime}{r D} \\
       D &=& 1 + \frac{S-V}{E+m}   \; .
\end{eqnarray}
\end{subequations}
$S$ and $V$ \footnote{
Note that the calculations in Ref. \cite{kelly96} using \texttt{LEA} neglected 
the $(S-V)$ term and replaced the momentum in Eq. 
(\ref{equation:SpinorDistortion}) by its asymptotic value, an approach later 
called EMA-noSV, where EMA denotes the Effective Momentum Approximation.
} 
are respectively the scalar and vector potential terms of the original 
four-component Dirac equation (see Ref \cite{kelly96}).  $D(r)$ is known as the
Darwin nonlocality factor and $U^C$ and $U^{LS}$ are the central and spin-orbit
potentials.  The Darwin potential $U^D$ is generally quite small.  The upper 
and lower components of the Dirac wave function are then obtained using
\begin{subequations}
\label{equation:Darwin}
\begin{eqnarray}
\label{equation:DarwinFactor}
 \psi &=& D^{1/2} \xi  \\
\label{equation:SpinorDistortion}
\zeta &=& \frac{\bm{\sigma}\cdot{\bm{p}}\hspace*{0.5mm}~\psi}{E+m+S-V}   \; .
\end{eqnarray}
\end{subequations}
This method is known as direct Pauli reduction \cite{udias95,hedayati95}.  A 
very similar approach is also employed by Meucci {\it et al.} \cite{meucci01}.
A somewhat similar approach based on the Eikonal Approximation (see the 
discussion of the \textsc{romea} calculations in Section \ref{sec:romea_rmsga}) 
has been employed by Radici {\it et al.} \cite{radici02,radici03}.

For our purposes, the two most important differences between relativistic and 
nonrelativistic \textsc{dwia} calculations are the suppression of the interior 
wave function by the Darwin factor in Eq. (\ref{equation:DarwinFactor}), 
and the dynamical enhancement of the lower components of the Dirac spinor (also
known as `spinor distortion') by the strong Dirac scalar and vector potentials 
in Eq. (\ref{equation:SpinorDistortion}).  

As demonstrated in Refs.  \cite{boffi87,jin94,udias95}, the Darwin factor tends
to increase the normalization factors deduced using an \textsc{rdwia} analysis.
Distortion of the bound-nucleon spinor destroys factorization and at large 
$p_{\rm miss}$ produces important oscillatory signatures in the interference 
response functions, $A_{LT}$, and recoil polarization -- see Refs. 
\cite{caballero98a,kelly99b,udias99,udias00,martinez02}.  The effect of spinor 
distortion within the Effective Momentum Approximation (EMA) has been studied 
by Kelly \cite{kelly99b}.  The \texttt{LEA} code has subsequently been upgraded
to evaluate Eq. (\ref{equation:Darwin}) without applying the EMA.  These two 
methods for constructing the ejectile distorted waves should be equivalent. The
predictions of the \texttt{LEA} and the Madrid codes given identical input are 
compared in Section \ref{sec:RDWIA-tests}.  

The approximations made by \textsc{dwia} violate current conservation and 
introduce gauge ambiguities.  The most common prescriptions
\begin{subequations}
\begin{eqnarray}
{\cal J}_q &\rightarrow& \frac{\omega}{q} {\cal J}_0 \\
{\cal J}_\mu &\rightarrow& {\cal J}_\mu + \frac{ {\cal J}\cdot q}{Q^2} q_\mu \\
{\cal J}_0 &\rightarrow& \frac{q}{\omega} {\cal J}_q
\end{eqnarray}
\end{subequations}
correspond to Coulomb, Landau, and Weyl gauges, respectively.  Typically, 
Gordon ambiguities and sensitivity to details of the off-shell extrapolation 
are largest in the Weyl gauge.  Although there is no fundamental preference for
any of these prescriptions, it appears that the data are in general least 
supportive of the Weyl gauge.  Further, the \textsc{cc1} operator is the most 
sensitive to spinor distortion while the \textsc{cc3} operator is the least.  
The intermediate \textsc{cc2} is chosen most often for \textsc{rdwia}.

Besides the interaction in the final state of the outgoing proton, in any 
realistic calculation with finite nuclei, the effect of the distortion of the 
electron wave function must be taken into account.  For relatively light 
nuclei and large kinetic energies, the EMA for electron distortion (the 
$q_{\rm eff}$ Approximation) is sufficient -- see Refs.
\cite{schiff56,giusti87}. In this approach, the electron current is approximated
by
\begin{equation}
j^\mu(\bm{q}_{\rm eff})
\approx
\frac{\bar{k}_i \bar{k}_f}{k_i k_f}
\bar{u}(\bar{\bm{k}}_f) \gamma^\mu u (\bar{\bm{k}}_i)~,
\end{equation}
where
$\bm{q}_{\rm eff}=\bar{\bm{k}}_i-\bar{\bm{k}}_f$ is the effective momentum
transfer based upon the effective wave numbers
\begin{equation}
\label{equation:eff_wave_nums}
\bar{\bm{k}}
=
{\bm{k}} + f_Z \frac{\alpha Z}{R_Z} \hat{\bm{k}}
\end{equation}
with $f_Z \approx 1.5$ and $R_Z \approx 1.2 A^{1/3}$.  
For all the \textsc{rdwia} calculations subsequently presented in this paper 
that are compared directly with data, this $q_{\rm eff}$ Approximation has been
used to account for electron Coulomb distortion.  Only the \textsc{rpwia} and 
\textsc{rdwia$_{\rm(U^{opt}=0)}$} comparison calculations shown in Fig. 
\ref{fig:pwia_dwia}, the baseline \textsc{rdwia} comparison calculations shown 
in Figs. \ref{fig:sigma_baseline} and \ref{fig:alt_baseline}, and the 
\textsc{rdwia} and \textsc{rmsga} comparison calculations shown in Fig.
\ref{fig:rdwia_rmsga} omit the effect of electron Coulomb distortion.  This is 
equivalent to setting $f_Z=0$ in Eq. (\ref{equation:eff_wave_nums}).

\subsubsection{Tests}
\label{sec:RDWIA-tests}
                                                                                
\begin{figure}
\resizebox{0.47\textwidth}{!}{\includegraphics{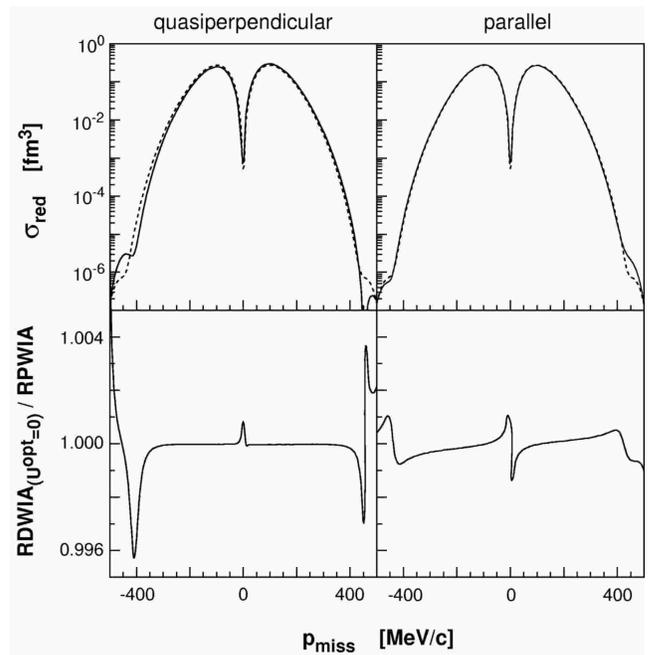}}
\caption{\label{fig:pwia_dwia}
Comparisons between \textsc{rpwia} and \textsc{rdwia$_{\rm(U^{opt}=0)}$}
calculations for the removal of protons from the $1p_{1/2}$-state of $^{16}$O 
as a function of $p_{\rm miss}$ for $E_{\rm beam}$ = 2.442 GeV.  In the upper 
panels, the solid curves represent the reduced cross sections for both the 
\textsc{rdwia$_{\rm(U^{opt}=0)}$} and the \textsc{rpwia} calculations (see text 
for details). The dashed curves correspond to the momentum distributions.  In 
the lower panels, \textsc{rdwia$_{\rm(U^{opt}=0)}$}$ 
\hspace*{0.5mm}$/$\hspace*{0.5mm}$\textsc{rpwia} reduced cross-section ratios 
are shown.  Agreement to much better than 1\% is obtained for both kinematics 
over the entire $p_{\rm miss}$ range.
}
\end{figure}

Fig. \ref{fig:pwia_dwia} illustrates a comparison between \textsc{rpwia} and 
\textsc{rdwia$_{\rm(U^{opt}=0)}$} calculations made by Kelly using \texttt{LEA} 
for the removal of protons from the $1p_{1/2}$-state of $^{16}$O as a function 
of $p_{\rm miss}$ for both quasiperpendicular and parallel kinematics for 
$E_{\rm beam}$ = 2.442 GeV.  
The \textsc{rdwia$_{\rm(U^{opt}=0)}$} calculations employed a partial-wave 
expansion of the second-order Dirac equation with optical potentials nullified 
and the target mass artificially set to 16001$u$ to minimize recoil corrections
and frame ambiguities.  The \textsc{rpwia} calculations (see Ref. 
\cite{amaldi67}) are based upon the Fourier transforms of the upper and lower 
components of the overlap function; that is, no partial-wave expansion is 
involved.  In the upper panels, the solid curves represent the reduced cross 
section for both the \textsc{rdwia$_{\rm(U^{opt}=0)}$} and \textsc{rpwia} 
calculations as the differences are indistinguishable on this scale.  The 
dashed curves show the momentum distributions.  In the lower panels, the ratios
between \textsc{rdwia$_{\rm(U^{opt}=0)}$} and \textsc{rpwia} reduced cross 
sections are shown.  With suitable choices for step size and maximum $\ell$ 
(here 0.05 fm and 80), agreement to much better than $1\%$ over the entire 
range of missing momentum is obtained, verifying the accuracy of \texttt{LEA} 
for plane waves.  Similar results are obtained with the Madrid code.  

The similarity between the reduced cross sections and the momentum 
distributions demonstrates that the violation of factorization produced by the 
distortion of the bound-state spinor is mild, but tends to increase with 
$p_{\rm miss}$.  Nevertheless, observables such as $A_{LT}$ that are sensitive 
to the interference between the lower and upper components are more strongly 
affected by violation of factorization.  

\begin{figure}
\resizebox{0.47\textwidth}{!}{\includegraphics{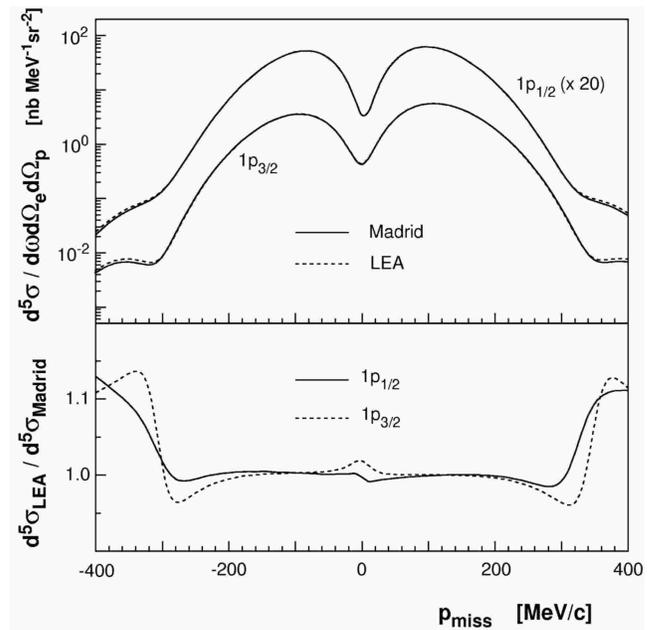}}
\caption{\label{fig:sigma_baseline}
Comparison baseline \textsc{rdwia} calculations by the Madrid Group and Kelly
(\texttt{LEA}) for the removal of protons from the 1$p$-shell of $^{16}$O as a
function of $p_{\rm miss}$ for $E_{\rm beam}$ = 2.442 GeV.  For the purposes of
this comparison, the input into both calculations was identical (see Table
\ref{table:baseline_options}).  Overall agreement is very good, and agreement
is excellent for $-$250 $<$ $p_{\rm miss}$ $<$ 250 MeV/$c$.
}
\end{figure}

\begin{figure}
\resizebox{0.47\textwidth}{!}{\includegraphics{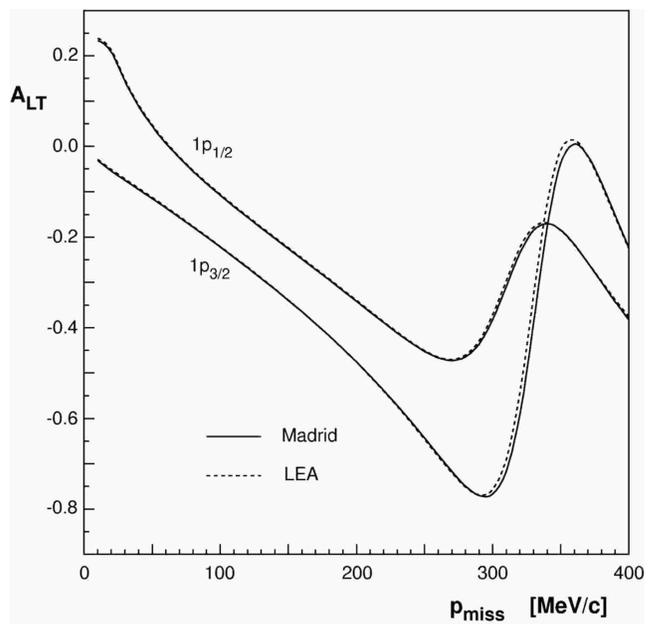}}
\caption{\label{fig:alt_baseline}
Comparison baseline \textsc{rdwia} calculations for the $A_{LT}$ asymmetry by 
the Madrid Group and Kelly (\texttt{LEA}) for the removal of protons from the 
1$p$-shell of $^{16}$O as a function of $p_{\rm miss}$ for $E_{\rm beam}$ $=$ 
2.442 GeV.  For the purposes of this comparison, the input into both 
calculations was identical (see Table \ref{table:baseline_options}).  Overall 
agreement is excellent over the entire $p_{\rm miss}$ range.
}
\end{figure}

Figs. \ref{fig:sigma_baseline} and \ref{fig:alt_baseline} compare 
calculations for the removal of protons from the $1p$-shell of $^{16}$O as a 
function of $p_{\rm miss}$ for $E_{\rm beam}$ = 2.442 GeV.  These predictions 
were made by Kelly \cite{kellytech01} using the \texttt{LEA} and the Madrid 
codes with identical input options, and are hereafter described as `baseline' 
calculations.  The baseline options are summarized in Table 
\ref{table:baseline_options} and were chosen to provide the most rigorous test 
of the codes rather than to be the optimal physics choices.  

\begin{table}
\caption{\label{table:baseline_options}
A summary of the basic \textsc{rdwia} options which served as input to the 
`baseline' comparison calculations of the Madrid Group and Kelly 
(\texttt{LEA}).  Results are shown in Figs. \ref{fig:sigma_baseline} and 
\ref{fig:alt_baseline}.
}
\begin{ruledtabular}
\begin{tabular}{rr}
            Input Parameter &          Option \\
\hline
bound-nucleon wave function &   \textsc{nlsh} -- Sharma {\it et al.} \cite{sharma93} \\
              Optical Model & \textsc{edai-o} -- Cooper {\it et al.} \cite{cooper93} \\
  nucleon spinor distortion &                                           relativistic \\
        electron distortion &                                                   none \\
           current operator &                                           \textsc{cc2} \\
       nucleon form factors &                                                 dipole \\
                      gauge &                                               Coulomb  \\
\end{tabular}
\end{ruledtabular}
\end{table}

Fig. \ref{fig:sigma_baseline} demonstrates that baseline cross-section 
calculations agree to better than $2\%$ for $p_{\rm miss}$ $<$ 250 MeV/$c$, but
that the differences increase to about $10\%$ by about 400 MeV/$c$.  
Nevertheless, Fig. \ref{fig:alt_baseline} shows that excellent agreement is 
obtained for $A_{LT}$ over this entire range of $p_{\rm miss}$, with only a 
very small observable shift.  The agreement of the strong oscillations in 
$A_{LT}$ for $p_{\rm miss}$ $\approx$ 300 MeV/$c$ predicted by both methods 
demonstrates that they are equivalent with respect to spinor distortion.  The 
small differences in the cross section for large $p_{\rm miss}$ appear to be 
independent of the input choices and probably arise from numerical errors in 
the integration of differential equations (perhaps due to initial conditions), 
but the origin has not yet been identified.  Regardless, it is remarkable to 
achieve this level of agreement between two independent codes under conditions 
in which the cross section spans three orders of magnitude.

\subsection{\label{sec:romea_rmsga}ROMEA$~$/$~$RMSGA}

In this subsection, an alternate relativistic model developed by the Ghent 
Group \cite{debruyne00,debruyne02a,debruyne02b,ryckebusch03} for $A(e,e'N)B$ 
processes is presented.  With respect to the construction of the
bound-nucleon wave functions and the nuclear-current operator, an approach 
similar to standard \textsc{rdwia} is followed.  The major differences lie in 
the construction of the scattering wave function.  The approach presented here 
adopts the relativistic Eikonal Approximation (EA) to determine the scattering 
wave functions and may be used in conjunction with either the Optical Model or 
the multiple-scattering Glauber frameworks for dealing with the FSI.

\subsubsection{Formalism}
\label{sec:romea_rmsga-formalism}

The EA belongs to the class of semi-classical approximations which are meant to
become `exact' in the limit of small de Broglie ($db$) wavelengths, 
$\lambda _{db} \ll a$, where $a$ is the typical range of the potential in which
the particle is moving.  For a particle moving in a relativistic (optical) 
potential consisting of scalar and vector terms, the scattering wave function 
takes on the EA form

\begin{equation}
\psi_{F}(\bm{r}\hspace*{0.5mm}) \sim
\left[
\begin{array}{c}
1 \\
\frac{1}{E+m+S-V}~ \bm{\sigma} \cdot \bm{p}
\end{array}
\right]
e^{i \bm{p} \cdot \bm{r}} {e^{i {\cal S} 
(\bm{r})}}
\chi_{m_{s}} \; .
\label{eq:thescatteringwave}
\end{equation}

\noindent
This wave function differs from a relativistic plane wave in two respects:  
first, there is a dynamical relativistic effect from the scalar ($S$) and 
vector ($V$) potentials which enhances the contribution from the lower 
components; and second, the wave function contains an eikonal phase which is 
determined by integrating the central ($U^C$) and spin-orbit ($U^{LS}$) terms 
of the distorting potentials along the (asymptotic) trajectory of the escaping 
particle.  In practice, this amounts to numerically calculating the integral 
($ \bm{r} \equiv (\bm{b},z)$)

\begin{eqnarray}
i {\cal S}(\bm{b},z)  & =  & - i \frac{m}{K} \int_{-\infty}^{z}
dz' \,
\biggl[
{U^{C} (\bm{b},z')} \nonumber \\ & &
+ {U^{LS} (\bm{b},z')}
[ \bm{\sigma} \cdot (\bm{b}
\times \bm{K} )- i Kz'\hspace*{0.25mm}] \biggr] \; ,
\label{eq:eikonalphase}
\end{eqnarray}

\noindent
where $\bm{K} \equiv \frac{1}{2} \left(\bm{p} + \bm{q}\hspace*{0.25mm}\right) $.

Within the \textsc{romea} calculation, the eikonal phase given by Eq.
(\ref{eq:eikonalphase}) is computed from the relativistic optical potentials as
they are derived from global fits to elastic proton-nucleus scattering data.  
It is worth stressing that the sole difference between the \textsc{romea} and 
the \textsc{rdwia} models is the use of the EA to compute the scattering wave 
functions.

For proton lab momenta exceeding 1 ~GeV/$c$, the highly inelastic nature of the
elementary nucleon-nucleon ($NN$) scattering process makes the use of a 
potential method for describing FSI effects somewhat artificial.  In this 
high-energy regime, an alternate description of FSI processes is provided by 
the Glauber Multiple-Scattering Theory.  A relativistic and unfactorized 
formulation of this theory has been developed by the Ghent Group 
\cite{debruyne02b,ryckebusch03}.  In this framework, the $A$-body wave function
in the final state reads

\begin{eqnarray}
\Psi ^{\bm{p}}_{A} \left( \bm{r}, \bm{r}_2,
\bm{r}_3, \ldots \bm{r}_A \right)  & \sim &
{\widehat{\mathcal{O}}}
\left[
\begin{array}{c}
1 \\
\frac{1}{E+m}~\bm{\sigma} \cdot \bm{p}
\end{array}
\right]
e^{i \bm{p} \cdot \bm{r}}
\chi_{m_{s}} \nonumber \\ & & \times \Psi _{B} \left( \bm{r}_2,
\ldots \bm{r}_A \right)  \; ,
\label{eq:glauber}
\end{eqnarray}
\noindent
where $\Psi _{B}$ is the wave function characterizing the state in which the 
$B$ nucleus is created.  In the above expression, the subsequent elastic or 
`mildly inelastic' collisions which the ejectile undergoes with `frozen' 
spectator nucleons are implemented through the introduction of the operator

\begin{displaymath}
{\widehat{\mathcal{O}} \left( \bm{r}, \bm{r}_2,
\bm{r}_3, \ldots \bm{r}_A \right) }
\equiv \prod _ {j=2} ^{A} \left[ 1 - \Gamma \left({p}, \bm{b} - \bm{b_j}
\right) \theta \left( z - z_j \right) \right]~,
\end{displaymath}

\noindent
where the profile function for $pN$ scattering is

\begin{eqnarray*}
\Gamma (p,\bm{b}) = \frac{{\sigma^{\rm tot}_{pN}}
(1-i {\epsilon_{pN}})}
{4\pi \beta_{pN}^{2}} \: \exp \left(
-\frac{b^{2}}{2\beta_{pN}^{2}} \right) \; .
\label{eq:profilefunction}
\end{eqnarray*}

\noindent

In practice, for the lab momentum of a given ejectile, the following 
input is required: the total proton-proton and proton-neutron cross 
section ${\sigma^{\rm tot}_{pN}}$, the slope parameters ${\beta_{pN}}$, 
and the ratio of the real-to-imaginary scattering amplitude ${\epsilon_{pN}}$.  
The parameters ${\sigma^{\rm tot}_{pN}}$, ${\beta_{pN}}$, and ${\epsilon_{pN}}$ 
are obtained through interpolation of the data base made available by 
the Particle Data Group \cite{pdg}.  The $A(e,e'N)B$ results obtained 
with a scattering state of the form of Eq. (\ref{eq:glauber}) are referred 
to as \textsc{rmsga} calculations.  It is worth stressing that in 
contrast to the \textsc{rdwia} and the \textsc{romea} models, all 
parameters entering the calculation of the scattering states in 
\textsc{rmsga} are directly obtained from the elementary proton-proton 
and proton-neutron scattering data.  
Thus, the scattering states are not subject to the $SV$ effects discussed
in Section \ref{sec:RDWIA-formalism}, which typically arise when relativistic 
potentials are employed.  However, the $SV$ effects are included for the 
bound-state wave function.

Note that for the kinematics of the $^{16}$O$(e,e'p)$ experiment presented in 
this paper, the de Broglie wavelength of the ejected proton is $\lambda _{db} 
\approx 1.3$~fm, and thus both the optical potential and the Glauber frameworks
may be applicable.  Indeed, for $T_p \approx 0.433$ GeV, various sets of 
relativistic optical potentials are readily available and $\lambda _{db}$ 
appears sufficiently small for the approximations entering the Glauber 
framework to be justifiable -- see Ref. \cite{debruyne02b}.

\subsubsection{Tests}
\label{sec:rmsga-tests}

As previously mentioned, the (\textsc{rdwia} and \textsc{romea}) and 
\textsc{rmsga} frameworks are substantially different in the way they address 
FSI.  While the \textsc{rdwia} and \textsc{romea} models are both essentially
one-body approaches in which all FSI effects are implemented through effective 
potentials, the \textsc{rmsga} framework is a full-fledged, multi-nucleon 
scattering model based on the EA and the concept of frozen spectators.  As 
such, when formulated in an unfactorized and relativistic framework, Glauber 
calculations are numerically involved and the process of computing the 
scattering state and the transition matrix elements involves numerical methods 
which are different from those adopted in \textsc{rdwia} frameworks.  For 
example, for $A(e,e^{\prime}N)B$ calculations in the \textsc{romea} and the
\textsc{rmsga}, partial-wave expansions are simply not a viable option.

The testing of the mutual consistency of the \textsc{rdwia} and `bare' 
\textsc{rmsga} (no MEC nor IC) calculations began by considering the special 
case of vanishing FSI.  In this limit, where all the Glauber phases are 
nullified in \textsc{rmsga} and (\textsc{rdwia} $\rightarrow$ \textsc{rpwia}), 
the two calculations were determined to reproduce one another to 4\% over the 
entire $p_{\rm miss}$ range, thereby establishing the validity of the numerics.
The Glauber phases were then enabled.  The basic options which then served as 
input to the comparison between the \textsc{rdwia} calculations and the 
\textsc{rmsga} calculations of the Ghent Group \cite{ryckebuschtech01} are 
presented in Table \ref{table:rdwia_rmsga_options}.
                                                                                
\begin{table}
\caption{\label{table:rdwia_rmsga_options}
A summary of the basic options which served as input to the comparison 
between the \textsc{rdwia} calculations and the `bare' \textsc{rmsga} (no MEC
nor IC) calculations of the Ghent Group.  Results are shown in Fig. 
\ref{fig:rdwia_rmsga}.
}
\begin{ruledtabular}
\begin{tabular}{rr}
            Input Parameter &                                    Option \\
\hline
bound-nucleon wave function & Furnstahl {\it et al.} \cite{furnstahl97} \\
              Optical Model &                           \textsc{edai-o} \\
  nucleon spinor distortion &                              relativistic \\
        electron distortion &                                      none \\
           current operator &                              \textsc{cc2} \\
       nucleon form factors &                                    dipole \\
                      gauge &                                   Coulomb \\
\end{tabular}
\end{ruledtabular}
\end{table}

\begin{figure}
\resizebox{0.47\textwidth}{!}{\includegraphics{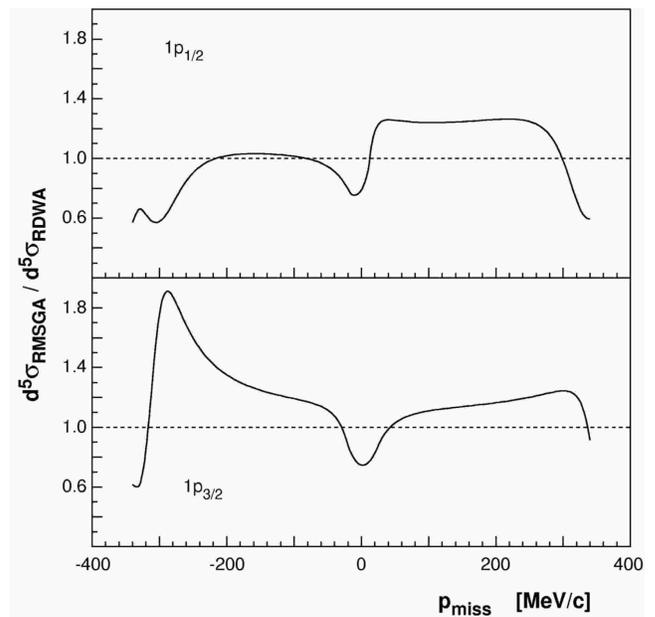}}
\caption{\label{fig:rdwia_rmsga}
\textsc{rdwia} calculations compared to `bare' (no MEC nor IC) \textsc{rmsga} 
calculations by the Ghent Group for the removal of protons from the 1$p$-shell 
of $^{16}$O as a function of $p_{\rm miss}$ for $E_{\rm beam}$ = 2.442 GeV. 
Both calculations employ the input presented in Table 
\ref{table:rdwia_rmsga_options}.  Apart from the treatment of FSI, all 
ingredients are identical.
}
\end{figure}

Fig. \ref{fig:rdwia_rmsga} shows the ratio of the bare \textsc{rmsga} 
calculations of the Ghent Group together with \textsc{rdwia} calculations for 
the removal of protons from the 1$p$-shell of $^{16}$O as a function of 
$p_{\rm miss}$ for $E_{\rm beam}$ = 2.442 GeV.  Apart from the treatment of 
FSI, all other ingredients to the calculations are identical (see Table 
\ref{table:rdwia_rmsga_options}).  For $p_{\rm miss}$ below the Fermi momentum, 
the variation between the predictions of the two approaches is at most 25\%, 
with the \textsc{rdwia} approach predicting a smaller cross section (stronger 
absorptive effects) than the \textsc{rmsga} model.  Not surprisingly, at larger
$p_{\rm miss}$ (correspondingly larger polar angles), the differences between 
the two approaches grow.

\section{\label{sec:results_this_work}Results for $\bm{Q^{2} \approx 0.8}$ (G$\bm{\textrm{e}}$V/$\bm{c}$)$\bm{^2}$} 

The data were interpreted in subsets corresponding to the 1$p$-shell and to the
1$s_{1/2}$-state and continuum.  The interested reader is directed to the works
of Gao {\it et al.} \cite{gao00} and Liyanage {\it et al.} \cite{liyanage01}, 
where these results have been briefly highlighted.  Note that when data are 
presented in the following discussion, statistical uncertainties only are 
shown.  A complete archive of the data, including systematic uncertainties, is 
presented in Appendix \ref{appendix:appendix_a}.

\subsection{\label{subsec:pshell}$\bm{1p}$-shell knockout}

\subsubsection{\label{subsubsec:sensitivities}Sensitivity to \textsc{rdwia} variations}

The consistency of the normalization factors $S_{\alpha}$ suggested by the
$1p$-shell data for $p_{\rm miss}$ $<$ 350 MeV/$c$ obtained in this 
measurement at 2.442 GeV (see Table \ref{table:pshellperpresults}) was examined
within the \textsc{rdwia} framework in a detailed study by the Madrid Group 
\cite{udiaspriv2}.  The study involved systematically varying a wide range of 
inputs to the \textsc{rdwia} calculations, and then performing least-squares 
fits of the predictions to the cross-section data.  The results of the study 
are presented in Table \ref{table:cc12specfacs}.

\begin{table*}
\caption{\label{table:cc12specfacs}
Normalization factors derived from the 2.442 GeV $1p$-shell cross-section data
of Table \ref{table:pshellperpresults} using the \textsc{cc1} and \textsc{cc2}
current operators.  The first term in each column is for the $1p_{1/2}$-state,
while the second term is for the $1p_{3/2}$-state.}
\begin{ruledtabular}
\begin{footnotesize}
\begin{tabular}{rrr|rrr|rrr|rrrrr|rrr|rrr|rr|rr|rr|rr}
            \multicolumn{3}{c|}{} &                             \multicolumn{3}{c|}{bound-} &      \multicolumn{3}{c|}{} &                                                  \multicolumn{5}{c|}{} &     \multicolumn{3}{c|}{nucleon} &        \multicolumn{3}{c|}{} &                                                 \multicolumn{4}{c|}{} &                                                 \multicolumn{4}{c}{} \\
            \multicolumn{3}{c|}{} &                            \multicolumn{3}{c|}{nucleon} &      \multicolumn{3}{c|}{} &                                           \multicolumn{5}{c|}{optical} &          \multicolumn{3}{c|}{FF} & \multicolumn{3}{c|}{doublet} &                                                 \multicolumn{4}{c|}{} &                                                 \multicolumn{4}{c}{} \\
\multicolumn{3}{c|}{prescription} &                       \multicolumn{3}{c|}{wavefunction} & \multicolumn{3}{c|}{gauge} &                                         \multicolumn{5}{c|}{potential} &       \multicolumn{3}{c|}{model} &    \multicolumn{3}{c|}{(\%)} &                                     \multicolumn{4}{c|}{S$_{\alpha}$} &                                       \multicolumn{4}{c}{$\chi^{2}$} \\
\hline
              fully &      & EMA- & \multicolumn{2}{c}{\textsc{nls}} &             &            &      &        &         \multicolumn{3}{c}{\textsc{eda}} &              &              &             &     & \textsc{gk}+ &                     &    &   &             \multicolumn{2}{c|}{} &             \multicolumn{2}{c|}{} &             \multicolumn{2}{c|}{} &             \multicolumn{2}{c}{} \\
                rel & proj & noSV &        \textsc{h} & \textsc{h-p} & \textsc{hs} &          C &    W &      L & \textsc{i-o} & \textsc{d1} & \textsc{d2} & \textsc{mrw} & \textsc{rlf} & \textsc{gk} &   d & \textsc{qmc} &                 100 & 50 & 0 & \multicolumn{2}{c|}{\textsc{cc1}} & \multicolumn{2}{c|}{\textsc{cc2}} & \multicolumn{2}{c|}{\textsc{cc1}} & \multicolumn{2}{c}{\textsc{cc2}} \\
\hline
                  * &      &               &                 * &              &             &          * &      &        &            * &             &             &              &              &           * &     &              &                   * &    &   &                       0.68 & 0.62 &                       0.74 & 0.67 &                        5.5 &  5.3 &                      2.0 &  31.0 \\
                    &    * &               &                 * &              &             &          * &      &        &            * &             &             &              &              &           * &     &              &                   * &    &   &                       0.78 & 0.73 &                       0.76 & 0.71 &                       17.0 & 79.0 &                      8.0 &  70.0 \\
                    &      &             * &                 * &              &             &          * &      &        &            * &             &             &              &              &           * &     &              &                   * &    &   &                       0.72 & 0.66 &                       0.75 & 0.69 &                        2.3 & 65.0 &                      2.2 &  65.0 \\
\hline
                  * &      &               &                   &            * &             &          * &      &        &            * &             &             &              &              &           * &     &              &                   * &    &   &                       0.60 & 0.52 &                       0.63 & 0.54 &                       10.0 & 97.0 &                     15.0 & 115.0 \\
                  * &      &               &                   &              &           * &          * &      &        &            * &             &             &              &              &           * &     &              &                   * &    &   &                       0.62 & 0.61 &                       0.65 & 0.65 &                       10.0 &  6.7 &                     18.0 &  41.0 \\
\hline
                  * &      &               &                 * &              &             &            &    * &        &            * &             &             &              &              &           * &     &              &                   * &    &   &                       0.63 & 0.59 &                       0.76 & 0.70 &                       25.0 &  9.2 &                      2.6 &  22.0 \\
                  * &      &               &                 * &              &             &            &      &      * &            * &             &             &              &              &           * &     &              &                   * &    &   &                       0.69 & 0.63 &                       0.73 & 0.67 &                        3.7 &  6.4 &                      2.5 &  34.0 \\
\hline
                  * &      &               &                 * &              &             &          * &      &        &              &           * &             &              &              &           * &     &              &                   * &    &   &                       0.64 & 0.60 &                       0.72 & 0.67 &                       29.0 & 12.0 &                      4.8 &   8.2 \\
                  * &      &               &                 * &              &             &          * &      &        &              &             &           * &              &              &           * &     &              &                   * &    &   &                       0.64 & 0.59 &                       0.71 & 0.65 &                       15.0 &  6.4 &                      0.7 &  15.0 \\
                  * &      &               &                 * &              &             &          * &      &        &              &             &             &            * &              &           * &     &              &                   * &    &   &                       0.62 & 0.60 &                       0.71 & 0.67 &                       35.0 & 11.0 &                      7.6 &   7.3 \\
                  * &      &               &                 * &              &             &          * &      &        &              &             &             &              &            * &           * &     &              &                   * &    &   &                       0.61 & 0.58 &                       0.70 & 0.65 &                       41.0 & 12.0 &                      6.1 &   7.9 \\
\hline
                  * &      &               &                 * &              &             &          * &      &        &            * &             &             &              &              &             &   * &              &                   * &    &   &                       0.69 & 0.63 &                       0.75 & 0.68 &                        4.8 &  5.9 &                      2.1 &  31.0 \\
                  * &      &               &                 * &              &             &          * &      &        &            * &             &             &              &              &             &     &            * &                   * &    &   &                       0.65 & 0.61 &                       0.72 & 0.66 &                       11.0 &  3.3 &                      0.5 &  16.0 \\
\hline
                  * &      &               &                 * &              &             &          * &      &        &            * &             &             &              &              &           * &     &              &                     &  * &   &                            & 0.64 &                            & 0.70 &                            &  6.1 &                          &  33.0 \\
                  * &      &               &                 * &              &             &          * &      &        &            * &             &             &              &              &           * &     &              &                     &    & * &                            & 0.66 &                            & 0.72 &                            &  7.4 &                          &  35.0 \\
\end{tabular}
\end{footnotesize}
\end{ruledtabular}
\end{table*}

Three basic approaches were considered: the fully relativistic approach,
the projected approach of Ud{\'\i}as {\it et al.} \cite{udias99,udias00}, and 
the EMA-noSV approach of Kelly \cite{kelly97,kelly99b}.  All three approaches 
included the effects of electron distortion.  While the fully relativistic 
approach involved solving the Dirac equation directly in configuration space, 
the projected approach included only the positive-energy components, and as a 
result, most (but not all) of the spinor distortion was removed from the wave 
functions.  Within the EMA-noSV approach, a relativized Schr\"{o}dinger 
equation was solved using the EMA, and all of the spinor distortion was 
removed.  This made the calculation similar to a factorized calculation, 
although spin--orbit effects in the initial and final states (which cause small
deviations from the factorized results) are included in EMA-noSV.  

\begin{figure}
\resizebox{0.48\textwidth}{!}{\includegraphics{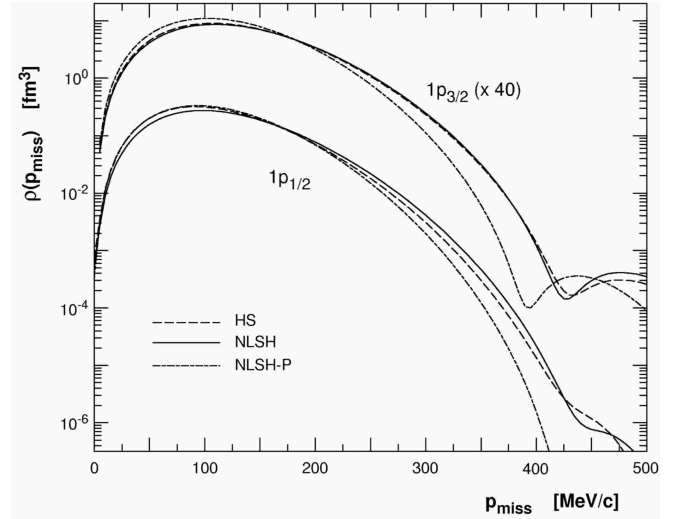}}
\caption{\label{fig:hsnlshnlshp}
Momentum distributions for the \textsc{hs}, \textsc{nlsh}, and \textsc{nlsh-p} 
models.  There is only a slight difference between \textsc{hs} and 
\textsc{nlsh} -- for the $1p_{3/2}$-state, \textsc{hs} is broader spatially and
thus drops off faster with increasing $p_{\rm miss}$.   On the other hand,
\textsc{nlsh-p} differs appreciably from both \textsc{hs} and \textsc{nlsh},
and is clearly distinguishable for $p_{\rm miss}$ $>$ 250 MeV/$c$ for both the
$1p_{1/2}$- and $1p_{3/2}$-states.  Note that both the \textsc{nlsh} and 
\textsc{nlsh-p} wave functions predict binding energies, single-particle 
energies, and a charge radius for $^{16}$O which are all in good agreement with
the data.
}
\end{figure}

The current operator was changed between \textsc{cc1} and \textsc{cc2}.  Three 
bound-nucleon wave functions (see Fig. \ref{fig:hsnlshnlshp}) derived from 
relativistic Lagrangians were considered: \textsc{hs} by Horowitz and Serot 
\cite{horowitz81,horowitz91}, \textsc{nlsh} by Sharma {\it et al.} 
\cite{sharma93}, and \textsc{nlsh-p} by Ud{\'\i}as {\it et al.} \cite{udias01} 
(which resulted from a Lagrangian fine-tuned to reproduce the Leuschner {\it et
al.} data).  Note that both the \textsc{nlsh} and \textsc{nlsh-p} wave 
functions predict binding energies, single-particle energies, and a charge 
radius for $^{16}$O which are all in good agreement with the data.  
The gauge prescription was changed between Coulomb, Weyl, and Landau.  The 
nucleon distortion was evaluated using three purely phenomenological $SV$ 
optical potentials (\textsc{edai-o}, \textsc{edad1}, and \textsc{edad2}) by 
Cooper {\it et al.} \cite{cooper93}, as well as \textsc{mrw} by McNeil 
{\it et al.} \cite{mcneil83} and \textsc{rlf} by Horowitz \cite{horowitz85} and
Murdock \cite{murdock87}.  The nucleon form-factor model was changed between 
\textsc{gk} by Gari and Kr\"{u}mpelmann \cite{gari85} and the dipole model.  
Further, the \textsc{qmc} model of Lu {\it et al.} \cite{lu98,lu99} predicts a 
density dependence for form factors that was calculated and applied to the 
\textsc{gk} form factors using the LDA (see Ref. \cite{kelly99a}).  

Note that 
the calculations for the $1p_{3/2}$-state include the incoherent contributions 
of the unresolved $2s_{1/2}1d_{5/2}$-doublet.  The bound-nucleon wave functions 
for these positive-parity states were taken from the parametrization of 
Leuschner {\it et al.} and normalization factors were fit to said data using 
\textsc{rdwia} calculations.  Factors for both states of 0.12(3) relative to 
full occupancy were determined.  The sensitivity of the present data to this 
incoherent admixture was evaluated by scaling the fitted doublet contribution 
using factors of 0.0, 0.5, and 1.0.

Qualitatively, the fully relativistic approach clearly did the best job of 
reproducing the data.  Fully relativistic results were shown to be much less 
gauge-dependent than the nonrelativistic results.  The \textsc{cc2} current 
operator was in general less sensitive to choice of gauge, and the data 
discouraged the choice of the Weyl gauge.  The different optical models had 
little effect on the shape of the calculations, but instead changed the overall
magnitude.  Both the \textsc{gk} and dipole nucleon form-factor models produced
nearly identical results.  The change in the calculated \textsc{gk+qmc} cross 
section was modest, being most pronounced in $A_{LT}$ for $p_{\rm miss}$ $>$ 
300 MeV/$c$.  The results were best for a 100\% contribution of the strength of
the $2s_{1/2}1d_{5/2}$-doublet to the $1p_{3/2}$-state, although the data were 
not terribly sensitive to this degree of freedom.

\begin{figure}
\resizebox{0.47\textwidth}{!}{\includegraphics{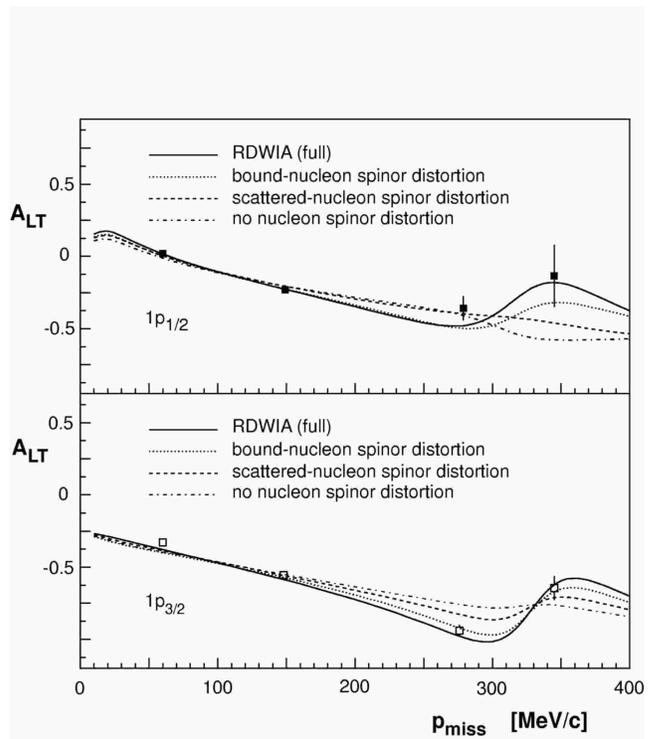}}
\caption{\label{fig:juncai2}
Left-right asymmetry $A_{LT}$ together with \textsc{rdwia} calculations for the
removal of protons from the 1$p$-shell of $^{16}$O as a function of 
$p_{\rm miss}$ for $E_{\rm beam}$ $=$ 2.442 GeV.  Uncertainties are statistical
(see Table \ref{table:pshellaltrlt} for the associated systematic 
uncertainties).  Note that the solid curves shown here are identical to those 
shown in Figs. \ref{fig:udiasalt} and \ref{fig:compare_onebody_alt}.
}
\end{figure}

Fig. \ref{fig:juncai2} shows the left-right asymmetry $A_{LT}$ together with
\textsc{rdwia} calculations for the removal of protons from the $1p$-shell of
$^{16}$O as a function of $p_{\rm miss}$ for $E_{\rm beam}$ $=$ 2.442 GeV.
The origin of the large change in the slope of $A_{LT}$ at $p_{\rm miss}$ 
$\approx$ 300 MeV/$c$ is addressed by the various calculations.  This 
`ripple' effect is due to the distortion of the bound-nucleon and ejectile 
spinors, as evidenced by the other three curves shown, in which the full 
\textsc{rdwia} calculations have been decomposed.  It is important to note 
that these three curves all retain the same basic ingredients, particularly 
the fully relativistic current operator and the upper components of the Dirac 
spinors.  Of the three curves, the dotted line resulted from a calculation 
where only the bound-nucleon spinor distortion was included, the dashed line 
resulted from a calculation where only the scattered-state spinor distortion 
was included, and the dashed-dotted line resulted from a calculation where 
undistorted spinors (essentially identical to a factorized calculation) were 
considered.  Clearly, the inclusion of the bound-nucleon spinor distortion is 
more important than the inclusion of the scattered-state spinor distortion, but
both are necessary to describe the data.

\begin{figure}
\resizebox{0.47\textwidth}{!}{\includegraphics{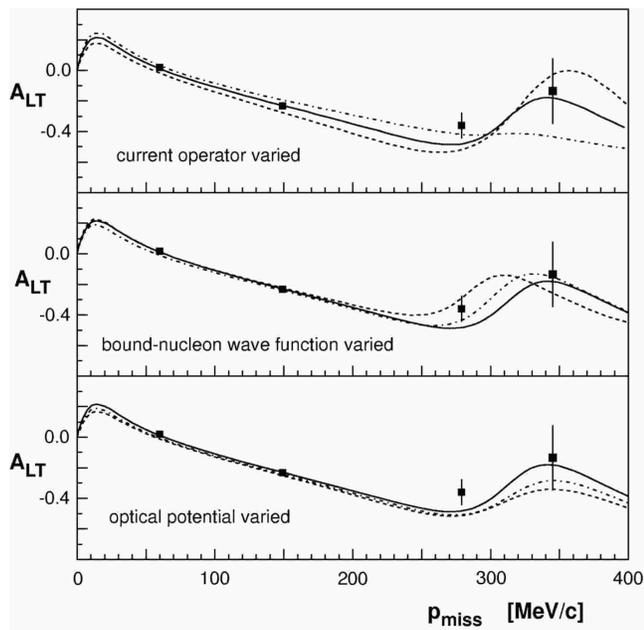}}
\caption{\label{fig:udiasalt}
Left-right asymmetry $A_{LT}$ together with \textsc{rdwia} calculations for the
removal of protons from the 1$p_{1/2}$-state of $^{16}$O as a function of
$p_{\rm miss}$ for $E_{\rm beam}$ $=$ 2.442 GeV.  Uncertainties are statistical
(see Table \ref{table:pshellaltrlt} for the associated systematic 
uncertainties).  The solid curves in all three panels are the same and are 
identical to those shown for the removal of protons from the 1$p_{1/2}$-state 
of $^{16}$O in Figs. \ref{fig:juncai2} and \ref{fig:compare_onebody_alt}.
}
\end{figure}

The effects of variations in the ingredients to the calculations of the
left-right asymmetry $A_{LT}$ for the $1p_{1/2}$-state only are shown in Fig.
\ref{fig:udiasalt}.  Note that the data are identical to those presented in 
Fig. \ref{fig:juncai2}, as are the solid curves.  In the top panel, the 
\textsc{edai-o} optical potential and \textsc{nlsh} bound-nucleon wave function
were used for all the calculations, but the choice of current operator was 
varied between \textsc{cc1} (dashed), \textsc{cc2} (solid), and \textsc{cc3} 
(dashed-dotted), resulting in a change in both the height and the 
$p_{\rm miss}$-location of the ripple in $A_{LT}$.  In the middle panel, the 
current operator \textsc{cc2} and \textsc{edai-o} optical potential were used 
for all the calculations, but the choice of bound-nucleon wave function was 
varied between \textsc{nlsh-p} (dashed), \textsc{nlsh} (solid), and \textsc{hs}
(dashed-dotted), resulting in a change in the $p_{\rm miss}$-location of the
ripple, but a relatively constant height.  In the bottom panel, the current
operator \textsc{cc2} and \textsc{nlsh} bound-nucleon wave function were used
for all the calculations, but the choice of optical potential was varied 
between \textsc{edad1} (dashed), \textsc{edai-o} (solid), and \textsc{edad2} 
(dashed-dotted), resulting in a change in the height of the ripple, but a 
relatively constant $p_{\rm miss}$-location.  More high-precision data, 
particularly for 150 $<$ $p_{\rm miss}$ $<$ 400 MeV/$c$, are clearly needed to 
accurately and simultaneously determine the current operator, the bound-state 
wave function, the optical potential, and of course the normalization factors.
This experiment has recently been performed in Hall A at Jefferson Lab by Saha 
{\it et al.} \cite{proposal00}, and the results are currently under analysis.

\subsubsection{\label{subsubsec:compare_onebody}Comparison to \textsc{rdwia}, \textsc{romea}, and \textsc{rmsga} calculations considering single-nucleon currents}

In this Section, the data are compared to \textsc{rdwia} and bare
\textsc{romea} and \textsc{rmsga} calculations (which take into consideration
single-nucleon currents only -- no MEC or IC).  The basic options employed in 
the calculations are summarized in Table \ref{table:rdwia_romea_rmsga_options}.
Note that both the EA-based calculations stop at $p_{\rm miss}$ $=$ 350 MeV/$c$
as the approximation becomes invalid.

\begin{table}
\caption{\label{table:rdwia_romea_rmsga_options}
A summary of the basic options which served as input to the single-nucleon 
current \textsc{rdwia}, \textsc{romea}, and \textsc{rmsga} comparison 
calculations.  Results are shown in Figs.  \ref{fig:juncai1} -- 
\ref{fig:juncai3}.
}
\begin{ruledtabular}
\begin{tabular}{rrr}
                            &                 &   \textsc{romea} \\
            Input Parameter &  \textsc{rdwia} & \& \textsc{rmsga}\\
\hline
bound-nucleon wave function &   \textsc{nlsh} &      \textsc{hs} \\
              Optical Model & \textsc{edai-o} &  \textsc{edai-o} \\
  nucleon spinor distortion &    relativistic &     relativistic \\
        electron distortion &             yes &              yes \\
           current operator &    \textsc{cc2} &     \textsc{cc2} \\
       nucleon form factors &     \textsc{gk} &           dipole \\
                      gauge &        Coulomb  &          Coulomb \\
\end{tabular}
\end{ruledtabular}
\end{table}

\begin{figure}
\resizebox{0.47\textwidth}{!}{\includegraphics{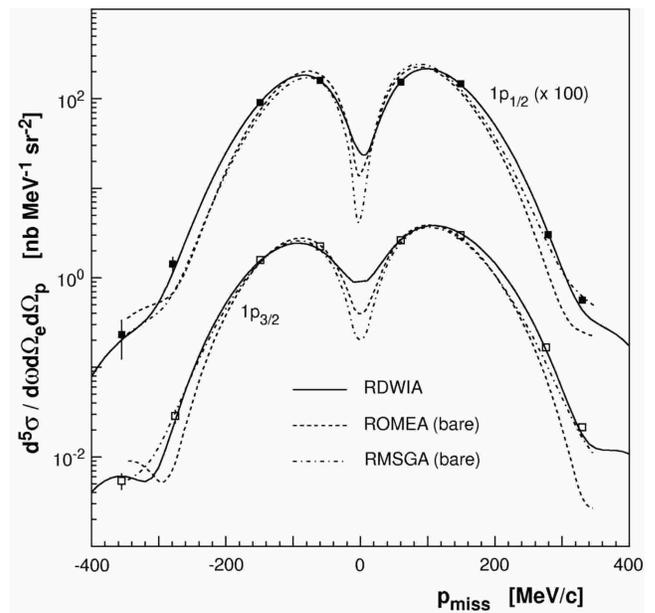}}
\caption{\label{fig:juncai1}
Measured cross-section data for the removal of protons from the 1$p$-shell of
$^{16}$O as a function of $p_{\rm miss}$ as compared to relativistic
calculations at $E_{\rm beam}$ = 2.442 GeV.  Uncertainties are statistical and,
on average, there is an additional $\pm$5.6\% systematic uncertainty (see Table 
\ref{table:pshellperpresults}) associated with the data.  The solid line is the
\textsc{rdwia} calculation, while the dashed and dashed-dotted lines are 
respectively the bare \textsc{romea} and \textsc{rmsga} calculations.
}
\end{figure}

Fig. \ref{fig:juncai1} shows measured cross-section data for the removal of 
protons from the 1$p$-shell of $^{16}$O as a function of $p_{\rm miss}$ as 
compared to relativistic calculations at $E_{\rm beam}$ = 2.442 GeV.  The solid
line is the \textsc{rdwia} calculation, while the dashed and dashed-dotted 
lines are respectively the bare \textsc{romea} and \textsc{rmsga} calculations.
The normalization factors for the \textsc{rdwia} calculations are 0.73 and 0.72
for the $1p_{1/2}$-state and $1p_{3/2}$-state, respectively.  For the 
\textsc{romea} and \textsc{rmsga} calculations, they are 0.6 and 0.7 for the 
$1p_{1/2}$-state and $1p_{3/2}$-state, respectively.  The \textsc{rdwia} 
calculations do a far better job of representing the data over the entire 
$p_{\rm miss}$ range.

\begin{figure}
\resizebox{0.47\textwidth}{!}{\includegraphics{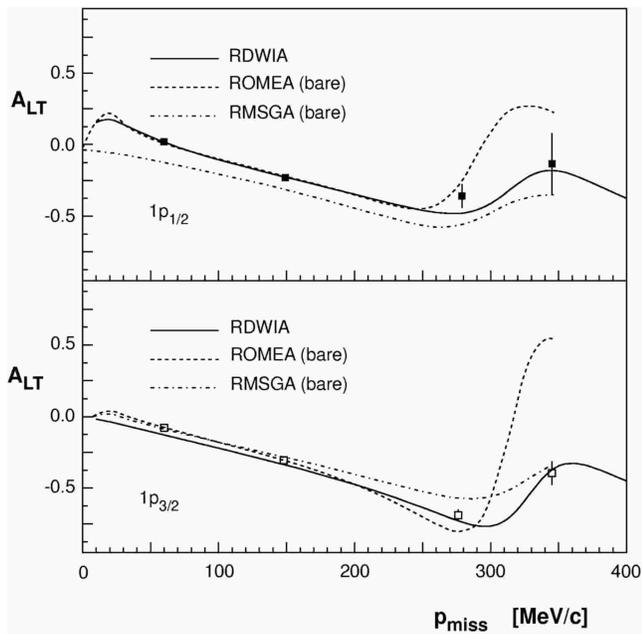}}
\caption{\label{fig:compare_onebody_alt}
Left-right asymmetry $A_{LT}$ together with relativistic calculations of the
$A_{LT}$ asymmetry for the removal of protons from the 1$p$-shell of $^{16}$O
as a function of $p_{\rm miss}$ for $E_{\rm beam}$ = 2.442 GeV.  Uncertainties
are statistical (see Table \ref{table:pshellaltrlt} for the associated 
systematic uncertainties).  The solid line is the \textsc{rdwia} calculation, 
while the dashed and dashed-dotted lines are respectively the bare 
\textsc{romea} and \textsc{rmsga} calculations.  Note that the solid curves 
shown here are identical to those shown in Figs. \ref{fig:juncai2} and 
\ref{fig:udiasalt}.
}
\end{figure}

Fig. \ref{fig:compare_onebody_alt} shows the left-right asymmetry $A_{LT}$
together with relativistic calculations for the removal of protons from the
1$p$-shell of $^{16}$O as a function of $p_{\rm miss}$ for $E_{\rm beam}$ = 
2.442 GeV.  The solid line is the \textsc{rdwia} calculation, while the dashed 
and dashed-dotted lines are respectively the bare \textsc{romea} and 
\textsc{rmsga} calculations.  Note again the large change in the slope of 
$A_{LT}$ at $p_{\rm miss}$ $\approx$ 300 MeV/$c$.  While all three calculations
undergo a similar change in slope, the \textsc{rdwia} calculation does the best
job of reproducing the data.  The \textsc{romea} calculation reproduces the 
data well for $p_{\rm miss}$ $<$ 300 MeV/$c$, but substantially overestimates 
$A_{LT}$ for $p_{\rm miss}$ $>$ 300 MeV/$c$. The \textsc{rmsga} calculation 
does well with the overall trend in the data, but struggles with reproducing 
the data for the $1p_{1/2}$-state.

\begin{figure}
\resizebox{0.47\textwidth}{!}{\includegraphics{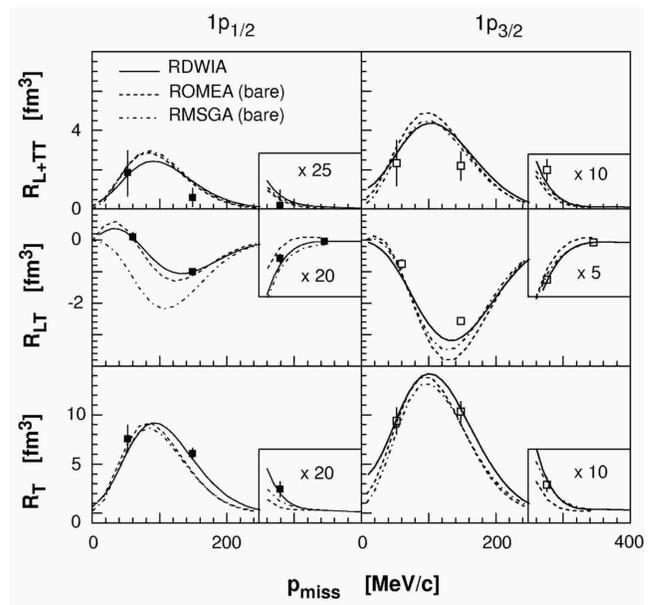}}
\caption{\label{fig:juncai3}
Data from this work together with relativistic calculations for the $R_{L+TT}$,
$R_{LT}$, and $R_{T}$ effective response functions for the removal of protons 
from the 1$p$-shell of $^{16}$O as a function of $p_{\rm miss}$.  Uncertainties
are statistical (see Tables \ref{table:pshellrlttrt}, \ref{table:pshellaltrlt},
and \ref{table:pshellrlrt} for the associated systematic uncertainties).  The 
solid line is the \textsc{rdwia} calculation, while the dashed and 
dashed-dotted lines are respectively the bare \textsc{romea} and \textsc{rmsga}
calculations.
}
\end{figure}

Fig. \ref{fig:juncai3} shows the $R_{L+TT}$, $R_{LT}$, and $R_{T}$ effective
response functions together with relativistic calculations for the removal of
protons from the 1$p$-shell of $^{16}$O as a function of $p_{\rm miss}$.  Note
that the data point located at $p_{\rm miss}$ $\approx$ 52 MeV/$c$ comes from 
the parallel kinematics measurements 
\footnote{
Strictly speaking, the effective longitudinal response function $R_{L}$ could 
not be separated from the quasiperpendicular kinematics data.  However, since 
both Kelly and Ud{\'\i}as {\it et al.} calculate the term 
$\frac{v_{TT}}{v_L}R_{TT}$ to be $<$ 10\% of $R_{L+TT}$ in these kinematics, 
$R_{L}$ and $R_{L+TT}$ responses are both presented on the same plot.
}
(see Table \ref{table:pshellrlrt}), while the other data points come from the
quasiperpendicular kinematics measurements (see Tables \ref{table:pshellrlttrt}
and \ref{table:pshellaltrlt}).  The solid line is the \textsc{rdwia} 
calculation, while the dashed and dashed-dotted lines are respectively the bare 
\textsc{romea} and \textsc{rmsga} calculations.  The agreement, particularly 
between the \textsc{rdwia} calculations and the data, is very good.  The spinor
distortions in the \textsc{rdwia} calculations which were required to predict 
the change in slope of $A_{LT}$ at $p_{\rm miss}$ $\approx$ 300 MeV/$c$ in Fig.
\ref{fig:juncai2} are also essential to the description of $R_{LT}$.  The 
agreement between the \textsc{rmsga} calculations and the data, particularly 
for $R_{LT}$, is markedly poorer.

Qualitatively, it should again be noted that none of the calculations presented
so far have included contributions from two-body currents.  The good agreement 
between the calculations and the data indicates that these currents are already
small at $Q^{2}$ $\approx$ 0.8 (GeV/$c$)$^{2}$.  This observation is supported 
by independent calculations by Amaro {\it et al.} \cite{amaro99,amaro03} which 
estimate the importance of such currents (which are highly dependent on 
$p_{\rm miss}$) to be large at lower $Q^{2}$, but only 2\% for the 
$1p_{1/2}$-state and 8\% for the $1p_{3/2}$-state in these kinematics.  It 
should also be noted that the \textsc{rdwia} results presented here are 
comparable with those obtained in independent \textsc{rdwia} analyses of our 
data by the Pavia Group -- see Meucci {\it et al.} \cite{meucci01}

\subsubsection{\label{subsubsec:compare_twobody}Comparison to \textsc{romea} and \textsc{rmsga} calculations including two-body currents}

In this Section, two-body current contributions to the \textsc{romea} and
\textsc{rmsga} calculations stemming from MEC and IC are presented.  These
contributions to the transition matrix elements were determined within the
nonrelativistic framework outlined by the Ghent Group in
\cite{ryckebusch99,ryckebusch01}.  Recall that the basic options employed in 
the calculations have been summarized in 
Table \ref{table:rdwia_romea_rmsga_options}.  Note again that both the EA-based
calculations stop at $p_{\rm miss}$ $=$ 350 MeV/$c$ as the approximation 
becomes invalid.

\begin{figure}
\resizebox{0.47\textwidth}{!}{\includegraphics{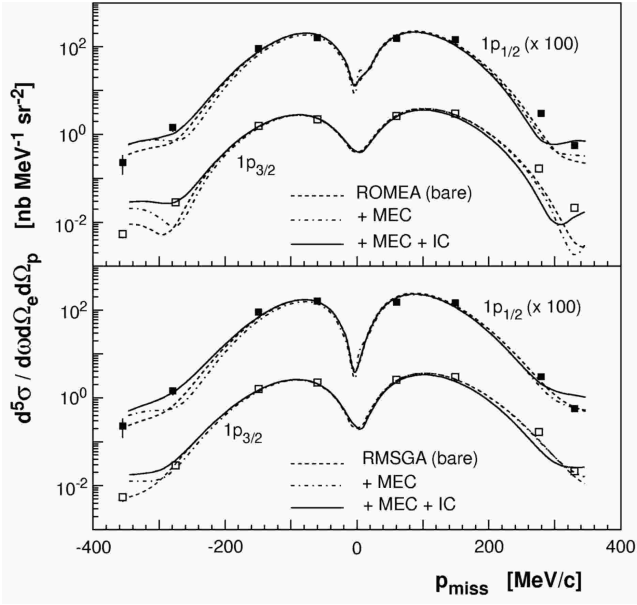}}
\caption{\label{fig:higher_order_cross}
Measured cross-section data for the removal of protons from the 1$p$-shell of 
$^{16}$O as a function of $p_{\rm miss}$ together with calculations by the 
Ghent Group at $E_{\rm beam}$ = 2.442 GeV.  Uncertainties are statistical and, 
on average, there is an additional $\pm$5.6\% systematic uncertainty (see Table 
\ref{table:pshellperpresults}). The curves labelled `bare' are identical to 
those shown in Fig. \ref{fig:juncai1}.
}
\end{figure}

Fig. \ref{fig:higher_order_cross} shows measured cross-section data for the 
removal of protons from the 1$p$-shell of $^{16}$O as a function of 
$p_{\rm miss}$ as compared to calculations by the Ghent Group which include MEC
and IC at $E_{\rm beam}$ = 2.442 GeV.  In the top panel, \textsc{romea} 
calculations are shown.  The dashed line is the bare calculation, the 
dashed-dotted line includes MEC, and the solid line includes both MEC and IC.  
In the bottom panel, \textsc{rmsga} calculations are shown.  The dashed line 
is the bare calculation, the dashed-dotted line includes MEC, and the solid 
line includes both MEC and IC.  Note that the curves labelled `bare' in this
figure are identical to those shown in Fig. \ref{fig:juncai1}.  The 
normalization factors are 0.6 and 0.7 for the $1p_{1/2}$-state and 
$1p_{3/2}$-state, respectively.  The impact of the two-body currents on the 
computed differential cross section for the knockout of 1$p$-shell protons from 
$^{16}$O is no more than a few percent for low $p_{\rm miss}$, but gradually 
increases with increasing $p_{\rm miss}$.  Surprisingly, explicit inclusion of 
the two-body current contributions to the transition matrix elements does not 
markedly improve the overall agreement between the calculations and the data.

\begin{figure}
\resizebox{0.47\textwidth}{!}{\includegraphics{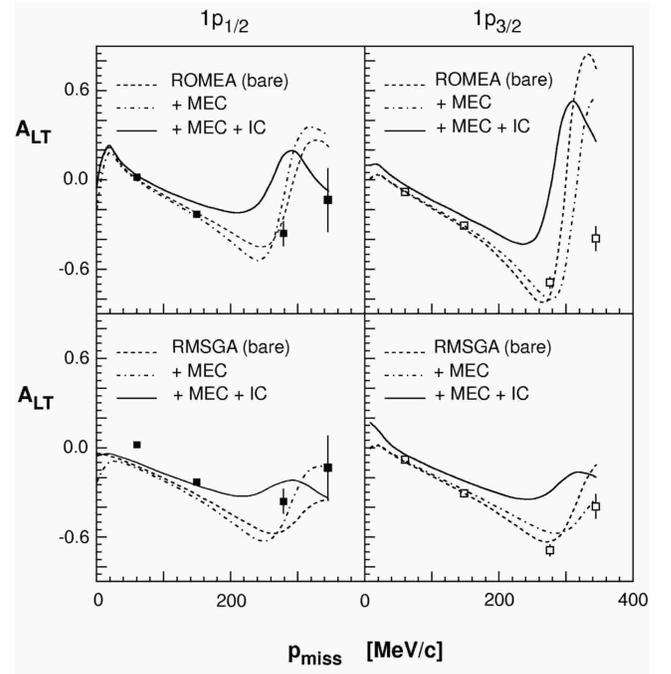}}
\caption{\label{fig:compare_twobody_alt}
Left-right asymmetry $A_{LT}$ together with calculations by the Ghent Group of 
the $A_{LT}$ asymmetry for the removal of protons from the 1$p$-shell of 
$^{16}$O as a function of $p_{\rm miss}$ for $E_{\rm beam}$ = 2.442 GeV.  Error
bars are statistical (see Table \ref{table:pshellaltrlt} for the associated 
systematic uncertainties).  The curves labelled `bare' are identical to 
those shown in Fig. \ref{fig:compare_onebody_alt}.
}
\end{figure}

Fig. \ref{fig:compare_twobody_alt} shows the left-right asymmetry $A_{LT}$ 
together with calculations by the Ghent Group for the removal of protons from 
the 1$p$-shell of $^{16}$O as a function of $p_{\rm miss}$ for $E_{\rm beam}$ 
$=$ 2.442 GeV.  In the top two panels, \textsc{romea} calculations are shown.  
The dashed lines are the bare calculations identical to those previously shown 
in Fig. \ref{fig:compare_onebody_alt}, the dashed-dotted line includes MEC, 
and the solid line includes both MEC and IC.  In the bottom panel, 
\textsc{rmsga} calculations are shown.  The dashed line is the bare 
calculation, the dashed-dotted line includes MEC, and the solid line includes 
both MEC and IC.  While all three calculations undergo a change in slope at 
$p_{\rm miss}$ $\approx$ 300 MeV/$c$, it is again clearly the bare calculations
which best represent the data.  Note that in general, the IC were observed to 
produce larger effects than the MEC.

\begin{figure}
\resizebox{0.47\textwidth}{!}{\includegraphics{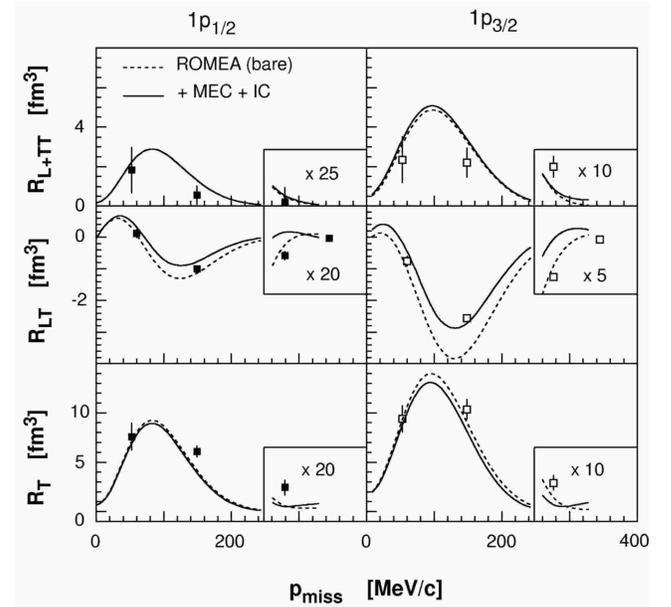}}
\caption{\label{fig:romea_struct}
Data from this work together with \textsc{romea} calculations by the Ghent 
Group for the $R_{L+TT}$, $R_{LT}$, and $R_{T}$ effective response functions 
for the removal of protons from the 1$p$-shell of $^{16}$O as a function of 
$p_{\rm miss}$.  Uncertainties are statistical (see Tables 
\ref{table:pshellrlttrt}, \ref{table:pshellaltrlt}, and \ref{table:pshellrlrt} 
for the associated systematic uncertainties).  The curves labelled `bare' are
identical to those shown in Fig. \ref{fig:juncai3}.
}
\end{figure}

\begin{figure}
\resizebox{0.47\textwidth}{!}{\includegraphics{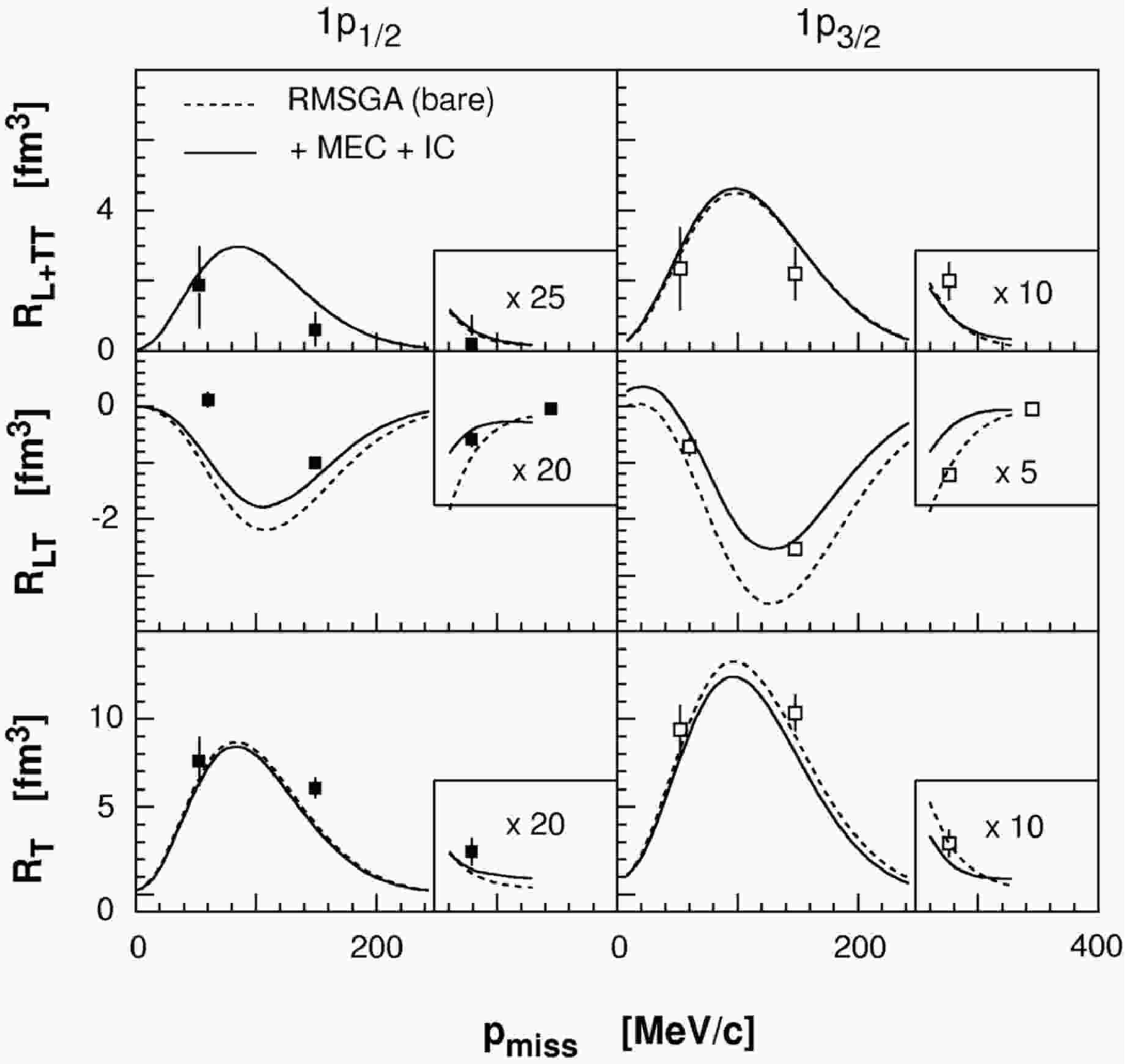}}
\caption{\label{fig:rmsga_struct}
Data from this work together with \textsc{rmsga} calculations by the Ghent 
Group for the $R_{L+TT}$, $R_{LT}$, and $R_{T}$ effective response functions 
for the removal of protons from the 1$p$-shell of $^{16}$O as a function of 
$p_{\rm miss}$.  Uncertainties are statistical (see Tables 
\ref{table:pshellrlttrt}, \ref{table:pshellaltrlt}, and \ref{table:pshellrlrt} 
for the associated systematic uncertainties).  The curves labelled `bare' are
identical to those shown in Fig. \ref{fig:juncai3}.
}
\end{figure}

Figs. \ref{fig:romea_struct} and \ref{fig:rmsga_struct} show the effective
$R_{L+TT}$, $R_{LT}$, and $R_{T}$ response functions together with 
\textsc{romea} and \textsc{rmsga} calculations by the Ghent Group for the 
removal of protons from the 1$p$-shell of $^{16}$O as a function of 
$p_{\rm miss}$.  The dashed lines are the bare \textsc{romea} and 
\textsc{rmsga} calculations identical to those previously shown in Fig. 
\ref{fig:juncai3}, while the solid lines include both MEC and IC.  In contrast 
to the cross-section (recall Fig. \ref{fig:higher_order_cross}) and $A_{LT}$ 
(recall Fig. \ref{fig:compare_twobody_alt}) situations, the agreement between 
the effective response-function data and the calculations improves with the 
explicit inclusion of the two-body current contributions to the transition 
matrix elements.

\subsection{Higher missing energies}

In this Section, \textsc{romea} calculations are compared to the 
higher-$E_{\rm miss}$ data.  The basic options employed in the calculations 
have been summarized in Table \ref{table:rdwia_romea_rmsga_options}.

\begin{figure*}
\resizebox{1.0\textwidth}{!}{\includegraphics{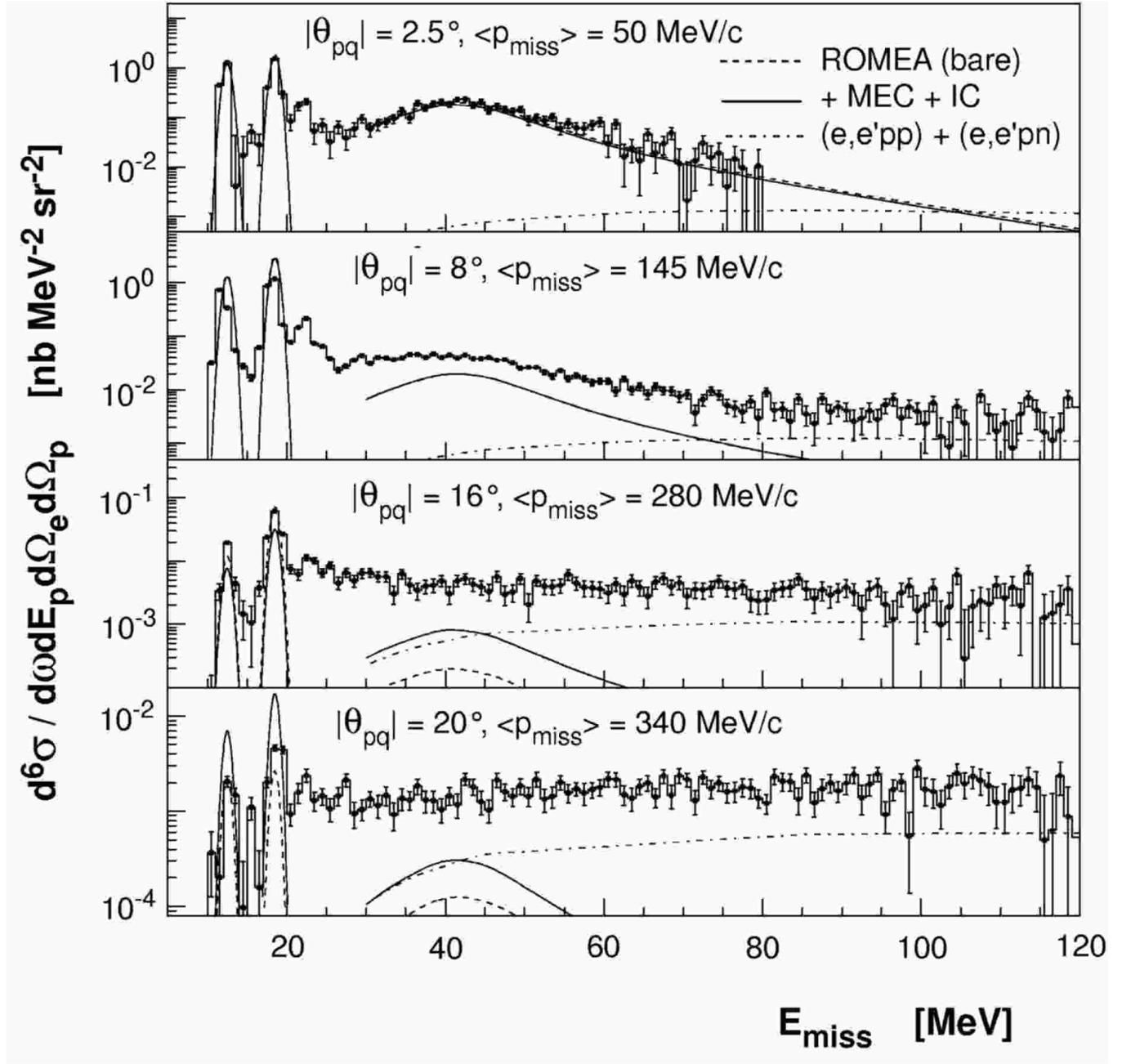}}
\caption{\label{fig:nilanga1}
Data from this work together with \textsc{romea} calculations by the Ghent
Group for the $E_{\rm miss}$-dependence of the cross section obtained at
$E_{\rm beam}$ = 2.442 GeV.  The data are the averaged cross section measured
on either side of $\bm{q}$ at each $\theta_{pq}$.  Normalization factors of
0.6, 0.7, and 1.0 have been used for the $1p_{1/2}$-, $1p_{3/2}$-, and 
$1s_{1/2}$-states, respectively.  Uncertainties are 
statistical and, on average, there is an additional $\pm$5.9\% systematic 
uncertainty (see Tables \ref{table:2442cont000deg} -- 
\ref{table:2442cont020deg}) associated with the data.  Also shown are 
calculations by the Ghent Group for the $(e,e^{\prime}pN)$ contribution.
}
\end{figure*}

Fig. \ref{fig:nilanga1} presents averaged measured cross-section data as a 
function of $E_{\rm miss}$ obtained at $E_{\rm beam}$ $=$ 2.442 GeV for four 
discrete HRS$_{h}$ angular settings ranging from 2.5$^{\circ}$ $<$ 
$\theta_{pq}$ $<$ 20$^{\circ}$, corresponding to average values of 
$p_{\rm miss}$ increasing from 50 to 340 MeV/$c$.  The cross-section values 
shown are the averaged values of the cross section measured on either side of 
$\bm{q}$ at each $\theta_{pq}$.  The strong peaks at $E_{\rm miss}$ $=$ 12.1 
and 18.3 MeV correspond to $1p$-shell proton removal from $^{16}$O.  As in
Section \ref{subsec:pshell}, the dashed curves corresponding to these peaks are
the bare \textsc{romea} calculations, while the solid lines include both MEC 
and IC.  The normalization factors remain 0.6 and 0.7 for the $1p_{1/2}$- and
$1p_{3/2}$-states, respectively.

For 20 $<$ $E_{\rm miss}$ $<$ 30 MeV, the spectra behave in a completely 
different fashion.  Appreciable strength exists which scales roughly with the 
$1p$-shell fragments and is not addressed by the present calculations of 
two-nucleon knockout.  The high-resolution experiment of Leuschner {\it et al.}
identified two additional $1p_{3/2}$ fragments and several positive-parity 
states in this region which are populated primarily by single-proton knockout 
from $2p2h$ components of the ground-state wave function.  Two-body currents 
and channel-coupling in the final state also contribute.  This strength has 
also been studied in $(\gamma,p)$ experiments, and has been interpreted by the 
Ghent Group \cite{ryckebusch92} as the post-photoabsorption population of 
states with a predominant $1p2h$ character via two-body currents.

For $E_{\rm miss}$ $>$ 30 MeV, in the top panel for $p_{\rm miss}$ $=$ 50
MeV/$c$, there is a broad and prominent peak centered at $E_{\rm miss}$ 
$\approx$ 40 MeV corresponding largely to the knockout of $1s_{1/2}$-state 
protons.  As can be seen in the lower panels, the strength of this peak 
diminishes with increasing $p_{\rm miss}$, and completely vanishes beneath a 
flat background by $p_{\rm miss}$ $=$ 280 MeV/$c$.  For $E_{\rm miss}$ 
$>$ 60 MeV and $p_{\rm miss}$ $\geq$ 280 MeV/$c$, the cross section decreases 
only very weakly as a function of $p_{\rm miss}$, and is completely independent
of $E_{\rm miss}$.

In order to estimate the amount of the cross section observed for 
$E_{\rm miss}$ $>$ 25 MeV that can be explained by the single-particle knockout
of protons from the $1s_{1/2}$-state, the data were compared to the 
\textsc{romea} calculations of the Ghent Group.  The dashed curves are the bare
calculations, while the solid lines include both MEC and IC.  A normalization 
factor of 1.0 for the $1s_{1/2}$-state single-particle strength was used.  The 
two calculations are indistinguishable for $p_{\rm miss}$ $\leq$ 145 MeV/$c$, 
and the agreement between these calculations and the measured cross-section 
data is reasonable (see the top two panels of Fig. \ref{fig:nilanga1} where 
there is an identifiable $1s_{1/2}$-state peak at $E_{\rm miss}$ $\approx$ 40 
MeV).  At higher $p_{\rm miss}$ (where there is no clear $1s_{1/2}$-state peak 
at $E_{\rm miss}$ $\approx$ 40 MeV), the data are substantially larger than the 
calculated bare cross section.  Inclusion of MEC and IC improves the agreement,
but there is still roughly an order-of-magnitude discrepancy.  The 
\textsc{rdwia} calculations demonstrate similar behavior.  Thus, the 
$p_{\rm miss}$ $\geq$ 280 MeV/$c$ data are not dominated by single-particle 
knockout.  Note that the magnitude of ($S_{T} - S_{L}$) is consistent with that
anticipated based on the measurements of Ulmer {\it et al.} at $Q^{2}$ = 0.14 
(GeV/$c$)$^2$ and Dutta {\it et al.} at $Q^{2}$ = 0.6 and 1.8 (GeV/$c$)$^2$.  
Together, these data suggest that transverse processes associated with the 
knockout of more than one nucleon decrease with increasing $Q^{2}$.

Also shown as dashed-dotted curves in Fig. \ref{fig:nilanga1} are the 
calculations by the Ghent Group \cite{janssen00} for the $(e,e^{\prime}pp)$ and
$(e,e^{\prime}pn)$ contributions to the $(e,e^{\prime}p)$ cross section 
performed within a Hartree-Fock framework.  This two-particle knockout cross 
section was determined using the Spectator Approximation, in a calculation 
which included MEC, IC, and both central short-range correlations (SRC) and 
tensor medium-range correlations.  Note that in these kinematics, this 
calculation performed with SRC alone produced only 2\% of the two-particle 
knockout cross section, while including both SRC and tensor correlations 
produced only 15\% of the two-particle knockout cross section.  The calculated 
two-particle knockout cross section is essentially transverse in nature, since 
the two-body currents are predominantly transverse.  The calculated strength 
underestimates the measured cross section by about 50\% but has the observed 
flat shape for $E_{\rm miss}$ $>$ 50 MeV.  It is thus possible that heavier 
meson exchange and processes involving three (or more) nucleons could provide
a complete description of the data.

\begin{figure}
\resizebox{0.47\textwidth}{!}{\includegraphics{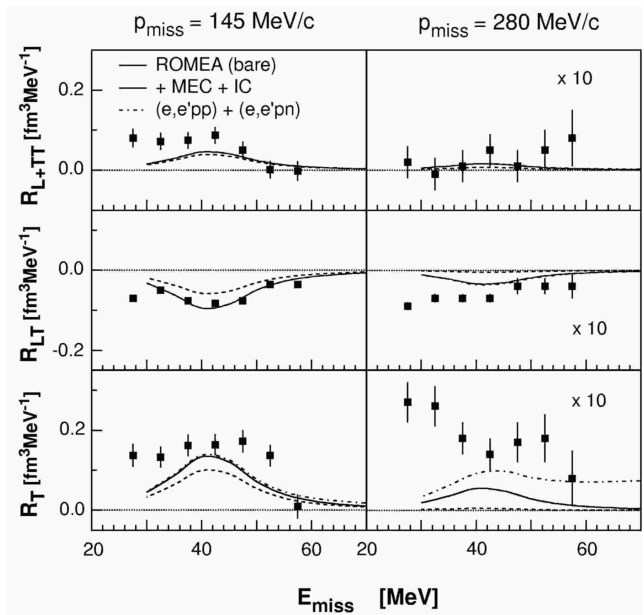}}
\caption{\label{fig:nilanga34}
Data from this work together with \textsc{romea} calculations by the Ghent
Group for the $E_{\rm miss}$-dependence of the $R_{L+TT}$, $R_{LT}$, and
$R_{T}$ effective response functions.  Uncertainites are statistical (see 
Tables \ref{table:contrlttrt}, \ref{table:contrlt}, and \ref{table:contrlrt} 
for the associated systematic uncertainties).  Also shown is the 
$(e,e^{\prime}pN)$ contribution.
}
\end{figure}

The measured effective response functions $R_{L+TT}$, $R_{LT}$, and $R_{T}$ 
together with \textsc{romea} calculations for $p_{\rm miss}$ $=$ 145 MeV/$c$ 
and $p_{\rm miss}$ $=$ 280 MeV/$c$ are presented in Fig. \ref{fig:nilanga34}.
Kinematic overlap restricted separations to $E_{\rm miss}$ $<$ 60 MeV.  The 
dashed curves are the bare \textsc{romea} calculations, while the solid curves 
include both MEC and IC.  Also shown as dashed-dotted curves are the incoherent
sum of these `full' calculations and the computed $(e,e^{\prime}pN)$ 
contribution.  In general, the data do not show the broad peak centered at 
$E_{\rm miss}$ $\approx$ 40 MeV corresponding to the knockout of 
$1s_{1/2}$-state protons predicted by the calculations.  At $p_{\rm miss}$ =
145 MeV/$c$, the bare calculation is consistently about 60\% of the magnitude 
of the data. Inclusion of MEC and IC does not appreciably change the calculated
$R_{L+TT}$, but does improve the agreement between data and calculation for 
$R_{LT}$ and $R_{T}$.  The measured response $R_{L+TT}$ (which is essentially 
equal to $R_{L}$ since $\frac{v_{TT}}{v_L}R_{TT}$ is roughly 7\% of $R_{L}$ in 
these kinematics -- see Ref. \cite{kelly99a}) is larger than the calculation 
for $E_{\rm miss}$ $<$ 50 MeV and smaller than the calculation for 
$E_{\rm miss}$ $>$ 50 MeV.  The agreement between the calculation and the data 
for $R_{LT}$ is very good over the entire $E_{\rm miss}$ range.  Since the 
measured response $R_{LT}$ is nonzero for $E_{\rm miss}$ $>$ 50 MeV, the 
measured response $R_{L}$ must also be nonzero.  The measured response $R_{T}$ 
is somewhat larger than the calculation for $E_{\rm miss}$ $<$ 60 MeV.

At $p_{\rm miss}$ = 280 MeV/$c$, the bare calculation does not reproduce the 
$E_{\rm miss}$-dependence of any of the measured effective response functions.
The inclusion of MEC and IC in the calculation substantially increases the 
magnitude of all three calculated response functions, and thus improves the 
agreement between data and calculation.  The measured $R_{L+TT}$ (which is 
dominated by $R_{L}$) is consistent with both the calculation and with zero.  
The measured $R_{LT}$ is about twice the magnitude of the calculation.  Since 
the measured $R_{LT}$ is nonzero over the entire $E_{\rm miss}$ range, the 
measured $R_{L}$ must also be nonzero.  The measured $R_{T}$ is significantly 
larger than both the calculations and nonzero out to at least $E_{\rm miss}$ 
$\approx$ 60 MeV.  The fact that the measured $R_{T}$ is much larger than the 
measured $R_{L}$ indicates the cross section is largely due to transverse 
two-body currents.  And finally, it is clear that $(e,e^{\prime}pN)$ accounts 
for a fraction of the measured transverse strength which increases dramatically
with increasing $p_{\rm miss}$.

\begin{figure*}
\resizebox{1.0\textwidth}{!}{\includegraphics{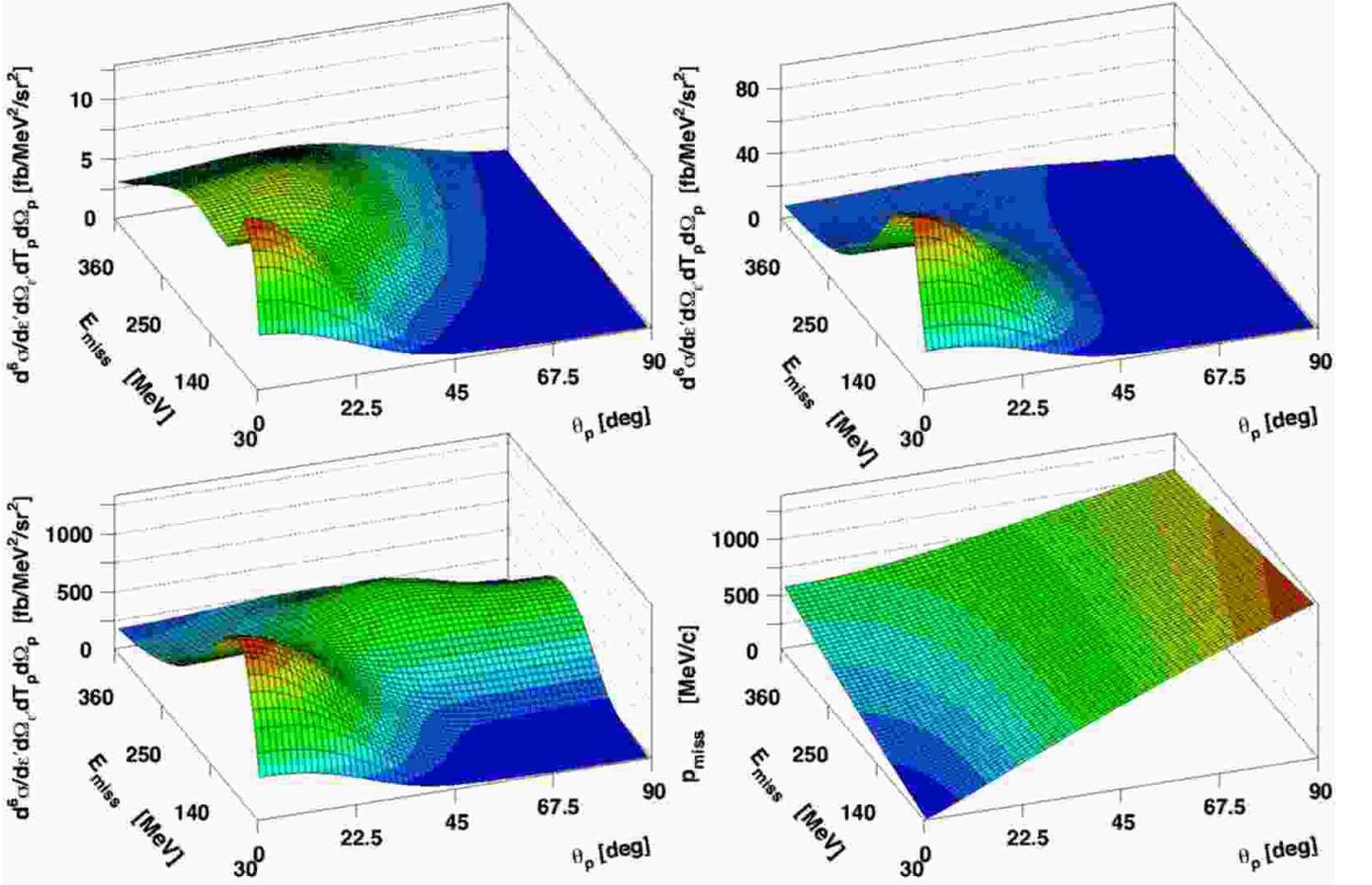}}
\caption{\label{fig:2n_contributions}
(Color online) Calculations by the Ghent Group of the contribution to the 
differential $^{16}$O$(e,e^{\prime}p)$ cross section from two-nucleon knockout 
as a function of $E_{\rm miss}$ and $\theta_{p}$ for $E_{\rm beam}$ = 2.442 
GeV.  The upper-left panel shows the contribution of central correlations.  The 
upper-right panel shows the combined contribution of central and tensor 
correlations.  The lower-left panel shows the combined contribution of central 
and tensor correlations (two-nucleon correlations) together with MEC and IC 
(two-body currents).  The relationship between the various kinematic quantities
is shown in the bottom-right panel.
}
\end{figure*}

Fig. \ref{fig:2n_contributions} shows the calculations by the Ghent Group 
\cite{ryckebuschtech01} of the contribution to the differential 
$^{16}$O$(e,e^{\prime}p)$ cross section from two-nucleon knockout as a function
of $E_{\rm miss}$ and $\theta_{p}$ for $E_{\rm beam}$ = 2.442 GeV.  The 
upper-left panel shows the contribution of central correlations.  The 
upper-right panel shows the combined contribution of central and tensor 
correlations. Tensor correlations are anticipated to dominate central 
correlations over the ranges of $E_{\rm miss}$ and $p_{\rm miss}$ investigated 
in this work.  The lower-left panel shows the combined contribution of central 
and tensor correlations (two-nucleon correlations) together with MEC and IC 
(two-body currents).  Two-body currents are anticipated to dominate two-nucleon
correlations over the ranges of $E_{\rm miss}$ and $p_{\rm miss}$ investigated 
in this work.  For convenience, the variation of $p_{\rm miss}$ with 
$E_{\rm miss}$ and $\theta_{p}$ is shown in the bottom-right panel.

\section{\label{sec:kin_consistency}Kinematic consistency of $\bm{1p}$-shell normalization factors} 

There has been longstanding discussion regarding the reliability of the 
spectroscopic factors determined for discrete states from single-nucleon 
electromagnetic knockout.  Recently, there has been speculation that these 
factors might appear to increase with $Q^{2}$ as a quasiparticle state is 
probed with finer resolution.  In this Section, Kelly \cite{kellyspecfacs} has 
used the \textsc{rdwia} to analyze the normalization factors fitted to the 
available $^{16}$O$(e,e^{\prime}p)$ data for the $1p_{1/2}$- and 
$1p_{3/2}$-states obtained in the experiments summarized in Table 
\ref{table:prevdat}.  If the \textsc{rdwia} model is accurate, these factors 
should be independent of the experimental kinematics.  

\begin{table*}
\caption{\label{table:prevdat}
A summary of the kinematic conditions for the data examined in the $^{16}$O$(e,e^{\prime}p)$ consistency study.}
\begin{ruledtabular}
\begin{tabular}{rr|rrrrrrr}
      &                                            &               & $T_{p}$ &         $Q^{2}$ &             &
   &                       \\
label &                                    authors &    kinematics &   (MeV) & (GeV/$c$)$^{2}$ &         $x$ & $2s_{1/2}1d_{5/2}$-doublet &                  data \\
\hline
    a &  Leuschner {\it et al.} \cite{leuschner94} &      parallel &      96 &          varied &      varied &
resolved \hspace*{1.1mm} &      reduced $\sigma$ \\
    b &      Spaltro {\it et al.} \cite{spaltro93} & perpendicular &      84 &            0.20 &        1.07 &
resolved \hspace*{1.1mm} & differential $\sigma$ \\
    c &      Chinitz {\it et al.} \cite{chinitz91} & perpendicular &     160 &            0.30 &        0.91 &  computed \footnotemark[1] & differential $\sigma$ \\
\hline
    d &                                  this work & perpendicular &     427 &            0.80 &        0.96 &  computed \footnotemark[2] & differential $\sigma$ \\
\hline
    e &    Bernheim {\it et al.} \cite{bernheim82} & perpendicular &     100 &            0.19 &        0.90 &  computed \footnotemark[2] &      reduced $\sigma$ \\
    f & Blomqvist1 {\it et al.} \cite{blomqvist95} &      parallel &      92 &            0.08 & 0.30 - 0.50 &
resolved \hspace*{1.1mm} &      reduced $\sigma$ \\
    g & Blomqvist2 {\it et al.} \cite{blomqvist95} & highly varied &     215 &     0.04 - 0.26 & 0.07 - 0.70 &
resolved \hspace*{1.1mm} &      reduced $\sigma$ \\
\end{tabular}
\end{ruledtabular}
\begin{minipage}[t]{\textwidth}
\footnotetext[1]{The $1p_{3/2}$-state data were corrected for the contamination
of the $1d_{5/2}$$2s_{1/2}$-doublet by Chinitz {\it et al}.}
\footnotetext[2]{The contamination of the $1p_{3/2}$-state by the $1d_{5/2}$$2s_{1/2}$-doublet was computed according to the method outlined in Section \ref{subsubsec:sensitivities}.}
\end{minipage}
\end{table*}

\subsection{\label{subsec:rdwia_norm_factors}$\textsc{RDWIA}$ analysis of the available $\bm{^{16}}$O$\bm{(e,e^{\prime}p)}$ data}

The \textsc{rdwia} calculations used in this procedure employed the 
$\bar{\Gamma}_2$ off-shell single-nucleon current operator with the 
\textsc{mmd} form factors of Mergell {\it et al.} \cite{mergell96} in the 
Coulomb gauge.  The partial-wave expansions were performed using the 
second-order Dirac equation, including spinor distortion.  Nucleon distortion 
was evaluated using the \textsc{edai-o} and \textsc{edad1} optical potentials, 
and electron distortion was evaluated in the $q_{\rm eff}$ Approximation.  No 
attempt to directly fit the overlap functions to the knockout data has been 
made here.  Instead, the recently developed wave functions \textsc{hs}, 
\textsc{nlsh}, and \textsc{nlsh-p} (recall Fig. \ref{fig:hsnlshnlshp}) were 
again tested, this time to see if they could satisfactorily reproduce the 
experimental $p_{\rm miss}$ distributions independently of $Q^2$.

The results are expressed in terms of normalization factors which compare a
\textsc{rdwia} calculation for a fully occupied subshell with experimental data
and are presented in Table \ref{table:kellyspecs}.  These factors were obtained
by least-squares fitting to the data in the range $p_{\rm miss}$ $<$ 200 
MeV/$c$ where the \textsc{rdwia} should be most reliable.  When experimentally 
unresolved, the contamination of the $1p_{3/2}$-state by the 
$1d_{5/2}$$2s_{1/2}$-doublet was included by the incoherent summation of the
parametrizations of Leuschner {\it et al.} as previously described.

\begin{table*}
\caption{\label{table:kellyspecs}
Normalization factors deduced for the data sets presented in Table \ref{table:prevdat}
for $p_{\rm miss}$ $<$ 200 MeV/$c$.  The first term in each column is for the
$1p_{1/2}$-state, while the second term is for the $1p_{3/2}$-state.}
\begin{ruledtabular}
\begin{tabular}{c|rr|rr|rr|rr|rr|rr|rr|rr|rr|rr|rr|rr}
  &
     \multicolumn{12}{c|}{\textsc{edai-o}} &
                                                  \multicolumn{12}{c}{\textsc{edad1}} \\
\cline{2-13}\cline{14-25}
  &                                 \multicolumn{4}{c|}{\textsc{hs}} &                               \multicolumn{4}{c|}{\textsc{nlsh}} &
      \multicolumn{4}{c|}{\textsc{nlsh-p}} &                                 \multicolumn{4}{c|}{\textsc{hs}} &                                \multicolumn{4}{c|}{\textsc{nlsh}} &                               \multicolumn{4}{c}{\textsc{nlsh-p}} \\
  & \multicolumn{2}{c}{$S_{\alpha}$} & \multicolumn{2}{c|}{$\chi^2$} & \multicolumn{2}{c}{$S_{\alpha}$} & \multicolumn{2}{c|}{$\chi^2$} & \multicolumn{2}{c}{$S_{\alpha}$} & \multicolumn{2}{c|}{$\chi^2$} & \multicolumn{2}{c}{$S_{\alpha}$} & \multicolumn{2}{c|}{$\chi^2$} & \multicolumn{2}{c}{$S_{\alpha}$} &  \multicolumn{2}{c|}{$\chi^2$} &  \multicolumn{2}{c}{$S_{\alpha}$} &  \multicolumn{2}{c}{$\chi^2$} \\
\hline
a &                      0.55 & 0.46 &                    0.9 &  4.2 &                      0.60 & 0.47 &                    2.5 &  6.0 &                      0.53 & 0.41 &                    1.0 &  1.5 &                      0.60 & 0.55 &                    0.8 &  2.4 &                      0.66 & 0.56 &
       2.3 &  3.7 &                       0.57 & 0.48 &                    0.8 &  1.4 \\
b &                      0.61 & 0.66 &                    2.7 &  6.3 &                      0.65 & 0.68 &                    5.4 &  8.0 &                      0.58 & 0.58 &                    2.3 &  3.6 &                      0.71 & 0.75 &                    2.3 &  5.1 &                      0.77 & 0.77 &
       4.2 &  6.4 &                       0.68 & 0.65 &                    2.2 &  4.1 \\
c &                      0.54 & 0.56 &                    8.7 & 17.9 &                      0.60 & 0.58 &                   24.8 & 25.3 &                      0.51 & 0.47 &                    9.0 & 15.0 &                      0.59 & 0.61 &                    8.0 & 18.2 &                      0.66 & 0.63 &
      16.7 & 22.1 &                       0.56 & 0.50 &                    7.2 & 23.8 \\
\hline
d &                      0.62 & 0.63 &                   30.8 &  4.7 &                      0.70 & 0.65 &                    0.5 &  6.7 &                      0.59 & 0.52 &                   19.6 & 15.3 &                      0.62 & 0.63 &                   32.5 &  2.4 &                      0.70 & 0.66 &
       1.2 &  3.9 &                       0.60 & 0.53 &                   20.0 & 14.4 \\
\hline
e &                      0.43 & 0.46 &                    1.0 &  1.9 &                      0.48 & 0.47 &                    2.2 &  2.4 &                      0.42 & 0.40 &                    1.0 &  1.1 &                      0.48 & 0.52 &                    1.0 &  1.5 &                      0.54 & 0.54 &
       1.5 &  1.9 &                       0.47 & 0.45 &                    1.0 &  1.3 \\
f &                      0.53 & 0.41 &                    3.0 &  4.2 &                      0.54 & 0.42 &                    5.0 &  5.8 &                      0.51 & 0.38 &                    2.7 &  1.9 &                      0.57 & 0.50 &                    3.2 &  3.7 &                      0.59 & 0.51 &
       5.8 &  5.0 &                       0.54 & 0.46 &                    2.7 &  2.1 \\
g &                      0.42 & 0.37 &                    2.0 &  1.4 &                      0.44 & 0.37 &                    4.7 &  2.0 &                      0.40 & 0.33 &                    2.5 &  5.9 &                      0.42 & 0.40 &                    1.8 &  1.9 &                      0.44 & 0.41 &
       6.6 &  2.8 &                       0.40 & 0.36 &                    1.9 &  5.4 \\
\end{tabular}
\end{ruledtabular}
\end{table*}
The data sets demonstrated a slight preference for the \textsc{edad1} optical 
potential over the \textsc{edai-o} optical potential.  This was concluded based
on the quality of the fits and the more consistent nature of the extracted 
normalization factors for low $Q^{2}$.  None of the variations considered in 
Table \ref{table:kellyspecs} (nor any of those considered in Table 
\ref{table:cc12specfacs} for that matter) were able to reproduce the 
$1p_{3/2}$-state for data set (b) in the range $50 < p_{\rm miss} < 120$ 
MeV/$c$.  This problem is also responsible for the discrepancy seen in Fig. 
\ref{fig:chinspal} for $R_{LT}$ at $Q^{2}=0.2$ (GeV/$c$)$^2$, and has not yet 
been explained satisfactorily. 

\begin{figure*}
\resizebox{0.86\textwidth}{!}{\includegraphics{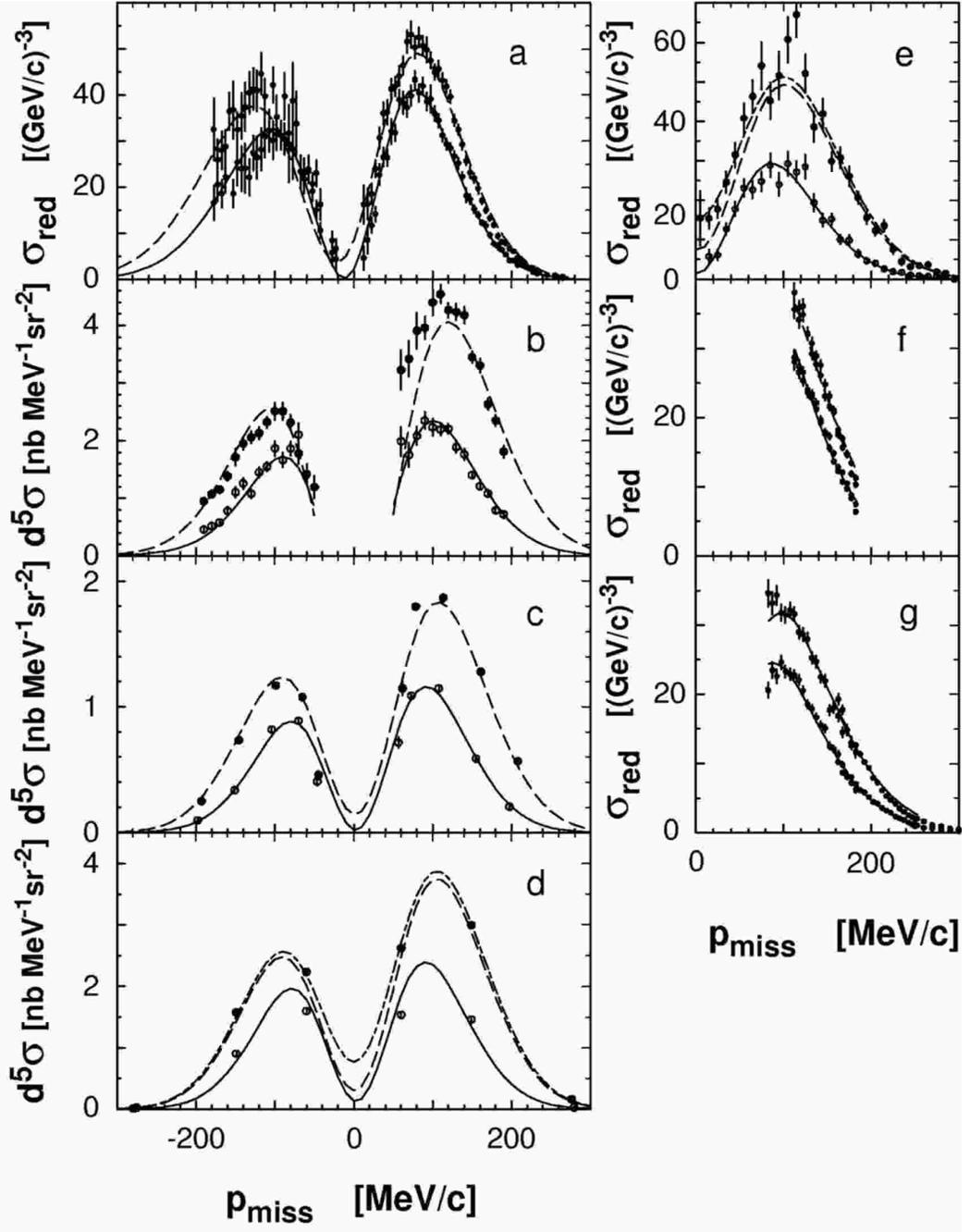}}
\caption{
\label{fig:hsedad1}
Fits to various $^{16}$O$(e,e^{\prime}p)$ data sets based on the \textsc{hs}
bound-nucleon wave function and the \textsc{edad1} optical potential.  See 
Table \ref{table:prevdat} for the key to the data-set labels.  Open points and 
solid lines pertain to the $1p_{1/2}$-state, while solid points and dashed 
lines pertain to the $1p_{3/2}$-state.  The dashed-dotted lines include the 
contributions of the positive parity $2s_{1/2}1d_{5/2}$-doublet to the 
$1p_{3/2}$-state.  Panel (d) shows the data from this work.
}
\end{figure*}

Unfortunately, none of the selected wave functions provided an optimal 
description of the experimental $p_{\rm miss}$ distributions independent of 
$Q^{2}$.  Fig. \ref{fig:hsedad1} shows a sample set of fits to the various 
$^{16}$O$(e,e^{\prime}p)$ data sets based on the \textsc{hs} bound-nucleon wave
function and the \textsc{edad1} optical potential.
From a $\chi^2$ perspective, it is clear that \textsc{nlsh} offered the best 
description of the data at $Q^2=0.8$ (GeV/$c$)$^2$, but that most of the lower 
$Q^2$ data are best described by either \textsc{nlsh-p} or \textsc{nlsh}; in 
fact, \textsc{hs} may be the best compromise currently available.  Data set (g)
from Mainz suggested a substantially different normalization.  

An estimate of the uncertainty in the normalization factors due to variations 
of the bound-nucleon wave function was made by comparing 
\textsc{nlsh}/\textsc{hs} and \textsc{nlsh-p}/\textsc{hs} normalization-factor 
ratios for each state holding the optical potential and other model input 
constant.  For the lowest-lying $1p_{1/2}$- and $1p_{3/2}$-states and averaged 
over all data sets, \textsc{nlsh}/\textsc{hs} ratios of 1.08 and 1.02 were 
obtained.  Similarly, \textsc{nlsh-p}/\textsc{hs} ratios of 0.96 and 0.87 were 
obtained.  These ratios are qualitatively consistent with the behavior of the 
data and the calculations near the peaks of the momentum distributions shown in
Fig. \ref{fig:hsnlshnlshp}.  Therefore, a cautious estimate of the 
uncertainty due to the bound-nucleon wave function is of order $\pm 10\%$.  
Further, by changing the optical potentials between \textsc{edai-o} and 
\textsc{edad1} and holding the bound-nucleon wave function and other model 
input constant, \textsc{edai-o}/\textsc{edad1} normalization-factor ratios for 
a given data set with $Q^2 < 0.4$ (GeV/$c$)$^2$ averaged to about 0.90.  This 
ratio became 0.98 at $Q^2 = 0.8$ (GeV/$c$)$^2$, where the attenuation in the 
potentials is practically identical.  Therefore, the uncertainty due to 
variations of the optical potential is at least $\pm5\%$ and would probably be 
larger if the sample of `reasonable' potentials were expanded.

The information presented in Table \ref{table:cc12specfacs} suggests that there
would be similar uncertainties in the normalization factors for the $Q^2 = 0.8$
(GeV/$c$)$^2$ data arising from Gordon and gauge ambiguities in the 
single-nucleon current operator.  Note that values of $S_\alpha$ for the same 
model are generally larger in Table \ref{table:cc12specfacs} than in Table 
\ref{table:kellyspecs} because the former summarizes a study of the entire 
$p_{\rm miss}$ range while the latter is limited to $p_{\rm miss}$ $<$ 200 
MeV/$c$, where the reaction model is likely to be most accurate.  Data for 
larger $p_{\rm miss}$ also tend to have a higher $\chi^2$.  These problems for 
large $p_{\rm miss}$ may arise from inaccuracies in the bound-nucleon wave 
functions above the Fermi momentum, neglecting two-body currents, neglecting 
channel coupling in the final state, or density-dependence in the form 
factors, to name a few.  Therefore, a realistic estimate of the model 
dependence of the normalization factors for $(e,e^\prime p)$ reactions should 
not be less than $\pm 15\%$.  This estimated precision is consistent with that 
suggested in Ref. \cite{kelly96} -- although the relativistic model improves 
our description of $A_{LT}$, recoil polarization, and other 
normalization-independent features of the reaction, the model dependencies that
affect the normalization uncertainty are not significantly improved.  Assuming 
that the reaction model is most reliable at large $Q^2$ and modest 
$p_{\rm miss}$, the six $1p_{1/2}$- and $1p_{3/2}$-state normalization factors 
for data set (d) in Table \ref{table:kellyspecs} (this work) were averaged to 
conclude that the normalization factors for the lowest $1p_{1/2}$- and 
$1p_{3/2}$-states in $^{15}$N are approximately 0.63(9) and 0.60(9).  

As previously mentioned, Leuschner {\it et al.} identified two additional 
$1p_{3/2}$-states with excitation energies between 9 and 13 MeV that together 
carry approximately $11\%$ of the strength of the lowest-energy fragment.  
However, those states were not resolved by the present experiment.  If the
assumption is made that the same ratio applies at $Q^{2} \approx 0.8$ 
(GeV/$c$)$^{2}$, then the total $1p_{3/2}$ strength below 15 MeV excitation is 
estimated to be approximately $67\%$ of full occupancy, and the total 
$1p$-shell spectroscopic strength below 15 MeV represents $3.9 \pm 0.6$ protons
or about $65\%$ of full occupancy.  This result remains $10-20\%$ below 
predictions from recent calculations of the hole spectral function by Barbieri 
and Dickhoff \cite{barbieri02}, but no experimental estimate for the 
additional $p$-shell strength that might lurk beneath the continuum is 
available.

To obtain more precise normalization factors, it would be necessary to apply a 
relativistic analysis to data in quasiperpendicular kinematics for several 
values of $Q^2$ larger than about 0.5 (GeV/$c$)$^2$ and with sufficient 
coverage of the $p_{\rm miss}$ distribution to fit the wave function,
requiring that the bound-nucleon wave function be independent of $Q^2$.  
Although such data do not yet exist for $^{16}$O, the recently completed 
experiment in Hall A at Jefferson Lab will provide substantially more data 
points for the critical $p_{\rm miss}$ $<$ 200 MeV/$c$ region for $Q^2=$ 0.9 
(GeV/$c$)$^2$.

\subsection{\label{subsec:discussion}$\bm{Q^{2}}$-dependence of normalization factors}

Lapik\'as {\it et al.} \cite{lapikas00} have performed a similar type of 
analysis of the $Q^2$-dependence of the normalization factors for the 
$^{12}$C$(e,e^\prime p)$ reaction.  In their work, several data sets with 
$Q^2 < 0.3$ (GeV/$c$)$^2$ were analyzed using a nonrelativistic \textsc{dwia} 
model.  For each data set, a normalization factor and the radius parameter for 
a Woods-Saxon binding potential were fitted to the reduced cross section for 
discrete states, and the potential depths were adjusted to fit the separation 
energies.  Consistent normalization factors were obtained for all data sets 
save those measured by Blomqvist {\it et al.} at Mainz \cite{blomqvist95b}.  A 
new experiment was thus performed at NIKHEF duplicating the Mainz kinematics.  
The new results were also consistent with all data sets save those from Mainz.
Lapik\'as {\it et al.} thus concluded that the Mainz data were normalized 
incorrectly.  Recall that similar doubts regarding the normalization of the 
companion $^{16}$O$(e,e^{\prime}p)$ experiment \cite{blomqvist95} at Mainz had 
been expressed earlier by Kelly \cite{kelly97}, but independent data 
duplicating the measurement are unfortunately not available.  After excluding 
the Mainz data, Lapik\'as {\it et al.} determined that the summed $1p$-shell 
strength for $^{12}$C could in fact be deduced from data for $Q^2 < 0.3$ 
(GeV/$c$)$^2$ with an uncertainty of $\pm 3\%$.  However, they did not consider
the effects of variations of the optical model or several other uncertain 
aspects of the reaction model.  As discussed previously, a more realistic 
estimate of the relative uncertainty in the normalization factors must be 
closer to $\pm15\%$ due to the inevitable model dependence of the 
\textsc{dwia}.  Furthermore, it is possible that variation of the Woods-Saxon 
radius might affect the resulting normalization factors.  If the overlap 
function is an intrinsic property of the nuclear wave function, it should not 
depend upon $Q^2$.  Further, it should be possible to fit a common radius to 
all data simultaneously; if not, the accuracy of the reaction model must be 
questioned.  And of course, it has been demonstrated in recent years that a 
relativistic \textsc{dwia} model is preferable to a nonrelativistic approach.

Lapik\'as {\it et al}. also used the bound-nucleon wave functions and 
normalization factors obtained from their nonrelativistic analysis at low $Q^2$
to analyze the transparency of $^{12}$C for $Q^2$ up to 7 (GeV/$c$)$^2$ and the
summed $1p$- and $1s$-shell spectroscopic amplitude.  They found the summed 
spectroscopic strength was approximately constant at 0.58 for $Q^2 \leq 
0.6$ (GeV/$c$)$^2$, but rose for larger $Q^2$ and appeared to approach the 
Independent-Particle Model limit of unity somewhere near $Q^2 \approx 10$ 
(GeV/$c$)$^2$.  They speculated that the apparent $Q^2$-dependence of this 
spectroscopic strength might be related to the resolution at which a 
quasiparticle is probed, with long-range correlations that deplete the 
single-particle strength becoming less important at higher $Q^2$ and finer 
resolution.  A subsequent analysis by Frankfurt {\it et al.} using Glauber 
calculations for heavier targets \cite{frankfurt01} supports this 
interpretation.

Little evidence is seen here for a systematic dependence in the normalization 
factors upon either $T_p$ or $Q^2$ for the lowest $1p$-states of $^{16}$O for 
the data that are presently available.  Unfortunately, these data do not reach 
high enough $Q^2$ to address the resolution hypothesis.  Furthermore, the 
normalization factors for two of the data sets appear to be anomalously low.  
A normalization problem might not be too surprising for data set (e) because it
comes from one of the earliest experiments on this reaction, but data set (g) 
comes from a fairly recent experiment at Mainz and uses an ejectile energy 
large enough for the reaction model to be reliable.  As discussed above for the
case of $^{12}$C, it is likely that data set (g) also has a normalization error
\footnote{
Difficulties associated with the reaction mechanism relatively far from QE 
kinematics may also be partly reponsible for the anomalously low normalization 
factors for these data.}.  
If these two data sets are disregarded, the remaining low $Q^2$ data are 
consistent with the normalization factors deduced from the current $Q^2 = 0.8$ 
(GeV/$c$)$^2$ data.

\section{\label{sec:conclude}Summary and Conclusions}

The $^{16}$O$(e,e^{\prime}p)$ reaction in QE, constant $(q,\omega)$ kinematics 
at $Q^{2}$ $\approx$ 0.8 (GeV/$c$)$^{2}$, $q \approx$ 1 GeV/$c$, and $\omega 
\approx$ 445 MeV was studied for 0 $<$ $E_{\rm miss}$ $<$ 120 MeV and 0 $<$ 
$p_{\rm miss}$ $<$ 350 MeV/$c$.  Five-fold differential cross-section data for 
the removal of protons from the $1p$-shell were obtained for 0 $<$ 
$p_{\rm miss}$ $<$ 350 MeV/$c$.  Six-fold differential cross-section data for 
0 $<$ $E_{\rm miss}$ $<$ 120 MeV were obtained for 0 $<$ $p_{\rm miss}$ $<$ 340
MeV/$c$.  These results were used to extract the $A_{LT}$ asymmetry and the 
$R_{L}$, $R_{T}$, $R_{L+TT}$, and $R_{LT}$ effective response functions over a 
large range of $E_{\rm miss}$ and $p_{\rm miss}$.

The data were interpreted in subsets corresponding to the 1$p$-shell and the 
1$s_{1/2}$-state and continuum, respectively.  1$p$-shell data were interpreted 
within three fully relativistic frameworks for single-particle knockout which 
do not include any two-body currents: \textsc{rdwia}, \textsc{romea}, and 
\textsc{rmsga}.  Two-body current contributions to the \textsc{romea} and 
\textsc{rmsga} calculations for the 1$p$-shell stemming from MEC and IC were 
also considered.  The 1$s_{1/2}$-state and continuum data were considered 
within the identical \textsc{romea} framework both before and after 
two-body current contributions due MEC and IC were included. $(e,e^{\prime}pN)$
contributions to these data were also examined.

Overall, the \textsc{rdwia} calculations provided by far the best description 
of the 1$p$-shell data.  Dynamic effects due to the inclusion of the lower 
components of the Dirac spinors in these calculations were necessary to 
self-consistently reproduce the $1p$-shell cross-section data, the $A_{LT}$ 
asymmetry, and the $R_{LT}$ effective response function over the entire 
measured range of $p_{\rm miss}$.  Within the \textsc{rdwia} framework, the 
four most important ingredients were the inclusion of both bound-nucleon and 
ejectile spinor distortion, the choice of current operator, the choice of 
bound-nucleon wave function, and the choice of optical potential.  Inclusion of
the spinor distortion resulted in a diffractive change in slope in $A_{LT}$ at 
$p_{\rm miss}$ $\approx$ 300 MeV/$c$ which agreed nicely with the data. A 
different choice of current operator either damped out or magnified this change
in slope.  A different choice of bound-nucleon wave function changed the 
$p_{\rm miss}$-location of the change in slope, but preserved the magnitude.  A
different choice of optical potential changed the magnitude of the change in 
slope but preserved the $p_{\rm miss}$-location.  

As anticipated, since $p_p$ $\approx$ 1 GeV/$c$, the \textsc{romea} 
calculations provided a reasonable description of the 1$p$-shell data.  For 
this energy range, optical models generally provide an overall better 
description of proton elastic scattering than does the Glauber model.  This is 
in part due to important medium modifications of the $NN$ interaction from 
Pauli blocking and spinor distortion.  Surprisingly, the unfactorized 
`out-of-the-box' \textsc{rmsga} calculation provided a fairly good description 
of the 1$p$-shell data already at this relatively low proton momentum.  Adding 
the contributions of two-body currents due to MEC and IC to the descriptions of
the 1$p$-shell data provided by the bare \textsc{romea} and \textsc{rmsga} 
calculations did not improve the agreement.

The \textsc{rdwia} calculation with single-nucleon currents was used to fit 
normalization factors to the data from this experiment and from several other 
experiments at lower $Q^{2}$.  Ignoring two experiments which appear to have 
normalization problems, normalization factors of 0.63(9) and 0.60(9) were 
obtained for the lowest $1p_{1/2}$- and $1p_{3/2}$-states with no significant 
dependence upon $Q^{2}$ or $T_p$.  The estimated uncertainties account for 
variations due to the choice of bound-nucleon wave functions, optical 
potentials, and other aspects of the model.  After accounting for other known 
but unresolved $1p_{3/2}$-states, the total $1p$-shell spectroscopic strength 
below about 15 MeV excitation is estimated to be about $0.65 \pm 0.10$ 
relative to full occupancy.

For 25 $<$ $E_{\rm miss}$ $<$ 50 MeV and $p_{\rm miss}$ $\leq$ 145 MeV/$c$, the 
reaction was dominated by the knockout of $1s_{1/2}$-state protons and the 
cross section and effective response functions were reasonably well-described 
by bare \textsc{romea} calculations which did not consider the 
contributions of two-body currents due to MEC and IC.  However, as 
$p_{\rm miss}$ increased beyond 145 MeV/$c$, the single-particle aspect of the 
reaction diminished.  Cross-section data and response functions were no longer 
peaked at $E_{\rm miss}$ $\approx$ 40 MeV, nor did they exhibit the Lorentzian 
$s$-shell shape.  Already at $p_{\rm miss}$ $=$ 280 MeV/$c$, the same bare 
\textsc{romea} calculations that did well describing the data for 
$p_{\rm miss}$ $<$ 145 MeV/$c$ underestimated the cross-section data by more 
than an order of magnitude.  Including the contributions of two-body currents 
due to MEC and IC improved the agreement for $E_{\rm miss}$ $<$ 50 MeV, but the
calculations still dramatically underpredict the data.

For 25 $<$ $E_{\rm miss}$ $<$ 120 MeV and $p_{\rm miss}$ $\geq$ 280 MeV/$c$, 
the cross-section data were almost constant as a function of both 
$p_{\rm miss}$ and $E_{\rm miss}$.  Here, the single-particle aspect of the 
$1s_{1/2}$-state contributed $<$10\% to the cross section.  Two-nucleon 
$(e,e^{\prime}pN)$ calculations accounted for only about 50\% of the magnitude 
of the cross-section data, but reproduced the shape well.  The model, which 
explained the shape, transverse nature, and 50\% of the measured cross section,
suggested that the contributions of the two-nucleon currents due to MEC and IC 
are much larger than those of the two-nucleon correlations.  The magnitude of 
the measured cross section that remains unaccounted for suggests additional 
currents and processes play an equally important role.

\begin{acknowledgments}
We acknowledge the outstanding support of the staff of the Accelerator and 
Physics Divisions at Jefferson Lab that made this experiment successful.  We 
thank T. W. Donnelly and J. W. Van Orden for valuable discussions.  This work 
was supported in part by the U. S. Department of Energy contract 
DE-AC05-84ER40150 under which the Southeastern Universities Research 
Association (SURA) operates the Thomas Jefferson National Accelerator Facility,
other Department of Energy contracts, the National Science Foundation, the 
Swedish Research Council (VR), the Italian Istituto Nazionale di Fisica 
Nucleare (INFN), the French Centre National de la Recherche Scientifique (CNRS)
and Commissariat \'{a} l'Energie Atomique, and the Natural Sciences and 
Engineering Research Council of Canada.
\end{acknowledgments}

\newpage

\appendix

\begin{widetext}

\section{Quasielastic results}
\label{appendix:appendix_a}
\subsection{Cross-section data}

The cross-section data are presented in this section.  Note that the data sets
obtained for $\theta_{pq}$ $=$ 0$^{\circ}$ presented in Tables 
\ref{table:pshellpararesults}, \ref{table:0843cont000deg}, 
\ref{table:1643cont000deg}, and \ref{table:2442cont000deg} do not truly 
represent parallel kinematics even though the HRS$_{h}$ was aligned along 
$\bm{q}$.  Since $\bm{{p}_{p}}$ and $\bm{q}$ had about the same magnitude, 
$\bm{{p}_{\rm miss}}$ arose from the slight angles between them, not from 
differences in their magnitudes.  However, since the distribution of 
$\bm{{p}_{\rm miss}}$ was symmetrical about $\bm{q}$, the conditions of 
parallel kinematics were closely approximated.

\subsubsection{$1p$-shell}

Cross-section data for QE proton knockout from the $1p$-shell of $^{16}$O are 
presented in Tables \ref{table:pshellpararesults} and 
\ref{table:pshellperpresults}.
 
\begingroup
\squeezetable
\begin{longtable*}{rr|rrrr|rrrr}
\caption{\label{table:pshellpararesults}
Measured cross-section data for QE proton knockout from the $1p$-shell of 
$^{16}$O for $<T_{p}>$ = 427 MeV.  The $p_{\rm miss}$ bins were 20 MeV/$c$ 
wide.  Cuts were applied to remove the radiative tail from $^{1}$H$(e,ep)$ such
that $<p_{\rm miss}>$ = 52.5 MeV/$c$ in each case.} \\
\hline \hline
               &               &                                                             \multicolumn{4}{r|}{$1p_{1/2}$-state} &                                                              \multicolumn{4}{r}{$1p_{3/2}$-state} \\
\cline{3-10}
$E_{\rm beam}$ & $\theta_{pq}$ &       $<Q^{2}>$ & $<\omega>$ & $d^{5}\sigma/d\omega d\Omega_{e} d\Omega_{p}$ &              (sys) &       $<Q^{2}>$ & $<\omega>$ & $d^{5}\sigma/d\omega d\Omega_{e} d\Omega_{p}$ &              (sys) \\
         (GeV) &  ($^{\circ}$) & (GeV/$c$)$^{2}$ &      (MeV) &                             (nb/MeV/sr$^{2}$) &               (\%) & (GeV/$c$)$^{2}$ &      (MeV) &                             (nb/MeV/sr$^{2}$) &               (\%) \\
\hline
         0.843 &           0.0 &           0.810 &      436.0 &                           0.0922 $\pm$ 0.0118 &                5.4 &           0.810 &      436.0 &                           0.1143 $\pm$ 0.0115 &                5.4 \\
\hline
         1.643 &           0.0 &           0.820 &      421.5 &                           0.5827 $\pm$ 0.0486 &                5.5 &           0.820 &      421.5 &                           0.7418 $\pm$ 0.0514 &                5.4 \\
\hline
         2.442 &           0.0 &           0.815 &      423.0 &                           1.5030 $\pm$ 0.1380 &                5.5 &           0.815 &      423.0 &                           1.8540 $\pm$ 0.1500 &                5.5 \\
\hline \hline
\end{longtable*}
\endgroup

\begingroup
\squeezetable
\begin{longtable*}{rr|rrr|rrr}
\caption{\label{table:pshellperpresults}
Measured cross-section data for QE proton knockout from the $1p$-shell of 
$^{16}$O for $<Q^{2}>$ $=$ 0.800 (GeV/$c$)$^{2}$, $<\omega>$ = 436 MeV, 
and $<T_{p}>$ = 427 MeV.  The $p_{\rm miss}$ bins were 20 MeV/$c$ wide.} \\
\hline \hline
               &               &                                                 \multicolumn{3}{r|}{$1p_{1/2}$-state} &                                                  \multicolumn{3}{r}{$1p_{3/2}$-state} \\
\cline{3-8}
$E_{\rm beam}$ & $\theta_{pq}$ & $<p_{\rm miss}>$ & $d^{5}\sigma/d\omega d\Omega_{e} d\Omega_{p}$ &              (sys) & $<p_{\rm miss}>$ & $d^{5}\sigma/d\omega d\Omega_{e} d\Omega_{p}$ &              (sys) \\
         (GeV) &  ($^{\circ}$) &        (MeV/$c$) &                             (nb/MeV/sr$^{2}$) &               (\%) &        (MeV/$c$) &                             (nb/MeV/sr$^{2}$) &               (\%) \\
\hline
         0.843 &           8.0 &            150.0 &                           0.0789 $\pm$ 0.0057 &                5.6 &            150.0 &                           0.1467 $\pm$ 0.0076 &                5.5 \\
               &          16.0 &            275.0 &                           0.0011 $\pm$ 0.0003 &                6.1 &            275.0 &                           0.0058 $\pm$ 0.0006 &                7.2 \\
\hline
         1.643 &          -8.0 &           -148.0 &                           0.2950 $\pm$ 0.0320 &                5.5 &           -146.0 &                           0.5160 $\pm$ 0.0370 &                5.5 \\
               &           8.0 &            148.0 &                           0.5250 $\pm$ 0.0310 &                5.5 &            146.0 &                           1.0390 $\pm$ 0.0360 &                5.4 \\
\hline \hline
\end{longtable*}
\endgroup

\subsubsection{Higher missing energies}

Cross-section data together with statistical uncertainties for QE proton 
knockout from $^{16}$O for $E_{\rm miss} >$ 20 MeV are presented in Tables 
\ref{table:0843cont000deg} -- \ref{table:2442cont020deg}.

\begingroup
\squeezetable
\begin{longtable*}{ccc|cc}
\caption{\label{table:0843cont000deg}
Measured cross-section data for QE proton knockout from $^{16}$O for 
$E_{\rm beam} =$ 0.843 GeV, $\theta_{pq} =$ 0.0$^{\circ}$, and $E_{\rm miss} 
>$ 25 MeV.  The $p_{\rm miss}$ bins were 5 MeV/$c$ wide.  Cuts were applied to 
remove the radiative tail from $^{1}$H$(e,ep)$.  There is an additional 
5.8\% systematic uncertainty associated with these results.} \\
\hline \hline
$E_{\rm miss}$ & $<\omega>$ &       $<Q^{2}>$ & $<p_{\rm miss}>$ & $d^{6}\sigma/d\omega dE_{p} d\Omega_{e} d\Omega_{p}$ \\
         (MeV) &      (MeV) & (GeV/$c$)$^{2}$ &        (MeV/$c$) &                              (nb/MeV$^{2}$/sr$^{2}$) \\
\hline
          27.5 &      445.2 &           0.796 &            38.7 &                                   0.0024 $\pm$ 0.0005 \\
          32.5 &      445.2 &           0.795 &            41.8 &                                   0.0085 $\pm$ 0.0008 \\
          37.5 &      445.8 &           0.794 &            45.0 &                                   0.0097 $\pm$ 0.0008 \\
          42.5 &      446.3 &           0.793 &            48.5 &                                   0.0113 $\pm$ 0.0009 \\
          47.5 &      447.7 &           0.790 &            51.3 &                                   0.0106 $\pm$ 0.0010 \\
          52.5 &      449.7 &           0.786 &            53.2 &                                   0.0065 $\pm$ 0.0010 \\
          57.5 &      451.7 &           0.782 &            55.5 &                                   0.0062 $\pm$ 0.0013 \\
\hline \hline
\end{longtable*}
\endgroup

\pagebreak

\begingroup
\squeezetable
\begin{longtable*}{ccc|cc}
\caption{\label{table:0843cont008deg}
Measured cross-section data for QE proton knockout from $^{16}$O for 
$E_{\rm beam} =$ 0.843 GeV, $\theta_{pq} =$ 8.0$^{\circ}$, and $E_{\rm miss} 
>$ 25 MeV.  The $p_{\rm miss}$ bins were 5 MeV/$c$ wide.  There is an 
additional 6.0\% systematic uncertainty associated with these results.} \\
\hline \hline
$E_{\rm miss}$ & $<\omega>$ &       $<Q^{2}>$ & $<p_{\rm miss}>$ & $d^{6}\sigma/d\omega dE_{p} d\Omega_{e} d\Omega_{p}$ \\
         (MeV) &      (MeV) & (GeV/$c$)$^{2}$ &        (MeV/$c$) &                              (nb/MeV$^{2}$/sr$^{2}$) \\
\hline
          25.5 &      444.9 &           0.796 &            143.3 &                                  0.0048 $\pm$ 0.0013 \\
          26.5 &      444.6 &           0.796 &            143.5 &                                  0.0019 $\pm$ 0.0010 \\
          27.5 &      445.2 &           0.795 &            143.8 &                                  0.0008 $\pm$ 0.0009 \\
          28.5 &      444.8 &           0.796 &            143.5 &                                  0.0041 $\pm$ 0.0011 \\
          29.5 &      444.6 &           0.796 &            143.9 &                                  0.0024 $\pm$ 0.0010 \\
\hline \hline
\end{longtable*}
\endgroup

\begingroup
\squeezetable
\begin{longtable*}{ccc|cc}
\caption{\label{table:0843cont016deg}
Measured cross-section data for QE proton knockout from $^{16}$O for 
$E_{\rm beam} =$ 0.843 GeV, $\theta_{pq} =$ 16.0$^{\circ}$, and $E_{\rm miss} 
>$ 25 MeV.  The $p_{\rm miss}$ bins were 5 MeV/$c$ wide.  There is an
additional 5.9\% systematic uncertainty associated with these results.} \\
\hline \hline
$E_{\rm miss}$ & $<\omega>$ &       $<Q^{2}>$ & $<p_{\rm miss}>$ & $d^{6}\sigma/d\omega dE_{p} d\Omega_{e} d\Omega_{p}$ \\
         (MeV) &      (MeV) & (GeV/$c$)$^{2}$ &        (MeV/$c$) &                              (nb/MeV$^{2}$/sr$^{2}$) \\
\hline
          27.5 &      446.0 &           0.795 &            281.6 &                                  0.0004 $\pm$ 0.0001 \\
          32.5 &      446.0 &           0.795 &            281.3 &                                  0.0003 $\pm$ 0.0001 \\
          37.5 &      445.9 &           0.795 &            281.2 &                                  0.0003 $\pm$ 0.0001 \\
          42.5 &      446.1 &           0.795 &            281.0 &                                  0.0002 $\pm$ 0.0001 \\
          47.5 &      448.1 &           0.790 &            281.8 &                                  0.0002 $\pm$ 0.0001 \\
          52.5 &      450.5 &           0.786 &            282.6 &                                  0.0003 $\pm$ 0.0001 \\
          57.5 &      452.8 &           0.781 &            283.6 &                                  0.0002 $\pm$ 0.0001 \\
\hline \hline
\end{longtable*}      
\endgroup

\begingroup
\squeezetable
\begin{longtable*}{ccc|cc}
\caption{\label{table:1643cont000deg}
Measured cross-section data for QE proton knockout from $^{16}$O for 
$E_{\rm beam} =$ 1.643 GeV, $\theta_{pq} =$ 0.0$^{\circ}$, and $E_{\rm miss}$ 
$>$ 25 MeV.   The $p_{\rm miss}$ bins were 5 MeV/$c$ wide.  Cuts were applied 
to remove the radiative tail from $^{1}$H$(e,ep)$.  There is an additional 
5.8\% systematic uncertainty associated with these results.} \\
\hline \hline
$E_{\rm miss}$ & $<\omega>$ &       $<Q^{2}>$ & $<p_{\rm miss}>$ & $d^{6}\sigma/d\omega dE_{p} d\Omega_{e} d\Omega_{p}$ \\
         (MeV) &      (MeV) & (GeV/$c$)$^{2}$ &        (MeV/$c$) &                              (nb/MeV$^{2}$/sr$^{2}$) \\
\hline
          27.5 &      450.0 &           0.795 &            62.5 &                                   0.0195 $\pm$ 0.0013 \\
          32.5 &      450.0 &           0.795 &            62.5 &                                   0.0293 $\pm$ 0.0012 \\
          37.5 &      450.0 &           0.795 &            62.5 &                                   0.0483 $\pm$ 0.0013 \\
          42.5 &      450.0 &           0.795 &            62.5 &                                   0.0534 $\pm$ 0.0014 \\
          47.5 &      450.0 &           0.795 &            62.5 &                                   0.0445 $\pm$ 0.0014 \\
          52.5 &      450.0 &           0.795 &            65.0 &                                   0.0263 $\pm$ 0.0015 \\
          57.5 &      450.0 &           0.795 &            67.5 &                                   0.0132 $\pm$ 0.0021 \\
\hline \hline
\end{longtable*}
\endgroup

\pagebreak

\begingroup
\squeezetable
\begin{longtable*}{ccc|cc|cr}
\caption{\label{table:1643cont008deg}
Measured cross-section data for QE proton knockout from $^{16}$O for 
$E_{\rm beam}$ $=$ 1.643 GeV, $|\theta_{pq}| =$ 8.0$^{\circ}$, and 
$E_{\rm miss}$ $>$ 25 MeV.  The $p_{\rm miss}$ bins were 5 MeV/$c$ wide.  There
is an additional 6.0\% systematic uncertainty associated with these results.} 
\\

\hline \hline
               &            &                 &                         \multicolumn{2}{r|}{$\theta_{pq}=+8.0^{\circ}$} &                          \multicolumn{2}{r}{$\theta_{pq}=-8.0^{\circ}$} \\
\cline{4-7}
$E_{\rm miss}$ & $<\omega>$ &       $<Q^{2}>$ & $<p_{\rm miss}>$ & $d^{6}\sigma/d\omega dE_{p} d\Omega_{e} d\Omega_{p}$ & $<p_{\rm miss}>$ & $d^{6}\sigma/d\omega dE_{p} d\Omega_{e} d\Omega_{p}$ \\
         (MeV) &      (MeV) & (GeV/$c$)$^{2}$ &        (MeV/$c$) &                              (nb/MeV$^{2}$/sr$^{2}$) &        (MeV/$c$) &                              (nb/MeV$^{2}$/sr$^{2}$) \\
\hline
          25.5 &      436.2 &           0.804 &            141.8 &                                  0.0361 $\pm$ 0.0036 &            153.2 &                                  0.0163 $\pm$ 0.0020 \\
          26.5 &      438.0 &           0.802 &            142.7 &                                  0.0231 $\pm$ 0.0031 &            152.2 &                                  0.0130 $\pm$ 0.0018 \\
          27.5 &      439.0 &           0.802 &            144.1 &                                  0.0193 $\pm$ 0.0029 &            151.7 &                                  0.0079 $\pm$ 0.0016 \\
          28.5 &      439.5 &           0.802 &            144.9 &                                  0.0138 $\pm$ 0.0026 &            150.5 &                                  0.0060 $\pm$ 0.0015 \\
\hline \hline
\end{longtable*}      
\endgroup

\pagebreak

\begingroup
\squeezetable
\begin{longtable*}{ccc|cc}
\caption{\label{table:2442cont000deg}
Measured cross-section data for QE proton knockout from $^{16}$O for 
$E_{\rm beam}$ $=$ 2.442 GeV, $\theta_{pq} =$ 0.0$^{\circ}$, and 
$E_{\rm miss} >$ 25 MeV.  The $p_{\rm miss}$ bins were 5 MeV/$c$ wide.  Cuts 
were applied to remove the radiative tail from $^{1}$H$(e,ep)$.  There is an
additional 5.8\% systematic uncertainty associated with these results.} \\
\hline \hline
$E_{\rm miss}$ & $<\omega>$ &       $<Q^{2}>$ & $<p_{\rm miss}>$ & $d^{6}\sigma/d\omega dE_{p} d\Omega_{e} d\Omega_{p}$ \\
         (MeV) &      (MeV) & (GeV/$c$)$^{2}$ &        (MeV/$c$) &                              (nb/MeV$^{2}$/sr$^{2}$) \\
\hline
          27.5 &      450.0 &           0.795 &            60.0 &                                   0.0552 $\pm$ 0.0042 \\
          32.5 &      450.0 &           0.795 &            60.0 &                                   0.0890 $\pm$ 0.0045 \\
          37.5 &      450.0 &           0.795 &            60.0 &                                   0.1387 $\pm$ 0.0050 \\
          42.5 &      450.0 &           0.795 &            60.0 &                                   0.1580 $\pm$ 0.0058 \\
          47.5 &      450.0 &           0.795 &            62.5 &                                   0.1348 $\pm$ 0.0062 \\
          52.5 &      450.0 &           0.795 &            65.0 &                                   0.0756 $\pm$ 0.0063 \\
          57.5 &      450.0 &           0.795 &            67.5 &                                   0.0402 $\pm$ 0.0084 \\
\hline \hline
\end{longtable*}
\endgroup

\begingroup
\squeezetable
\begin{longtable*}{ccc|cc}
\caption{\label{table:2442cont0025deg}
Measured cross-section data for QE proton knockout from $^{16}$O
for $E_{\rm beam}$ $=$ 2.442 GeV, $\theta_{pq} =$ $-2.5^{\circ}$, and
$E_{\rm miss}$ $>$ 25 MeV.  The $p_{\rm miss}$ bins were 5 MeV/$c$ wide.
There is an additional 5.9\% systematic uncertainty associated with these 
results.} \\
\hline \hline
$E_{\rm miss}$ & $<\omega>$ &       $<Q^{2}>$ & $<p_{\rm miss}>$ & $d^{6}\sigma/d\omega dE_{p} d\Omega_{e} d\Omega_{p}$ \\
         (MeV) &      (MeV) & (GeV/$c$)$^{2}$ &        (MeV/$c$) &                              (nb/MeV$^{2}$/sr$^{2}$) \\
\hline
          25.5 &      445.0 &           0.823 &            46.5 &                                  0.0330 $\pm$ 0.0182 \\
          26.5 &      446.5 &           0.822 &            46.8 &                                  0.0661 $\pm$ 0.0185 \\
          27.5 &      449.5 &           0.821 &            46.7 &                                  0.0388 $\pm$ 0.0173 \\
          28.5 &      449.3 &           0.820 &            45.3 &                                  0.0590 $\pm$ 0.0176 \\
          29.5 &      450.4 &           0.823 &            46.2 &                                  0.0955 $\pm$ 0.0192 \\
\hline \hline
\end{longtable*}
\endgroup

\pagebreak

\begingroup
\squeezetable
\begin{longtable*}{ccc|cc|cr}
\caption{\label{table:2442cont008deg}
Measured cross-section data for QE proton knockout from $^{16}$O for 
$E_{\rm beam}$ $=$ 2.442 GeV, $|\theta_{pq}|$ $=$ 8.0$^{\circ}$, and 
$E_{\rm miss}$ $>$ 25 MeV.  The $p_{\rm miss}$ bins were 5 MeV/$c$ wide.  There
is an additional 6.0\% systematic uncertainty associated with these results.} \\
\hline \hline
               &            &                 &                         \multicolumn{2}{r|}{$\theta_{pq}=+8.0^{\circ}$} &                          \multicolumn{2}{r}{$\theta_{pq}=-8.0^{\circ}$} \\
\cline{4-7}
$E_{\rm miss}$ & $<\omega>$ &       $<Q^{2}>$ & $<p_{\rm miss}>$ & $d^{6}\sigma/d\omega dE_{p} d\Omega_{e} d\Omega_{p}$ & $<p_{\rm miss}>$ & $d^{6}\sigma/d\omega dE_{p} d\Omega_{e} d\Omega_{p}$ \\
         (MeV) &      (MeV) & (GeV/$c$)$^{2}$ &        (MeV/$c$) &                              (nb/MeV$^{2}$/sr$^{2}$) &        (MeV/$c$) &                              (nb/MeV$^{2}$/sr$^{2}$) \\
\hline
          25.5 &      438.1 &           0.793 &            150.0 &                                  0.0595 $\pm$ 0.0058 &            160.0 &                                  0.0174 $\pm$ 0.0030 \\
          26.5 &      439.6 &           0.791 &            152.5 &                                  0.0310 $\pm$ 0.0051 &            160.0 &                                  0.0165 $\pm$ 0.0027 \\
          27.5 &      440.5 &           0.791 &            155.0 &                                  0.0371 $\pm$ 0.0049 &            155.2 &                                  0.0190 $\pm$ 0.0026 \\
          28.5 &      441.7 &           0.789 &            155.0 &                                  0.0474 $\pm$ 0.0050 &            157.9 &                                  0.0249 $\pm$ 0.0027 \\
\hline \hline
\end{longtable*}
\endgroup

\begingroup
\squeezetable
\begin{longtable*}{ccc|cc|cr}
\caption{\label{table:2442cont016deg}
Measured cross-section data for QE proton knockout from $^{16}$O for 
$E_{\rm beam}$ $=$ 2.442 GeV, $|\theta_{pq}|$ $=$ 16.0$^{\circ}$, and 
$E_{\rm miss}$ $>$ 25 MeV.  The $p_{\rm miss}$ bins were 5 MeV/$c$ wide.  There
is an additional 5.8\% systematic uncertainty associated with these results.} \\
\hline \hline
               &            &                 &                        \multicolumn{2}{r|}{$\theta_{pq}=+16.0^{\circ}$} &                         \multicolumn{2}{r}{$\theta_{pq}=-16.0^{\circ}$} \\
\cline{4-7}
$E_{\rm miss}$ & $<\omega>$ &       $<Q^{2}>$ & $<p_{\rm miss}>$ & $d^{6}\sigma/d\omega dE_{p} d\Omega_{e} d\Omega_{p}$ & $<p_{\rm miss}>$ & $d^{6}\sigma/d\omega dE_{p} d\Omega_{e} d\Omega_{p}$ \\
         (MeV) &      (MeV) & (GeV/$c$)$^{2}$ &        (MeV/$c$) &                              (nb/MeV$^{2}$/sr$^{2}$) &        (MeV/$c$) &                              (nb/MeV$^{2}$/sr$^{2}$) \\
\hline
          25.5 &      439.9 &           0.794 &            268.9 &                                  0.0078 $\pm$ 0.0008 &            282.1 &                                  0.0095 $\pm$ 0.0020 \\
          26.5 &      440.0 &           0.788 &            273.8 &                                  0.0076 $\pm$ 0.0008 &            278.1 &                                  0.0014 $\pm$ 0.0019 \\
          27.5 &      443.6 &           0.799 &            274.9 &                                  0.0053 $\pm$ 0.0007 &            280.4 &                                  0.0078 $\pm$ 0.0020 \\
          28.5 &      444.8 &           0.788 &            275.8 &                                  0.0054 $\pm$ 0.0007 &            276.0 &                                  0.0043 $\pm$ 0.0020 \\
\hline \hline
\end{longtable*}
\endgroup

\begingroup
\squeezetable
\begin{longtable*}{ccc|cc|cr}
\caption{\label{table:2442cont020deg}
Measured cross-section data for QE proton knockout from $^{16}$O for 
$E_{\rm beam}$ $=$ 2.442 GeV, $|\theta_{pq}|$ $=$ 20.0$^{\circ}$, and 
$E_{\rm miss}$ $>$ 25 MeV.  The $p_{\rm miss}$ bins were 5 MeV/$c$ wide.  There
is an additional 5.9\% systematic uncertainty associated with these results.} \\
\hline \hline
               &            &                 &                        \multicolumn{2}{r|}{$\theta_{pq}=+20.0^{\circ}$} &                         \multicolumn{2}{r}{$\theta_{pq}=-20.0^{\circ}$} \\
\cline{4-7}
$E_{\rm miss}$ & $<\omega>$ &       $<Q^{2}>$ & $<p_{\rm miss}>$ & $d^{6}\sigma/d\omega dE_{p} d\Omega_{e} d\Omega_{p}$ & $<p_{\rm miss}>$ & $d^{6}\sigma/d\omega dE_{p} d\Omega_{e} d\Omega_{p}$ \\
         (MeV) &      (MeV) & (GeV/$c$)$^{2}$ &        (MeV/$c$) &                              (nb/MeV$^{2}$/sr$^{2}$) &        (MeV/$c$) &                              (nb/MeV$^{2}$/sr$^{2}$) \\
\hline
          25.5 &      440.2 &           0.787 &            339.4 &                                  0.0019 $\pm$ 0.0003 &            349.6 &                                  0.0002 $\pm$ 0.0005 \\
          26.5 &      439.3 &           0.790 &            340.7 &                                  0.0019 $\pm$ 0.0003 &            348.4 &                                  0.0011 $\pm$ 0.0006 \\
          27.5 &      441.4 &           0.796 &            346.8 &                                  0.0028 $\pm$ 0.0003 &            347.5 &                                  0.0015 $\pm$ 0.0006 \\
          28.5 &      441.7 &           0.792 &            345.5 &                                  0.0012 $\pm$ 0.0003 &            347.6 &                                  0.0007 $\pm$ 0.0005 \\
\hline \hline
\end{longtable*}
\endgroup

\subsection{Effective response functions and asymmetries}
\subsubsection{$1p$-shell}

Effective response functions and asymmetries for QE proton knockout from 
$^{16}$O are presented in Tables \ref{table:pshellrlrt}, 
\ref{table:pshellrlttrt}, and \ref{table:pshellaltrlt}.

\begingroup
\squeezetable
\begin{longtable*}{r|r|r|r|r|r}
\caption{\label{table:pshellrlrt}
Effective response functions $R_{L}$ and $R_{T}$ for QE proton knockout from 
the $1p$-shell of $^{16}$O for $<T_{p}>$ $=$ 427 MeV.  The $p_{\rm miss}$ bins 
were 20 MeV/$c$ wide.  Cuts were applied to remove the radiative tail from 
$^{1}$H$(e,ep)$ such that $<p_{\rm miss}>$ $=$ 52.5 MeV/$c$ in each case.  Note
that the $\omega$ acceptance of the three measurements was shifted by about 3\%
in order to keep the phase-space acceptance flat.  The resulting change in the 
domain was taken as an additional 4\% systematic uncertainty.} \\
\hline \hline
                                     \multicolumn{3}{r|}{$1p_{1/2}$-state} &                                       \multicolumn{3}{r}{$1p_{3/2}$-state} \\
\hline
$<p_{\rm miss}>$ &         $R_{L}$ (fm$^{3}$) &         $R_{T}$ (fm$^{3}$) & $<p_{\rm miss}>$ &         $R_{L}$ (fm$^{3}$) &         $R_{T}$ (fm$^{3}$) \\
       (MeV/$c$) &           (stat)$~~~$(sys) &           (stat)$~~~$(sys) &        (MeV/$c$) &           (stat)$~~~$(sys) &           (stat)$~~~$(sys) \\
\hline
            52.5 & 1.82 $\pm$ 1.17 $\pm$ 0.59 & 7.58 $\pm$ 1.42 $\pm$ 0.72 &             52.5 & 2.35 $\pm$ 1.18 $\pm$ 0.73 & 9.40 $\pm$ 1.39 $\pm$ 0.85 \\
\hline \hline
\end{longtable*}
\endgroup

\begingroup
\squeezetable
\begin{longtable*}{r|r|r|r|r|r}
\caption{\label{table:pshellrlttrt}
Effective response functions $R_{L+TT}$ and $R_{T}$ for QE proton knockout from
the $1p$-shell of $^{16}$O for $<Q^{2}>$ $=$ 0.800 (GeV/$c$)$^{2}$, 
$<\omega>$ $=$ 436 MeV, and $<T_{p}>$ $=$ 427 MeV.  The $p_{\rm miss}$ bins 
were 20 MeV/$c$ wide.} \\
\hline \hline
                                     \multicolumn{3}{r|}{$1p_{1/2}$-state} &                                        \multicolumn{3}{r}{$1p_{3/2}$-state} \\
\hline
$<p_{\rm miss}>$ &      $R_{L+TT}$ (fm$^{3}$) &         $R_{T}$ (fm$^{3}$) & $<p_{\rm miss}>$ &     $R_{L+TT}$  (fm$^{3}$) &          $R_{T}$ (fm$^{3}$) \\
       (MeV/$c$) &           (stat)$~~~$(sys) &           (stat)$~~~$(sys) &        (MeV/$c$) &           (stat)$~~~$(sys) &            (stat)$~~~$(sys) \\
\hline
           149.0 & 0.56 $\pm$ 0.49 $\pm$ 0.12 & 6.08 $\pm$ 0.61 $\pm$ 0.24 &            149.0 & 2.20 $\pm$ 0.75 $\pm$ 0.30 & 10.35 $\pm$ 1.04 $\pm$ 0.43 \\
           279.0 & 0.01 $\pm$ 0.03 $\pm$ 0.01 & 0.12 $\pm$ 0.04 $\pm$ 0.01 &            276.0 & 0.20 $\pm$ 0.06 $\pm$ 0.03 &  0.29 $\pm$ 0.08 $\pm$ 0.01 \\
\hline \hline
\end{longtable*}
\endgroup

\begingroup
\squeezetable
\begin{longtable*}{r|r|r|r|r|r}
\caption{\label{table:pshellaltrlt}
The asymmetry $A_{LT}$ and effective response functions $R_{LT}$ for QE proton 
knockout from the $1p$-shell of $^{16}$O for $<Q^{2}>$ $=$ 0.800 
(GeV/$c$)$^{2}$, $<\omega>$ $=$ 436 MeV, and $<T_{p}>$ $=$ 427 MeV.  The 
$p_{\rm miss}$ bins were 20 MeV/$c$ wide.  Save for the data labelled [*] which
were obtained at $E_{\rm beam}$ = 1.643 GeV, the beam energy was 2.442 GeV.} \\
\hline \hline
                                                                \multicolumn{3}{r|}{$1p_{1/2}$-state} &                                                \multicolumn{3}{r}{$1p_{3/2}$-state} \\
\hline
                     $<p_{\rm miss}>$ &                     $A_{LT}$ &            $R_{LT}$ (fm$^{3}$) &     $<p_{\rm miss}>$ &                    $A_{LT}$ &            $R_{LT}$ (fm$^{3}$) \\
                            (MeV/$c$) &             (stat)$~~~$(sys) &             (stat)$~~~~~$(sys) &            (MeV/$c$) &            (stat)$~~~$(sys) &             (stat)$~~~~~$(sys) \\
\hline
                  60.0\hspace*{2.9mm} &  0.02 $\pm$ 0.02 $\pm $ 0.02 &  0.117 $\pm$ 0.134 $\pm$ 0.037 &  60.0\hspace*{2.8mm} & -0.08 $\pm$ 0.02 $\pm$ 0.02 & -0.754 $\pm$ 0.165 $\pm$ 0.084 \\
                            148.0 [*] & -0.25 $\pm$ 0.02 $\pm $ 0.03 & -1.198 $\pm$ 0.235 $\pm$ 0.085 &            147.0 [*] & -0.31 $\pm$ 0.04 $\pm$ 0.03 & -2.820 $\pm$ 0.292 $\pm$ 0.183 \\
                 149.0\hspace*{2.9mm} & -0.23 $\pm$ 0.02 $\pm $ 0.03 & -0.999 $\pm$ 0.066 $\pm$ 0.077 & 148.0\hspace*{2.9mm} & -0.31 $\pm$ 0.01 $\pm$ 0.03 & -2.560 $\pm$ 0.096 $\pm$ 0.173 \\
                 279.0\hspace*{2.9mm} & -0.36 $\pm$ 0.08 $\pm $ 0.04 & -0.029 $\pm$ 0.007 $\pm$ 0.002 & 276.0\hspace*{2.8mm} & -0.69 $\pm$ 0.04 $\pm$ 0.04 & -0.250 $\pm$ 0.013 $\pm$ 0.019 \\
\hline \hline
\end{longtable*}
\endgroup

\subsubsection{Higher missing energies}

Effective response functions for QE proton knockout from $^{16}$O for 
$E_{\rm miss} >$ 25 MeV are presented in Tables \ref{table:contrlrt} --
\ref{table:contrlt}.

\begingroup
\squeezetable
\begin{longtable*}{r|r|r}
\caption{\label{table:contrlrt}
Effective response functions $R_{L}$ and $R_{T}$ for QE proton knockout from 
$^{16}$O at $\theta_{pq}$ = 0$^{\circ}$ for $E_{\rm miss} >$ 25 MeV.  The 
$p_{\rm miss}$ bins were 5 MeV/$c$ wide.} \\
\hline \hline
$<E_{\rm miss}>$ &         $R_{L}$ (fm$^{3}$/MeV) &        $R_{T}$ (fm$^{3}$/MeV) \\
           (MeV) &             (stat)$~~~~~$(sys) &            (stat)$~~~~~$(sys) \\
\hline
            27.5 &  0.067 $\pm$ 0.050 $\pm$ 0.020 & 0.203 $\pm$ 0.065 $\pm$ 0.021 \\
            32.5 &  0.075 $\pm$ 0.044 $\pm$ 0.030 & 0.346 $\pm$ 0.056 $\pm$ 0.035 \\
            37.5 &  0.141 $\pm$ 0.040 $\pm$ 0.052 & 0.523 $\pm$ 0.049 $\pm$ 0.051 \\
            42.5 &  0.138 $\pm$ 0.037 $\pm$ 0.063 & 0.600 $\pm$ 0.044 $\pm$ 0.066 \\
            47.5 &  0.100 $\pm$ 0.039 $\pm$ 0.054 & 0.517 $\pm$ 0.047 $\pm$ 0.051 \\
            52.5 &  0.054 $\pm$ 0.044 $\pm$ 0.049 & 0.304 $\pm$ 0.054 $\pm$ 0.028 \\
            57.5 & -0.047 $\pm$ 0.111 $\pm$ 0.030 & 0.253 $\pm$ 0.145 $\pm$ 0.049 \\
\hline \hline
\end{longtable*}
\endgroup

\begingroup
\squeezetable
\begin{longtable*}{r|r|r|r|r}
\caption{\label{table:contrlttrt}
Effective response functions $R_{L+TT}$ and $R_{T}$ for QE proton knockout from
$^{16}$O at $\theta_{pq}$ $=$ 8$^{\circ}$ and $\theta_{pq}$ $=$ 16$^{\circ}$ 
for $E_{\rm miss}$ $>$ 25 MeV.  The $p_{\rm miss}$ bins were 5 MeV/$c$ wide.} \\
\hline \hline
                                   \multicolumn{3}{r|}{$\theta_{pq}$=8$^{\circ}$} &                 \multicolumn{2}{r}{$\theta_{pq}$=16$^{\circ}$} \\
\hline
$<E_{\rm miss}>$ &      $R_{L+TT}$ (fm$^{3}$/MeV) &        $R_{T}$ (fm$^{3}$/MeV) &      $R_{L+TT}$ (fm$^{3}$/MeV) &        $R_{T}$ (fm$^{3}$/MeV) \\
       (MeV) &                 (stat)$~~~~~$(sys) &            (stat)$~~~~~$(sys) &             (stat)$~~~~~$(sys) &            (stat)$~~~~~$(sys) \\
\hline
            27.5 &  0.080 $\pm$ 0.023 $\pm$ 0.010 & 0.137 $\pm$ 0.029 $\pm$ 0.008 &  0.002 $\pm$ 0.004 $\pm$ 0.003 & 0.027 $\pm$ 0.005 $\pm$ 0.005 \\
            32.5 &  0.072 $\pm$ 0.021 $\pm$ 0.008 & 0.133 $\pm$ 0.026 $\pm$ 0.008 & -0.001 $\pm$ 0.004 $\pm$ 0.001 & 0.026 $\pm$ 0.005 $\pm$ 0.001 \\
            37.5 &  0.075 $\pm$ 0.021 $\pm$ 0.008 & 0.162 $\pm$ 0.027 $\pm$ 0.010 &  0.001 $\pm$ 0.004 $\pm$ 0.001 & 0.018 $\pm$ 0.004 $\pm$ 0.001 \\
            42.5 &  0.087 $\pm$ 0.021 $\pm$ 0.013 & 0.164 $\pm$ 0.026 $\pm$ 0.010 &  0.005 $\pm$ 0.004 $\pm$ 0.001 & 0.014 $\pm$ 0.004 $\pm$ 0.001 \\
            47.5 &  0.050 $\pm$ 0.022 $\pm$ 0.008 & 0.172 $\pm$ 0.028 $\pm$ 0.010 &  0.001 $\pm$ 0.004 $\pm$ 0.001 & 0.017 $\pm$ 0.005 $\pm$ 0.001 \\
            52.5 &  0.001 $\pm$ 0.022 $\pm$ 0.006 & 0.137 $\pm$ 0.027 $\pm$ 0.009 &  0.005 $\pm$ 0.005 $\pm$ 0.001 & 0.018 $\pm$ 0.006 $\pm$ 0.002 \\
            57.5 & -0.002 $\pm$ 0.024 $\pm$ 0.004 & 0.010 $\pm$ 0.030 $\pm$ 0.006 &  0.008 $\pm$ 0.007 $\pm$ 0.001 & 0.008 $\pm$ 0.007 $\pm$ 0.001 \\
\hline \hline
\end{longtable*}
\endgroup

\pagebreak

\begingroup
\squeezetable
\begin{longtable*}{r|r|r}
\caption{\label{table:contrlt}
Effective response functions $R_{LT}$ for QE proton knockout from $^{16}$O
at $\theta_{pq}$ $=$ 8$^{\circ}$ and $\theta_{pq}$ $=$ 16$^{\circ}$ for 
$E_{\rm miss}$ $>$ 25 MeV.  The $p_{\rm miss}$ bins were 5 MeV/$c$ wide.  Note 
that the data presented for $\theta_{pq}$ $=$ 8$^{\circ}$ represent the average
of the data obtained at $E_{\rm beam}$ = 1.643 GeV and $E_{\rm beam}$ = 2.442 
GeV.} \\
\hline \hline
   \multicolumn{2}{r|}{$\theta_{pq}$=8$^{\circ}$} & \multicolumn{1}{r}{$\theta_{pq}$=16$^{\circ}$} \\
\hline
$<E_{\rm miss}>$ &        $R_{LT}$ (fm$^{3}$/MeV) &                        $R_{LT}$ (fm$^{3}$/MeV) \\
           (MeV) &             (stat)$~~~~~$(sys) &                             (stat)$~~~~~$(sys) \\
\hline
            27.5 & -0.070 $\pm$ 0.006 $\pm$ 0.005 &                 -0.009 $\pm$ 0.001 $\pm$ 0.001 \\
            32.5 & -0.050 $\pm$ 0.005 $\pm$ 0.004 &                 -0.007 $\pm$ 0.001 $\pm$ 0.001 \\
            37.5 & -0.076 $\pm$ 0.005 $\pm$ 0.010 &                 -0.007 $\pm$ 0.001 $\pm$ 0.002 \\
            42.5 & -0.083 $\pm$ 0.006 $\pm$ 0.039 &                 -0.007 $\pm$ 0.001 $\pm$ 0.001 \\
            47.5 & -0.077 $\pm$ 0.006 $\pm$ 0.010 &                 -0.004 $\pm$ 0.002 $\pm$ 0.001 \\
            52.5 & -0.036 $\pm$ 0.006 $\pm$ 0.015 &                 -0.004 $\pm$ 0.002 $\pm$ 0.001 \\
            57.5 & -0.036 $\pm$ 0.006 $\pm$ 0.005 &                 -0.004 $\pm$ 0.003 $\pm$ 0.001 \\
\hline \hline
\end{longtable*}
\endgroup

\end{widetext}

\section{A `dip'-region investigation}

A small portion of the beam time allocated to the measurement discussed in the 
main body of this article was used for an exploratory investigation of the 
`dip' located in the energy-transfer region between the QE peak and the 
$\Delta(1232)$-resonance.  For this investigation, $E_{\rm beam}$ $=$ 1.643 GeV
was employed, and the HRS$_{e}$ position and central momentum were fixed at 
$\theta_{e}$ $=$ 37.17$^{\circ}$ and $p_{e}$ $=$ 1056 MeV/$c$, respectively.  
This resulted in $q$ $\approx$ 1.026 GeV/$c$, $\omega$ $\approx$ 589 MeV, and 
$Q^{2}$ $\approx$ 0.706 (GeV/$c$)$^{2}$
\footnote{
The quantity $y$ (which is the minimum value of the initial momentum of the 
nucleon) is generally used to label non-QE kinematics.  According to Day 
{\it et al.} \cite{day90}, 
$y = [(m_{A}+\omega)\sqrt{\Lambda^{2}-m^{2}_{B}W^{2}}-q \Lambda]/W^{2}$,
where $W = \sqrt{(m_{A}+\omega)^{2}-q^{2}}$ and 
$\Lambda = (m^{2}_{B}-m^{2}_{N}+W^{2})/2$. In QE kinematics, $y$ $=$ 0.
$y$ $=$ 0.16 in these dip-region kinematics.
}.  
The HRS$_{h}$ was then positioned at $\theta_{h}$ $=$ 38.45$^{\circ}$ 
($\theta_{pq}$ $=$ 0$^{\circ}$) and its central momentum varied from 828 
MeV/$c$ to 1190 MeV/$c$ in five steps of $\Delta p_{p}$ $\approx$ 70 MeV/$c$ 
per step.  These momentum settings were close enough to each other that there 
was adequate acceptance overlap between them to allow for radiative corrections
to be performed.  The configuration of the experimental apparatus and 
data-acquisition system was identical in all aspects to that used for the QE 
measurement.  The data analysis was also identical to that performed on the QE 
data, save for an additional cut to remove H$(e,e^{\prime}p)\pi^{0}$ events.

\begin{figure}
\resizebox{0.47\textwidth}{!}{\includegraphics{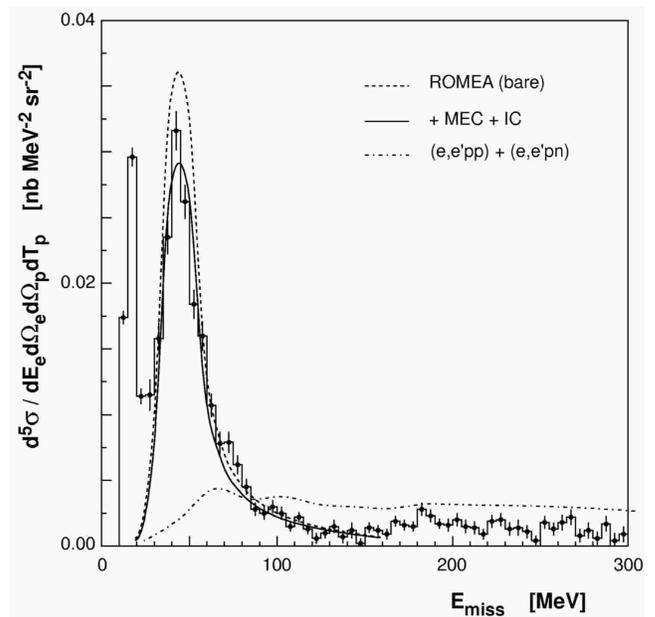}}
\caption{\label{fig:ryckedip}
Data from this work together with calculations by the Ghent Group for the 
$E_{\rm miss}$-dependence of the cross-section data obtained in dip-region 
kinematics for $E_{\rm beam}$ $=$ 1.643 GeV.  Uncertainties are statistical 
and, on average, there is an additional $\pm$5.9\% systematic uncertainty (see 
Table \ref{table:dipcross}) associated with the data.
}
\end{figure}

Fig. \ref{fig:ryckedip} shows the measured cross-section data for the dip 
region as a function of $E_{\rm miss}$ compared to calculations by the Ghent 
Group for $E_{\rm beam}$ = 1.643 GeV (see Table \ref{table:dipcross}).  The
dashed curve is the bare \textsc{romea} calculation for proton knockout 
from the $1s_{1/2}$-state of $^{16}$O and the solid curve is the same 
calculation including the effects of MEC and IC (see the main text of this 
article for further details).   A normalization factor of 1.0 was employed for
these calculations.  The dashed-dotted curve illustrates the calculated 
$(e,e^{\prime}pN)$ contribution.  

In contrast to the QE energy region, the bare calculation actually 
overestimated the $1s_{1/2}$-state strength in these kinematics.  Also in 
contrast to the QE energy region, the inclusion of MEC and IC decrease the 
magnitude of the calculated cross section and improve the agreement.  Finally, 
while the $(e,e^{\prime}pN)$ calculations have the measured flat shape for 
$E_{\rm miss}$ $>$ 100 MeV, they are twice as large as the cross-section data.

\squeezetable
\begin{table}
\caption{\label{table:dipcross}
Measured cross-section data for proton knockout from $^{16}$O for 
$E_{\rm beam}$ $=$ 1.643 GeV in dip-region kinematics for $E_{\rm miss}$ $>$ 
25 MeV.  There is an additional 5.9\% systematic uncertainty associated with 
these results.}
\begin{ruledtabular}
\begin{tabular}{rrr}
$E_{\rm miss}$ & $<p_{\rm miss}>$ & $d^{6}\sigma/d\omega dE_{p} d\Omega_{e} d\Omega_{p}$ \\
         (MeV) &        (MeV/$c$) &                              (nb/MeV$^{2}$/sr$^{2}$) \\
\hline
          27.5 &            163.6 &                                  0.0115 $\pm$ 0.0012 \\
          32.5 &            152.1 &                                  0.0158 $\pm$ 0.0012 \\
          37.5 &            138.8 &                                  0.0235 $\pm$ 0.0013 \\
          42.5 &            132.8 &                                  0.0316 $\pm$ 0.0015 \\
          47.5 &            128.3 &                                  0.0262 $\pm$ 0.0013 \\
          52.5 &            119.4 &                                  0.0184 $\pm$ 0.0011 \\
          57.5 &            112.1 &                                  0.0160 $\pm$ 0.0011 \\
          62.5 &            107.5 &                                  0.0107 $\pm$ 0.0009 \\
          67.5 &             95.0 &                                  0.0078 $\pm$ 0.0009 \\
          72.5 &             95.0 &                                  0.0079 $\pm$ 0.0008 \\
          77.5 &             95.0 &                                  0.0062 $\pm$ 0.0007 \\
          82.5 &             90.0 &                                  0.0045 $\pm$ 0.0006 \\
          87.5 &             84.0 &                                  0.0028 $\pm$ 0.0005 \\
          92.5 &             76.0 &                                  0.0025 $\pm$ 0.0005 \\
          97.5 &             71.7 &                                  0.0030 $\pm$ 0.0005 \\
         102.5 &             68.0 &                                  0.0025 $\pm$ 0.0005 \\
         107.5 &             60.0 &                                  0.0015 $\pm$ 0.0004 \\
         112.5 &             61.4 &                                  0.0022 $\pm$ 0.0004 \\
         117.5 &             58.0 &                                  0.0013 $\pm$ 0.0004 \\
         122.5 &             51.0 &                                  0.0006 $\pm$ 0.0004 \\
         127.5 &             47.5 &                                  0.0010 $\pm$ 0.0004 \\
         132.5 &             50.0 &                                  0.0015 $\pm$ 0.0005 \\
         137.5 &             60.0 &                                  0.0007 $\pm$ 0.0005 \\
         142.5 &             45.0 &                                  0.0012 $\pm$ 0.0006 \\
         147.5 &             39.0 &                                  0.0002 $\pm$ 0.0004 \\
         152.5 &             39.3 &                                  0.0014 $\pm$ 0.0004 \\
         157.5 &             48.8 &                                  0.0012 $\pm$ 0.0004 \\
         162.5 &             51.7 &                                  0.0009 $\pm$ 0.0004 \\
         167.5 &             55.0 &                                  0.0019 $\pm$ 0.0004 \\
         172.5 &             62.0 &                                  0.0016 $\pm$ 0.0004 \\
         177.5 &             65.9 &                                  0.0015 $\pm$ 0.0004 \\
         177.5 &             65.9 &                                  0.0015 $\pm$ 0.0004 \\
         182.5 &             69.0 &                                  0.0028 $\pm$ 0.0005 \\
         187.5 &             77.2 &                                  0.0023 $\pm$ 0.0005 \\
         192.5 &             94.0 &                                  0.0017 $\pm$ 0.0004 \\
         197.5 &             97.7 &                                  0.0016 $\pm$ 0.0005 \\
         202.5 &             97.7 &                                  0.0020 $\pm$ 0.0005 \\
         207.5 &            103.0 &                                  0.0015 $\pm$ 0.0005 \\
         212.5 &            114.1 &                                  0.0014 $\pm$ 0.0005 \\
         217.5 &            118.3 &                                  0.0009 $\pm$ 0.0004 \\
         222.5 &            122.3 &                                  0.0019 $\pm$ 0.0005 \\
         227.5 &            125.9 &                                  0.0020 $\pm$ 0.0005 \\
         232.5 &            135.0 &                                  0.0013 $\pm$ 0.0005 \\
         237.5 &            150.5 &                                  0.0014 $\pm$ 0.0006 \\
         242.5 &            157.7 &                                  0.0011 $\pm$ 0.0005 \\
         247.5 &            157.7 &                                  0.0004 $\pm$ 0.0004 \\
         252.5 &            167.0 &                                  0.0018 $\pm$ 0.0005 \\
         257.5 &            170.0 &                                  0.0013 $\pm$ 0.0005 \\
         262.5 &            173.2 &                                  0.0018 $\pm$ 0.0006 \\
         267.5 &            179.5 &                                  0.0022 $\pm$ 0.0006 \\
         272.5 &            185.9 &                                  0.0008 $\pm$ 0.0005 \\
         277.5 &            188.0 &                                  0.0012 $\pm$ 0.0006 \\
         282.5 &            210.0 &                                  0.0006 $\pm$ 0.0005 \\
         287.5 &            200.0 &                                  0.0017 $\pm$ 0.0006 \\
         292.5 &            200.0 &                                  0.0004 $\pm$ 0.0006 \\
         297.5 &            205.0 &                                  0.0009 $\pm$ 0.0006 \\
\end{tabular}
\end{ruledtabular}
\end{table}

\newpage

\bibliography{fissum_etal}

\end{document}